\documentclass[11pt]{article}

\usepackage[margin=1in]{geometry} 
\usepackage{amsthm,amsmath,amssymb}
\usepackage{tikz}
\usetikzlibrary{shapes,decorations,arrows,calc,arrows.meta,fit,positioning}
\tikzset{
    -Latex,auto,node distance =1 cm and 1 cm,semithick,
    state/.style ={ellipse, draw, minimum width = 0.7 cm, fill = yellow!25},
    point/.style = {circle, draw, inner sep=0.04cm,fill,node contents={}},
    bidirected/.style={Latex-Latex,dashed},
    el/.style = {inner sep=2pt, align=left, sloped}
}

\usepackage[linesnumbered,ruled,vlined]{algorithm2e}
\SetKwInOut{KwIn}{Input}
\SetKwInOut{KwOut}{Output}

\usepackage{multirow}

\usepackage{apacite}
\usepackage{natbib}
\usepackage{booktabs,float,makecell}
\usepackage[colorlinks,citecolor=black,linkcolor=black]{hyperref}

\newtheorem{assumption}{Assumption}
\newtheorem{theorem}{Theorem}
\newtheorem{lemma}{Lemma}
\newtheorem{remark}{Remark}
\newtheorem{example}{Example}

\newcommand{\pr}{\mathbb{P}} 
\newcommand{\var}{\mathrm{var}}

\newcommand{\E}{\mathbb{E}}

\newcommand{\train}{\mathrm{train}}
\newcommand{\test}{\mathrm{test}}
\newcommand{\ate}{\mathrm{ATE}}

\title{Causal Inference on Sequential Treatments via Tensor Completion}

\author{
  Chenyin Gao\thanks{Department of Statistics, North Carolina State University, Raleigh, NC} 
  \and
  Han Chen\thanks{Department of Statistical Science, Duke University, Durham, NC} 
  \and
  Anru R. Zhang\thanks{Departments of Biostatistics \& Bioinformatics and Computer Science, Duke University, Durham, NC} 
  \and
  Shu Yang\thanks{Department of Statistics, North Carolina State University, Raleigh, NC}
}

\date{}

\begin{document}

\maketitle

\begin{abstract}
 Marginal Structural Models (MSMs) are popular for causal inference of sequential treatments in longitudinal observational studies, which however are sensitive to model misspecification. To achieve flexible modeling, we envision the potential outcomes to form a three-dimensional tensor indexed by subject, time, and treatment regime and propose a tensorized history-restricted MSM (HRMSM). The semi-parametric tensor factor model allows us to leverage the underlying low-rank structure of the potential outcomes tensor and exploit the pre-treatment covariate information to recover the counterfactual outcomes. We incorporate the inverse probability of treatment weighting in the loss function for tensor completion to adjust for time-varying confounding. Theoretically, a non-asymptotic upper bound on the Frobenius norm error for the proposed estimator is provided. Empirically, simulation studies show that the proposed tensor completion approach outperforms the parametric HRMSM and existing matrix/tensor completion methods.  Finally, we illustrate the practical utility of the proposed approach to study the effect of ventilation on organ dysfunction from the Medical Information Mart for Intensive Care database. 

\end{abstract}

\section{Introduction}\label{sec:intro}

Randomized trial studies have been recognized as the gold standard for learning causality due to the randomization of treatments that ensure the patients exposed to one treatment resemble those exposed to another in all possible pre-treatment covariates. However, inferring causality in longitudinal studies becomes more challenging due to the presence of sequential treatments and time-varying confounding. As explained by \cite{robins2000marginal2}, standard regression methods, whether or not adjusting for confounders, are fallible when coping with sequential treatments. In response, \textit{Marginal Structural Models} (MSMs) \citep{robins2000marginal} and \textit{Structural Nested Mean Models} (SNMMs) \citep{robins1994correcting} have been proposed for causal inference in longitudinal studies, targeting different causal questions. MSMs are tailored to estimate the population-level (or marginal) causal effects, whereas SNMMs allow causal effects to be dynamically influenced by time-varying confounders. In particular, MSMs are well-suited for assessing the average impact of treatments on populations, which is the primary focus of our paper. However, these frameworks are often stringently parametric, as the standard estimation (e.g., G-computation \citep{robins1986new}) typically relies on specific model assumptions for the outcomes, making them sensitive to model misspecification. In addition, they require a fixed time horizon setting and do not scale well to a prolonged time horizon. {There are several advanced, flexible machine learning methods built on encoder-decoder architectures to capture long-range, complex dependencies between outcomes and time-varying confounders (e.g., counterfactual recurrent network (CRN) \cite{bica2020time}, G-Net \citep{li2020g}, and Causal Transformer (CT) \citep{melnychuk2022causal}). However, most of them only account for the instantaneous treatment effect (i.e., the effect of the current treatment on the current outcome). For example, CT uses an adversarial objective to produce time-varying representations that are balanced only on the current treatment assignment, which neglects the lagged treatment effect often widely seen in medical practice and, therefore, impedes their deployment when a delayed treatment effect is present.}

\begin{figure}[!ht]
    \centering
\begin{tikzpicture}[-]
\fill[yslant=1,xslant=-1, gray!20] (2.47,2.00) rectangle +(.36,1);
\fill[yslant=1,xslant=-1, gray!20] (2.85,1.00) rectangle +(.37,1);
\fill[yslant=1,xslant=-1, gray!20] (3.25,2.00) rectangle +(.35,1);
\fill[yslant=1,xslant=-1, gray!20] (3.63,0.00) rectangle +(.36,1);

\fill[gray!20] (-1.54,-1.54) rectangle +(1,1);
\fill[gray!20] (-0.54,1.46) rectangle +(1,1);
\fill[gray!20] (1.47,0.46) rectangle +(1,1);
\fill[gray!20] (0.46,-0.54) rectangle +(1,1);

\fill[yslant=1,xslant=0, gray!20] (2.45,-2.00) rectangle +(.4,1);
\fill[yslant=1,xslant=0, gray!20] (3.65,-1.02) rectangle +(.36,1);

\foreach \x in{0,...,4}
{   \draw (0,\x ,4) -- (4,\x ,4);
    \draw (\x ,0,4) -- (\x ,4,4);
    \draw (4,\x ,4) -- (4,\x ,0);
    \draw (\x ,4,4) -- (\x ,4,0);
    \draw (4,0,\x ) -- (4,4,\x );
    \draw (0,4,\x ) -- (4,4,\x );
}
\node[] at (2,-.5,4) {treatment regime};
\node[] at (2,-1,4) {$l\in\{(0,0),(0,1), (1,0), (1,1)\}$};
\node[] at (-1,2,4) {subject};
\node[] at (-1.5,1.5,4) {$i\in\{1,2,3,4\}$};
\node[] at (0,5,4) {time};
\node[] at (-1,4.5,4) {$t\in\{1,2,3,4\}$};
\end{tikzpicture}
\caption{A illustration of potential outcomes tensor: a (subject$\times$time$\times$treatment regime) tensor of dimension $\mathbb{R}^{4\times4\times 2^2}$ with $N=4$, $T=4$ and $k=2$; 
the gray cells represent the observed entries, whereas the white cells represent the missing entries. \label{fig:tensor_illustration}}
\end{figure}
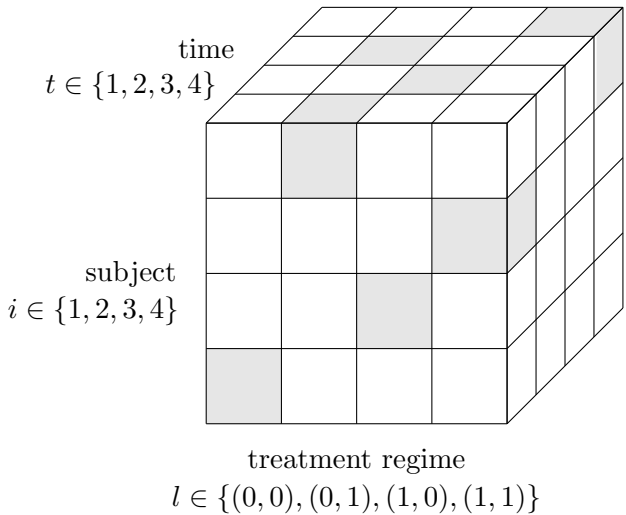

As causal inference is essentially a missing data analysis, it can be closely related to a problem of matrix or tensor completion, which aims at recovering the underlying low-rank or approximately low-rank structure from the observed samples. In the longitudinal observational studies with sequential treatments, we develop a novel tensorized history-restricted Marginal Structural Model (HRMSM), an extension of MSMs assuming the sequential treatments only influence the outcomes on a short-term basis \citep{neugebauer2007causal}. {In practice, such as in critical care settings like
MIMIC-III, clinical interventions often have effects that are most pronounced in the short
term due to rapidly changing patient states.} Our new perspective conceptualizes the potential outcomes as a three-dimensional tensor, with each tensor mode corresponding to the subject, time, and treatment regime (see Figure \ref{fig:tensor_illustration} for an illustration). Thus, the sequential treatment effect estimation problem is formulated as a tensor completion problem, allowing for the use of tensor structural information such as low rankness. In this paper, we substantiate the low rankness by the tensor factor model, i.e., the Tucker decomposition \citep{tucker1966some}. Hence, our proposal is model-free,  utilizing the data-adaptive latent factors to recover the missing potential outcomes (i.e., counterfactuals) and making it less susceptible to model misspecification. Furthermore, our model scales well to a large sample size and a prolonged time horizon. For instance, if the outcomes remain stable across subjects over time, the latent factors associated with time can be exploited to enhance the precision of the estimated causal effects \citep{abadie2010synthetic}.

Tensor completion is one of the most actively studied problems in tensor-related literature and has attracted broad attention in a wide range of applications  \citep{xu2013block, semerci2014tensor, xie2016topicsketch, mandal2019weighted}. Despite the intrinsic connection between causal inference and missing data analysis, most tensor completion methods cannot be directly employed as they often assume that the missingness occurs completely at random \citep{liu2012tensor,jain2014provable,barak2016noisy,xia2021statistically,cai2021nonconvex} or specially designed \citep{zhang2019cross}. However, the sequential treatment assignment in longitudinal studies can be an endogenous process, as its value may be affected by some time-dependent prognostic factors, rendering the independent Bernoulli missingness implausible. To properly handle the treatment endogeneity, we adopt sequential ignorability in Assumption \ref{assum:SRA}, where treatment is sequentially randomized given the historical information. Under the assumption of sequential ignorability, an inverse probability of treatment weighted (IPTW) tensor completion algorithm is proposed to impute the counterfactuals; see Section \ref{subsubsec:IPTW} for more details. Certainly, other types of ignorability assumptions are available to handle treatment endogeneity. One possible assumption is latent ignorability, where the potential outcomes are conditional independent of the treatments given the observable and the latent factors. Therefore, the relationships between the subject-specific latent factors can be leveraged to impute the other missing outcomes \citep{agarwal2022synthetic, agarwal2023causal}. Typically, they require that subject-specific latent factors satisfy the well-supported condition, which stipulates that the latent factors associated with one subject are expressible as a linear combination of other subjects. This concept is akin to the incoherence condition usually required for tensor completion, where each tensor entry contains a similar amount of information, enabling the recovery of any missingness from the observed tensor. However, the latent ignorability assumption might not be reasonable in our context, as this independence is ascertained given all time-related latent factors, including the future ones.

 Intuitively, the tensor completion problem can be solved with well-studied matrix completion algorithms by unfolding tensors into matrices \citep{tomioka2010estimation, gandy2011tensor, liu2012tensor}. Yet, such unfolding-based methods might discard the multi-way structure of the tensor and therefore lose additional correlational information among modes. To directly explore the tensor structures, \cite{yuan2016tensor, yuan2017incoherent} propose to use tensor nuclear norm minimization, which is guaranteed to be more efficient. However, the computation of tensor nuclear norm minimization is NP-hard in general, which may prevent its usage for real-data applications. One possible remedy is to restrict the feasible solutions in a sub-class of tensors, e.g., a class of low multi-linear rank tensors. Followed by a low-rank tensor space optimization algorithm, a polynomial-time computable estimator can be obtained \citep{xia2017polynomial, xia2021statistically,cai2021nonconvex,zhen2024nonnegative}.  Furthermore, most works consider the low-rank completion based on the outcomes tensor only, whereas much fewer works focus on incorporating the baseline covariates to assist tensor completion with exceptions including \cite{acar2011all}, \cite{zhou2017tensor}, \cite{ibriga2022covariate} and  \cite{zhou2021partially}. In our paper, we incorporate the baseline covariates into the proposed tensorized HRMSM using the sieve approximation to capture their relationship with the potential outcomes, motivated by the semi-parametric tensor factor model in \cite{chen2020semiparametric}. Hence, instead of searching for the optimum in the unrestricted low-rank tensor space, we minimize the weighted loss function via the projected gradient descent optimizer in the covariate-relevant functional space with dimensions endowed by the order of sieve basis, which is faster and more stable.

Our contributions are summarized as follows: 1) we propose a tensorized HRMSM framework based on the Tucker decomposition for modeling the potential outcomes, which does not rely on stringent parametric assumptions, allows the incorporation of the baseline covariates for tensor completion, and scales well to a large sample size and an infinite time horizon setting; 2) theoretically, we provide a non-asymptotic upper bound on the Frobenius norm error for the completed potential outcomes tensor, taking into account that the uncertainty in estimating the IPTW; and 3) empirically, we provide several synthetic experiments and a real-data application to highlight the benefits of the proposed framework comparing with existing competitors.

The rest of the paper is organized as follows. Section \ref{sec:CI-sequnetial} layouts the longitudinal observational study and a set of essential assumptions. Section \ref{sec:Tensor-completion} illustrates the low-rank tensor completion paradigm and describes the proposed tensorized HRMSM in detail. Section \ref{sec:Statistical-Guarantee} establishes the theoretical properties of our proposed method under the potential outcomes framework. A number of experiments are presented in Section \ref{sec:Simulations} with a real-data application in Section \ref{sec:Application}.
All the technical proofs and details are provided in the Appendix.

\section{Causal inference with sequential treatments}\label{sec:CI-sequnetial}
\subsection{Notation and causal assumptions}\label{subsec:assumption}

Let $X_{i,0}$ be a $d_{0}$-vector of pre-treatment covariates, and $(X_{i,t},Y_{i,t},A_{i,t})$
be the vector of time-varying covariates, an outcome variable and a binary treatment for subject
$i=1,\cdots, N$ at time $t=1,\cdots, T$. Next, we use the colon in
the indices to select a slice (both ends inclusive) of a vector; e.g., $a_{1:T}=(a_{1},a_{2},\ldots,a_{T})$. Following \cite{robins2000marginal}, let $X_{i,t}^{a_{1:T}}$ and $Y_{i,t}^{a_{1:T}}$ be the potential
 time-varying covariates and the potential outcome at $t$ had the subject $i$ followed the treatment trajectory $a_{1:T}\in \mathbb{A}_T$, where $\mathbb{A}_T$ includes all the possible treatment histories over $T$ time points. The full data for subject $i$ is summarized by $\mathbb{V}_{i,t}=\left\{(X_{i,0},X_{i,1:t}^{a_{1:T}}, Y_{i,1:t}^{a_{1:T}}):a_{1:T}\in\mathbb{A}_T\right\}$, and $\mathbb{V}_{i,t}$ is null for $t\leq0$ by convention. By the nature of the data collection procedure, in which $X_{i,t-1}$ actualizes before $A_{i,t}$, and $A_{i,t}$ actualizes before $Y_{i,t}$, we assume $X_{i,1:t}^{a_{1:T}}=X_{i,1:t}^{a_{1:t}}$ and $Y_{i,t}^{a_{1:T}}=Y_{i,t}^{a_{1:t}}$, that is, the future treatments will not affect the present variables. Also, as many treatments influence each subject on a short-term basis, the entire treatment history prior to an event occurrence may not be all relevant \citep{feldman2004administration,heron2010ecological}.
Following \cite{neugebauer2007causal}, we assume that the potential outcomes at time $t$ depends on the current treatment $a_t$ and the previous $(k-1)$ treatment assignments for $k\geq2$.

 \begin{assumption}[$k$-history restricted potential outcomes] \label{assum:HR} $Y_{i,t}^{a_{1:t}}=Y_{i,t}^{a_{(t-k+1):t}}$, where $k>0$ is a integer, for all $a_{1:t}$, $i=1,\cdots,N$ and $t=1,\cdots,T$.
 \end{assumption}

The fundamental problem in causal inference is that not all potential outcomes can be observed for a particular subject. Thus, the identification of causal effects requires further assumptions. Following \cite{van2005history}, we make the following assumptions.
 
\begin{assumption}[Causal consistency]
\label{assum:SUTVA} $X_{i,1:t}=X_{i,1:t}^{A_{i,1:t}}$ and $Y_{i,t}=Y_{i,t}^{A_{i,1:t}}$ for $i=1,\cdots,N$ and $t=1,\cdots,T$.
\end{assumption}

\begin{assumption}[Sequential treatment randomization]
$$
A_{i,t} 
\perp 
\left\{\left(X_{i, 0}, X_{i, 1:t}^{(A_{1: (t-1)}, a_{t: T})}, Y_{i, 1: t}^{A_{1: (t-1)}, a_{t: T}}\right): a_{t: T} \in \mathbb{A}_{T-t+1}\right\} \mid H_{i, t},
$$ where $H_{i,t}=(X_{i,0:(t-1)}$, $A_{i,1:(t-1)}$, $Y_{i,1:(t-1)})$,  for $i=1,\cdots,N$ and $t=1,\cdots,T$.\label{assum:SRA}
\end{assumption}

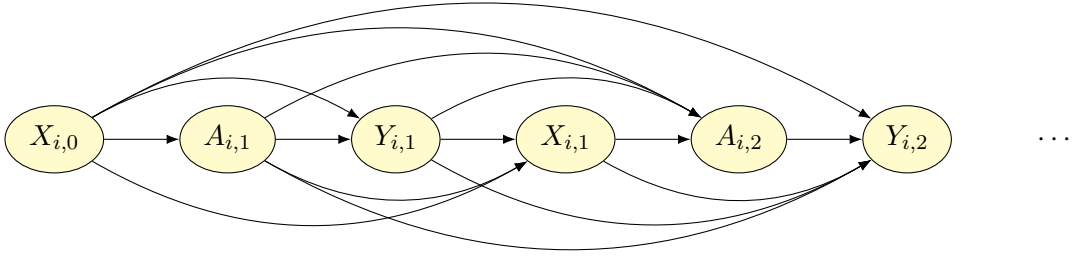
\begin{figure}[!t]
\centering
\begin{tikzpicture}
    \node[state] (X1) at (0,0) {$X_{i,0}$};
    \node[state] (A1) [right =of X1] {$A_{i,1}$}; 
    \node[state] (Y1) [right =of A1] {$Y_{i,1}$};
\node[state] (X2) [right =of Y1] {$X_{i,1}$};
\node[state] (A2) [right =of X2] {$A_{i,2}$};
\node[state] (Y2) [right =of A2] {$Y_{i,2}$};
\node[] () [right =of Y2] {$\cdots$};
    \path (X1) edge (A1);
    \path (A1) edge (Y1);
    \path (Y1) edge (X2);
    \path (X2) edge (A2);
    \path (A2) edge (Y2);
    
    \path (X1) [bend left=30] edge (Y1);
    \path (X1) [bend right=30] edge (X2);
    \path (X1) [bend left=30] edge (A2);
    \path (X1) [bend left=30] edge (Y2);
    
    \path (A1) [bend right=30] edge (X2);
    \path (A1) [bend left=30] edge (A2);
    \path (A1) [bend right=30] edge (Y2);
    
    \path (Y1) [bend left=30] edge (A2);
    \path (Y1) [bend right=30] edge (Y2);
    
    \path (X2) [bend right=30] edge (Y2);
\end{tikzpicture}
\caption{A directed acyclic graph illustrates the data-generating process of $(X_{i,0:T}$, $A_{i,1:T}$, $Y_{i,1:T})$ for individual $i$ under Assumption \ref{assum:SRA}.
\label{fig:causal_graph}}
\end{figure}

\begin{assumption}
\label{assum:positivity} Let $e(H_{i,t})=\mathbb{P}(A_{i,t}=1\mid H_{i,t})$
and there exists a constant $\delta<0.5$ such that we have $0<\delta<$ $e(H_{i,t})<1-\delta<1$ for $t=0,\cdots,T$
almost surely.
\end{assumption}

Assumption \ref{assum:SUTVA} is necessary to consistently map the observed data with the potential outcomes without any interference between subjects. 
Assumption \ref{assum:SRA} implies that the treatment assignment
at any time $t$ is randomized among the subjects sharing the same value
of the observed history accrued up
to time $t$, also known as treatment ignorability or
no unmeasured confounders.
In the observational studies, Assumption \ref{assum:SRA} holds if the observed history contains all predictors of treatment assignment and outcome; see an example of a causal graph satisfying Assumption \ref{assum:SRA} in Figure \ref{fig:causal_graph}.
Assumption \ref{assum:positivity} is also known as
the uniform overlap condition \citep{rosenbaum1983central}, which ensures an adequate overlap between the treatment and control covariate distributions and allows each subject to have probability at least $\delta$ of receiving either treatment at each time point $t$. The parameter $\delta$ will play an important role in our main theoretical results. To provide some intuition, the potential outcomes under any treatment regime will have a positive probability of being observed. Thus, the recovery of any unobserved potential outcomes will be possible.

\subsection{Marginal structural models}\label{subsec:MSM}

\textit{Marginal structural models} (MSMs) are first established in \citet{robins1986new} and rapidly become one of the most important types of models
for causal inference in the presence of time-varying confounding. The MSMs impose parametric structural models for $Y_{i,t}^{a_{1:t}}$, which describe the marginal effect of time-varying treatments and possibly
adjust for baseline treatment modification effects. Under Assumptions
\ref{assum:HR}-\ref{assum:positivity}, we have 
\begin{equation}
\E\{Y_{i,t}^{a_{(t-k+1):t}}\mid X_{i,0}\}=\mu(a_{(t-k+1):t},X_{i,0}),\label{eq:MSM}
\end{equation}
where $\mu(\cdot)$ is an unknown function of the past $k$ treatments $a_{(t-k+1):t}\in \mathbb{A}_k$
and the baseline covariate $X_{i,0}$, termed as the history-restricted MSM (HRMSM), where $\mathbb{A}_k$ includes all the possible treatment regimes over $k$ time points \citep{neugebauer2007causal}. Unlike the traditional MSMs, HRMSM only considers the treatments assigned between time $t-k+1$ and $t$, instead of the entire treatment history $a_{1:t}$.

Let the time-specific IPTW estimating function be $\Psi(H_{t}\mid\mu,\eta_{t})$ with a parametric structural model $\mu(a_{(t-k+1):t},X_{i,0};\eta_{t})$ characterized by a fix-dimensional unknown parameter $\eta_{t}$. Under Assumptions \ref{assum:HR}-\ref{assum:positivity}, the IPTW can be used to consistently estimate $\eta_{t}$ as in
\begin{equation}
\Psi(H_{i,t}\mid\mu,\eta_{t})=w_{i,t}\frac{\partial\mu(A_{(t-k+1):t},X_{i,0};\eta_{t})}{\partial\eta_{t}}\left\{ Y_{i,t}-\mu(A_{(t-k+1):t},X_{i,0};\eta_{t})\right\} ,\label{eq:MSM_estimating_equation}
\end{equation}
with weights $w_{i,t}=\left[\prod_{j=t-k+1}^{t}e(H_{i,j})^{A_{i,j}}
\{1-e(H_{i,j})\}^{1-A_{i,j}}\right]^{-1}$. Therefore, the weight $w_{i,t}$ can be perceived as the inverse probability that subject $i$ received his own observed treatment regime $A_{i,(t-k+1):t}$
given his accrued history information $H_{i,t}$. By accounting for the statistic endogeneity of the treatment process $A_{i,t}$ with time-dependent confounders $X_{i,t}$, one can estimate the potential outcomes under any series of treatment regimes and establish a valid causal relationship. Example \ref{rmk:MSM_model} is provided for an illustration.
\begin{example}\label{rmk:MSM_model}
Suppose that $X_{i,0}\in \mathbb{R}^{d}$, $\mathbb{E}\{Y_{i, t}^{a_{(t-k+1): t}} \mid X_{i, 0}\}=X_{i, 0}^{\intercal}\eta_{1}+\mathrm{cum}(a_{(t-k+1): t}) \eta_{2}+t \eta_{3}$, where $\mathrm{cum}(a_{(t-k+1): t})$ is the number of active treatments of the regimes $a_{(t-k+1): t}$, the HRMSM can be posited by $\mu(a_{(t-k+1): t}, X_{i, 0} ; \eta)=\{X_{i, 0}^{\intercal}, \mathrm{cum}(a_{(t-k+1): t}), t\} \eta$ with $\eta = (\eta_{1}, \eta_{2}, \eta_{3})$ for $t=1, \cdots, T$.
\end{example}

Despite the appealing strengths of the HRMSM, its estimation error heavily hinges on the outcome model specification,
which is usually restrictive and error-prone in practice. Although many other robust variants of MSMs have been formulated
\citep{van2003unified,mortimer2005application,yu2006double}, most of them exploit the latent factors over time in a stringent manner and thus do not scale well to large $T$. 
On the other hand, flexible machine learning techniques such as tensor completion tools have been proposed to handle missing data.  Given the close relationship between causal inference and missing data analysis, one is motivated to see whether a flexible and efficient estimation can be achieved via low-rank tensor completion.

\section{Tensor completion to causal inference}\label{sec:Tensor-completion}

\subsection{Preliminaries}

Before we delve into the details of tensor completion, we introduce
some basic notations, preliminaries, and tensor algebra, which can
be helpful for our subsequent illustration. We use lowercase letters,
e.g., $x,y$, to denote scalars, and the bold lowercase letters, e.g.,
$\boldsymbol{x},\boldsymbol{y}$, are used to denote vectors. For any
two vectors $\boldsymbol{x}\in\mathbb{R}^{p_{1}},\boldsymbol{y}\in\mathbb{R}^{p_{2}}$,
let $\boldsymbol{x}\otimes\boldsymbol{y}\in\mathbb{R}^{p_{1}\times p_{2}}$
denotes their outer product. The bold uppercase letter such as $\boldsymbol{X},\boldsymbol{Y}$
are used to denote matrices, and the calligraphic letters $\mathcal{X},\mathcal{Y}$
are used to denote higher-order tensors. For any two tensors, say
$\mathcal{X},\mathcal{Y}\in\mathbb{R}^{p_{1}\times p_{2}\times p_{3}}$
, let $\mathcal{X}\odot\mathcal{Y}$ denotes their entry-wise product
$(\mathcal{X}\mathcal{Y})_{i_{1},i_{2},i_{3}}=\mathcal{X}_{i_{1},i_{2},i_{3}}\mathcal{Y}_{i_{1},i_{2},i_{3}}$,
and $\langle\mathcal{X},\mathcal{Y}\rangle$ denotes their inner product
$\langle\mathcal{X},\mathcal{Y}\rangle=\sum_{i_{1},i_{2},i_{3}}\mathcal{X}_{i_{1},i_{2},i_{3}}\mathcal{Y}_{i_{1},i_{2},i_{3}}$. Fibers and slices are two subarrays of tensors formed by fixing a subset of the indices. In particular, we define the fibers by $\mathcal{X}_{:,i_2,i_3}\in\mathbb{R}^{p_1},\mathcal{X}_{i_1,:,i_3}\in\mathbb{R}^{p_2},\mathcal{X}_{i_1,i_2,:}\in\mathbb{R}^{p_3}$,
and the slices by $
\mathcal{X}_{i_1,:,:}\in\mathbb{R}^{p_2\times p_3},\mathcal{X}_{:,i_2,:}\in\mathbb{R}^{p_1\times p_3},\mathcal{X}_{:,:,i_3}\in\mathbb{R}^{p_1\times p_2}$.

Then, we introduce the unfolding (or the matricization) operator $\mathcal{M}_{k}(\cdot)$
that transforms tensors to matrices on mode-$k$. For example, after
unfolding the tensor $\mathcal{X}$ on its first mode, we have $\mathcal{M}_{1}(\mathcal{X})\in\mathbb{R}^{p_{1}\times p_{2}p_{3}}$, where $[\mathcal{M}_{1}(\mathcal{X})]_{i_{1},i_{2}+p_{2}(i_{3}-1)}=\mathcal{X}_{i_{1},i_{2},i_{3}}$. Also, we introduce the mode-$k$ tensor-matrix products. For example,
let $\boldsymbol{U}_{1}\in\mathbb{R}^{r_{1}\times p_{1}}$, the mode-$1$
tensor-matrix multiplication is $\mathcal{X}\times_{1}\boldsymbol{U}_{1}\in\mathbb{R}^{r_{1}\times p_{2}\times p_{3}}$, where $(\mathcal{X}\times_{1}\boldsymbol{U}_{1})_{i_{1},i_{2},i_{3}}=\sum_{j_{1}=1}^{p_{1}}\mathcal{X}_{j_{1},i_{2},i_{3}}\boldsymbol{U}_{i_{1},j_{1}}$.
General mode-$k$ matricization and mode-$k$ tensor-matrix products
can be defined in a similar manner. The multi-linear rank of $\mathcal{X}$
is defined as the vector $(r_{1},r_{2},r_{3})=\{\text{rank}(\mathcal{M}_{1}(\mathcal{X}))$,
$\text{rank}(\mathcal{M}_{2}(\mathcal{X}))$, $\text{rank}(\mathcal{M}_{3}(\mathcal{X}))\}$.
For any $(r_{1},r_{2},r_{3})$ multi-linear rank tensor $\mathcal{X}$
in general, it can be factorized by the Tucker decomposition \citep{tucker1966some}
into a small core tensor $\mathcal{G}\in\mathbb{R}^{r_{1}\times r_{2}\times r_{3}}$
and three matrices $\boldsymbol{U}_{i}\in\mathbb{R}^{p_{i}\times r_{i}}$
with orthogonal columns, denoted as
\begin{equation}
\mathcal{X}=\mathcal{G}\times_{1}\boldsymbol{U}_{1}\times_{2}\boldsymbol{U}_{2}\times_{3}\boldsymbol{U}_{3}=[\![\mathcal{G};\boldsymbol{U}_{1},\boldsymbol{U}_{2},\boldsymbol{U}_{3}]\!],\label{eq:Tucker}
\end{equation}
where $\mathcal{G}$ can be considered as the principal components
with $\boldsymbol{U}_{i}$ being the mode-$i$ loading matrices.
Next, as tensors are generalizations of matrices, several important
matrix norms can be generalized under the tensor formulation \citep{landsberg2012tensors,hogben2013handbook}.
Let the spectral norm of a tensor $\mathcal{X}$ be $\|\mathcal{X}\|=\max\{\langle\mathcal{X},\boldsymbol{a}\otimes\boldsymbol{b}\otimes\boldsymbol{c}\rangle:\boldsymbol{a}\in\mathbb{R}^{p_{1}},\boldsymbol{b}\in\mathbb{R}^{p_{2}},\boldsymbol{c}\in\mathbb{R}^{p_{3}},\|\boldsymbol{a}\|=\|\boldsymbol{b}\|=\|\boldsymbol{c}\|=1\}$, where we use $\|\cdot\|$, with a little abuse of notations, to denote
the spectral norm of a tensor and a matrix and the Euclidean norm
of a vector. Likewise, the nuclear norm is $\|\mathcal{X}\|_{*}=\max\{\langle\mathcal{X},\mathcal{Y}\rangle:\mathcal{Y}\in\mathbb{R}^{p_{1}\times p_{2}\times p_{3}},\|\mathcal{Y}\|=1\}$. Besides, let $\|\mathcal{X}\|_{F}=\sqrt{\langle\mathcal{X},\mathcal{X}\rangle}$
and $\|\mathcal{X}\|_{\max}=\max_{i_{1},i_{2},i_{3}}|\mathcal{X}_{i_{1},i_{2},i_{3}}|$
be the Frobenius norm and max norm of a tensor, respectively; see
\citet{kolda2009tensor} for a more comprehensive review. We use $c_{0},C_{0}$
to denote generic positive constants, whose actual values may vary
from line to line. 

\subsection{Tensor and model formulation}

\subsubsection{Low-rank tensorized HRMSM with baseline covariates}\label{subsubsec:low-rank}

To address these concerns, we envision the potential outcomes for $N$ subjects at $T$ times to form an $N\times T\times K$ tensor of three modes with entries $\mathcal{Y}_{i,t,l}=Y_{i,t}^{l_{(b)}}$, where $K=2^k$ and $l_{(b)}$ converts $l$ from decimal to binary. Likewise, $\{a_{i,(t-k+1):t}\}_{(10)}=l$
is for decimal conversion. Take $k=5$ as an example, we have $(3)_{(b)}=00011$
and $(00101)_{(10)}=5$. Therefore, the potential outcomes can be represented as a tensor of three modes, i.e., subject $\times$
time $\times$ treatment regime. Next, we assume that $\mathcal{Y}$ admits the Tucker decomposition in
(\ref{eq:Tucker}) with noises, and the low-rank tensorized HRMSM is formulated by 
\begin{equation}
\mathcal{Y}=
\mathcal{Y}^* + \mathcal{E}=
[\![\mathcal{G}^{*};\boldsymbol{U}_{1}^{*},\boldsymbol{U}_{2}^{*},\boldsymbol{U}_{3}^{*}]\!]+\mathcal{E},\label{eq:Tucker-Y}
\end{equation}
where $\mathcal{G}^{*}\in\mathbb{R}^{r_{1}\times r_{2}\times r_{3}}$,
$\boldsymbol{U}_{1}^{*}\in\mathbb{R}^{N\times r_{1}}$, $\boldsymbol{U}_{2}^{*}\in\mathbb{R}^{T\times r_{2}}$,
$\boldsymbol{U}_{3}^{*}\in\mathbb{R}^{K\times r_{3}}$, and all entries
of $\mathcal{E}$ are independent mean-zero sub-Gaussian random variables with $\var(\mathcal{E}_{i,t,l})\leq\sigma^{2}$. Various causal estimands can be easily formulated with the potential outcomes tensor $\mathcal{Y}$ to assess the treatment effects. Let the causal estimand of our interest be the average treatment effect (ATE) between treatments $l_{(b)}$ and $l'_{(b)}$, defined as ${\tau}_{\ate}^{l,l'}=(NT)^{-1}\E[\sum_{i=1}^N \sum_{t=1}^T \{Y_{i,t}^{l_{(b)}}-Y_{i,t}^{l'_{(b)}}\}]$. It can be readily formulated by
$\tau_{\ate}^{l,l'} = \mathcal{Y}^* \times_1 N^{-1}\mathbf{1}_N 
\times_2 T^{-1}\mathbf{1}_T
\times_3 \{e_l(K) - e_{l'}(K)\}$ with the tensor representation, where { $e_l(K)$ is a $K$-dimensional zero vector except its $l$-entry being 1.} 

Intuitively, the Tucker decomposition can be interpreted as the
cluster heterogeneity or latent factors along each mode of $\mathcal{Y}$. For starters, the heterogeneity among subjects may be attributed to $r_{1}$ hidden factors, e.g., genetic or demographic factors. Second, the temporal variability may be dictated by a few time categories since diagnosis (e.g., early, middle, and late stages). Lastly, the variation of the sequential treatment regimes may be summarized into a few key factors, such as the number of cumulated treatments in the past. These latent factors, $\boldsymbol{U}_{1}$, $\boldsymbol{U}_{2}$, and $\boldsymbol{U}_{3}$, are constructed in a data-adaptive way, making them more flexible and robust. For instance, suppose that $\mathcal{Y}^*$ depends on some unknown function of the covariates $X_{0}$. Our tensorized HRMSM can still accurately approximate the potential outcomes through the inferred factors $\boldsymbol{U}_1$, without knowing the specific functional form, whereas the traditional MSMs are fallible if misspecified; see our simulation studies for more empirical evidence. {To clarify our choice of tensor decomposition, we briefly compare Tucker decomposition with two widely used alternatives: the Canonical Polyadic (CP) and Tensor Train (TT) decompositions. CP decomposition imposes a common rank across all modes, which is overly restrictive for our data, where the latent structures underlying patients, time points, and treatment regimes differ substantially in complexity. TT decomposition is most suitable for higher-order tensors whose modes themselves admit an inherent sequential ordering. In our case, although the time mode has its own internal index order, the tensor is third-order and the three modes (time, treatment regimes, subjects) do not possess a meaningful sequence among the modes, so this requirement is not satisfied. In contrast, Tucker decomposition offers the flexibility of assigning different ranks to each mode and produces interpretable mode-specific latent factors. These properties make Tucker decomposition well aligned with both the structure of our tensor representation and the scientific goals of our analysis.} 

\begin{figure}[!ht]
\centering\includegraphics[width=0.9\linewidth]{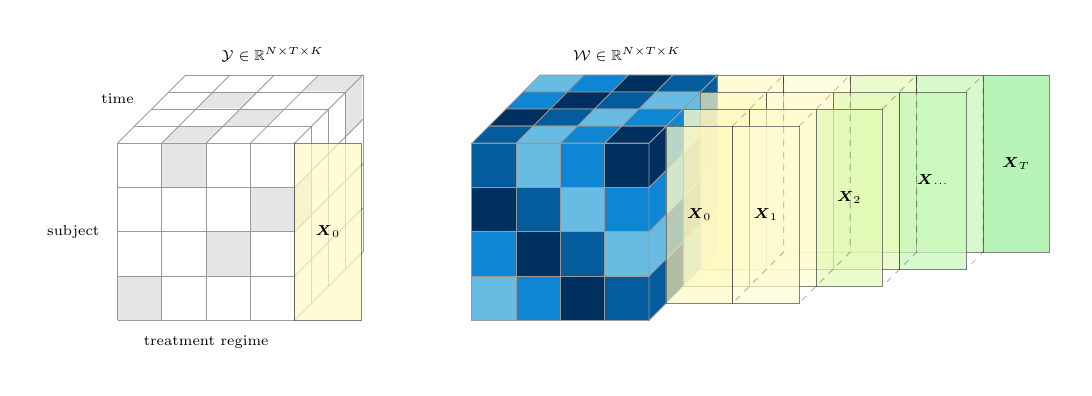}
\caption{\label{fig:Illustration} (Left) The potential outcomes tensor $\mathcal{Y}\in\mathbb{R}^{N\times T\times K}$
coupled with baseline covariates $\boldsymbol{X}_{0}$; (Right) The weight
tensor $\mathcal{W}\in\mathbb{R}^{N\times T\times K}$ coupled with
baseline covariates $\boldsymbol{X}_{0}$ and time-varying covariates
$\boldsymbol{X}_{1:T}$; the gray cells represent the observed entries
and the shades of blue cells represent the weight intensities.}
\end{figure}

Besides, the individual baseline covariates, such as age, gender, race, and general
health status stored in $\boldsymbol{X}_{0}\in\mathbb{R}^{N\times d_{0}}$ may assist tensor completion as it provides the covariate characteristic
on the subject mode, leading to a covariate-assisted tensorized HRMSM; see Figure \ref{fig:Illustration} (left). For concreteness, we assume the latent components associated with the subject-mode loading matrix $\boldsymbol{U}_{1}^{*}$ can be partially explained by the
matrix $\boldsymbol{X}_{0}$, or more generally by $G^{*}(\boldsymbol{X}_{0})$
based on an unknown function $G^{*}:\mathbb{R}^{N\times d_{0}}\rightarrow\mathbb{R}^{N\times r_{1}}$
. Equivalently, we have $\boldsymbol{U}_{1}^{*}=G^{*}(\boldsymbol{X}_{0})+\boldsymbol{\Gamma}$,
where $\boldsymbol{\Gamma}$ is the residual component of $\boldsymbol{U}_{1}^{*}$
after accounting for $G^{*}(\boldsymbol{X}_{0})$, {and $\{G^{*}(\boldsymbol{X}_{0})\}^\intercal \boldsymbol{\Gamma} = 0$.} {Since both $G^{*}(\boldsymbol{X}_{0})$ and $\boldsymbol{\Gamma}$ have orthonormal columns, we have $(\boldsymbol{U}_{1}^{*})^\intercal \boldsymbol{U}_{1}^{*} = I_N$ by construction.} This semi-parametric model is found in the
matrix/tensor factor analysis literature \citep{bai2012statistical,fan2016projected,chen2020semiparametric}. The key idea for estimating the unknown $G^{*}(\boldsymbol{X}_{0})$
relies on the sieve approximation \citep{chen2007large}. Let $\Phi(\boldsymbol{X}_{0})\in\mathbb{R}^{N\times d_{\Phi}}$
be the user-specified basis functions (e.g., B-spline, polynomial series), where each row corresponds to the subject and each column corresponds to a sieve basis. Let $\boldsymbol{B}\in\mathbb{R}^{d_{\Phi}\times r_{1}}$ be the unknown sieve coefficients 
matrix and $\boldsymbol{R}\in\mathbb{R}^{N\times r_{1}}$
be the residual matrix, we have $G^{*}(\boldsymbol{X}_{0})=\Phi(\boldsymbol{X}_{0})\boldsymbol{B}+\boldsymbol{R}$. {Importantly, the error terms $\boldsymbol{\Gamma}$ and $\boldsymbol{R}$ arise due to different reasons and cannot be absorbed into one single term, where $\boldsymbol{\Gamma}$ represents the system error unexplained by $\boldsymbol{X}_{0}$ while $\boldsymbol{R}$ represents the approximation error originating from the sieve basis approximation.}
Thus, the proposed covariate-assisted tensorized HRMSM does not involve restrictive parametric assumptions. We modify the running example in Example \ref{rmk:MSM_model} for an illustration. 
\begin{example}\label{rmk:t-HRMSM}
Suppose that $X_{i,0}\in \mathbb{R}^{d}$, $\mathbb{E}\{Y_{i, t}^{a_{(t-k+1): t}} \mid X_{i, 0}\}=X_{i, 0}^{\intercal}\eta_{1}+(X_{i, 0}^2)^\intercal \eta_{2}+ 
    t \eta_{3} + \operatorname{cum}(a_{(t-k+1): t}) \eta_4 +
    \operatorname{cum}(a_{(t-k+1): t}) X_{i, 0}^\intercal\eta_{5}${, where $X_{i,0}^2$ is the coordinate-wise square of $X_{1,0}$}. Let the basis function be constructed via polynomials up to the 2-nd order as $\Phi(X_{i, 0})=\{1, X_{i, 0}^\intercal, (X_{i, 0}^2)^{\intercal}\}^{\intercal}$. Then, we have
$$
\mathbb{E}\left\{Y_{i, t}^{a_{(t-k+1): t}} \mid X_{i, 0}\right\}=\mathbb{E}\left(\mathcal{Y}_{i, t, l} \mid X_{i, 0}\right)=\mathcal{G} \times_1 \Phi\left(X_{i, 0}\right)^{\intercal} \times_2(1, t) \times_3\left\{1, \operatorname{cum}(l_{(b)})\right\},
$$
where
$$
\mathcal{G}_{1, ;,:}=\left(\begin{array}{ll}
0 & \eta_4 \\
\eta_3 & 0
\end{array}\right), \quad \mathcal{G}_{1+j, ;,:}=\left(\begin{array}{ll}
\eta_{1,j} & \eta_{5,j} \\
0 & 0
\end{array}\right), \quad \mathcal{G}_{1+d+j, ;,:}=\left(\begin{array}{ll}
\eta_{2,j} & 0 \\
0 & 0
\end{array}\right),\quad 
j=1,\cdots, d,
$$
and therefore is encompassed by our tensorized HRMSM in (\ref{eq:Tucker-Y}). 
\end{example}

\subsubsection{Tensor completion and inverse probability treatment weighting}\label{subsubsec:IPTW}

Due to the fundamental problem in the causal inference that only the potential outcomes under the actualized treatment can be observed, only $N\times T$ realizations of the potential outcomes are observed. To link our setup to the tensor completion
literature, we define the observation pattern tensor as $\Omega$,
where $\Omega_{i,t,l}=1$ if $l=\{A_{i,(t-k+1):t}\}_{(10)}$
and zero otherwise. Hence, only the potential outcomes $\mathcal{Y}_{i,t,l}$
whose associated subjects have actually received the treatment regime $l_{(b)}$ are observed. Despite recent advances in the tensor completion literature \citep{xia2021statistically,cai2021nonconvex}, most of them typically assume uniform missingness, that is, each entry of the observation pattern $\Omega$ follows independent and identical Bernoulli distributions. However, this assumption is unrealistic in causal inference with observational studies where the missingness of potential outcomes is tied to the actual treatment uptake, and the time-varying confounders can affect the treatment assignment.

Let 
$\mathcal{W}$ be the propensity weight tensor with entries $\mathcal{W}_{i,t,l}=\mathbb{P}\{A_{i,(t-k+1):t}=l_{(b)}\mid H_{i,t}\}^{-1}$, which is the inverse product of a series of propensity scores. To obtain consistent causal estimation via tensor completion, the IPTW can be used to create a pseudo-population, where the missingness occurs randomly. In particular, we propose to recover the potential outcomes tensor by  minimizing the following weighted loss function:
\begin{equation}
\min_{\mathcal{G},\boldsymbol{U}_{1,X},\boldsymbol{U}_{2},\boldsymbol{U}_{3}} L(\mathcal{G};\boldsymbol{U}_{1,X},\boldsymbol{U}_{2},\boldsymbol{U}_{3};\mathcal{W})= \min_{\mathcal{G},\boldsymbol{U}_{1,X},\boldsymbol{U}_{2},\boldsymbol{U}_{3}}{ \frac{1}{2}}\|\sqrt{{\mathcal{W}}}\odot\boldsymbol{P}_{\Omega}(\mathcal{Y}-[\![\mathcal{G};\boldsymbol{U}_{1,X},\boldsymbol{U}_{2},\boldsymbol{U}_{3}]\!])\|_{F}^{2},\label{eq:loss_1}
\end{equation}
where { $\sqrt{\mathcal{W}}$ is the entry-wise square root of $\mathcal{W}$,} $\boldsymbol{U}_{1,X}$ lies in the column space of $\Phi(\boldsymbol{X}_{0})$ and
$\boldsymbol{P}_{\Omega}(\cdot)$ is the projection operator of any tensor
onto the subspace of tensors whose entries vanish if $\Omega_{i,t,l}=0$. 

In practice, the weight tensor $\mathcal{W}$ is often unknown and is replaced by the estimated weight tensor $\widehat{\mathcal{W}}$, yielded by the inverse product of the estimated propensity scores. As the time-varying confounders for each time point $t$ are coupled with the weight tensor $\mathcal{W}$ along the subject mode; see Figure \ref{fig:Illustration} (right), the collected historical information $H_{i,t}$ can be utilized to assist in reasoning about the sequential treatment mechanism, i.e., the propensity weights; see Remark \ref{rmk:estimating_PS} for an example of weight tensor estimation.
\begin{remark}
\label{rmk:estimating_PS} 
The most popular propensity score estimation is to use maximum likelihood estimation
(MLE) at each time period. However,
the dimension of $H_{i,t}$ keeps increasing as time $t$
proceeds, which may lead to unstable estimation. Toward this end, a regularized maximum likelihood estimation can be adopted with a proper
penalty, e.g., the Lasso penalty  
and
smoothly clipped absolute deviation (SCAD) penalty. 
The penalized negative log-likelihood under the logistic modeling $e(H_{i,t};\alpha_t)$ is
\begin{align}
\widehat{\alpha}_t=& \min_{\alpha_t=(\alpha_{0,t},\alpha_{1,t})} \frac{1}{N}\sum_{i=1}^{N}\left\{ \log(1+e^{\alpha_{0,t}+\alpha_{1,t}H_{i,t}})-A_{i,t}(\alpha_{0,t}+\alpha_{1,t}H_{i,t})\right\} +p_{\lambda_{\alpha}}(|\alpha_{1,t}|)\nonumber \\
 & =L_w(\alpha_{t})+p_{\lambda_{\alpha}}(|\alpha_{1,t}|),\label{eq:penalized_Q_PS}
\end{align}
where $L_w(\alpha_{t})$ can be replaced with other loss function, e.g.,
the generalized quasi-likelihood function \citep{ning2020robust} and the covariate imbalance loss function \citep{lee2021improving},
and $p_{\lambda_{\alpha}}(\cdot)$ is some penalization function gauged by
the tuning parameter $\lambda_{\alpha}$. Hence, the estimated weight tensor $\widehat{\mathcal{W}}=(\widehat{\mathcal{W}}_{i,t,l})\in\mathbb{R}^{N\times T\times K}$ is
$$\widehat{\mathcal{W}}_{i,t,l}=\left[\prod_{j=t-k+1}^{t}e(H_{i,j};\widehat{\alpha}_j)^{A_{i,j}}
\{1-e(H_{i,j};\widehat{\alpha}_j)\}^{1-A_{i,j}}\right]^{-1},$$
where $l=\{A_{i,(t-k+1):t}\}_{(10)}$.
\end{remark}

\subsubsection{Algorithm details}\label{subsubsec:algorithm}
For implementing the minimization problem in (\ref{eq:loss_1})  with $\mathcal{W}$ replaced by $\widehat{\mathcal{W}}$, we will use the projected gradient descent (PGD) algorithm to find the optimal solution $(\widehat{\mathcal{G}},\widehat{\boldsymbol{U}}_{1,X},\widehat{\boldsymbol{U}}_{2},\widehat{\boldsymbol{U}}_{3})$. 
The idea of PGD is to repeatedly update the current solution in the opposite direction of the partial gradients of the function $L(\cdot)$ while projecting the updated solution onto a restricted space to satisfy any desired constraints. In particular, we first compute the partial gradients of $L(\cdot)$ on $(\mathcal{G},\boldsymbol{U}_{1,X},\boldsymbol{U}_{2},\boldsymbol{U}_{3})$ at each iteration step $I$:
\begin{align*}
 & \frac{\partial L}{\partial\mathcal{G}}=\nabla L\times_{1}\boldsymbol{U}_{1,X}^{\intercal}\times_{2}\boldsymbol{U}_{2}^{\intercal}\times_{3}\boldsymbol{U}_{3}^{\intercal},\quad \frac{\partial L}{\partial\boldsymbol{U}_{1,X}}=\mathcal{M}_{(1)}(\nabla L)(\boldsymbol{U}_{3}\otimes\boldsymbol{U}_{2})\mathcal{M}_{(1)}(\mathcal{G})^{\intercal},\\
 & \frac{\partial L}{\partial\boldsymbol{U}_{2}}=\mathcal{M}_{(2)}(\nabla L)(\boldsymbol{U}_{3}\otimes\boldsymbol{U}_{1,X})\mathcal{M}_{(2)}(\mathcal{G})^{\intercal},\quad \frac{\partial L}{\partial\boldsymbol{U}_{3}}=\mathcal{M}_{(3)}(\nabla L)(\boldsymbol{U}_{2}\otimes\boldsymbol{U}_{1,X})\mathcal{M}_{(3)}(\mathcal{G})^{\intercal},
\end{align*}
where $\nabla L=\nabla L(\mathcal{X})=\mathcal{\widehat{W}}\odot\Omega\odot(\mathcal{X}-\mathcal{Y})$ and $\mathcal{X}=\mathcal{G}\times_1 \boldsymbol{U}_{1,X}\times_2 \boldsymbol{U}_{2}\times_3 \boldsymbol{U}_{3}$.  Next, we update $(\mathcal{G}^{(I)},\boldsymbol{U}_{1,X}^{(I)},\boldsymbol{U}_{2}^{(I)},\boldsymbol{U}_{3}^{(I)})$ by subtracting $\eta \cdot \partial L/\partial(\mathcal{G}^{(I)},\boldsymbol{U}_{1,X}^{(I)},\boldsymbol{U}_{2}^{(I)},\boldsymbol{U}_{3}^{(I)})$, which moves the current iterate towards the opposite direction of partial gradients with step size $\eta$, denoted by $(\mathcal{G}^{(I+1)},\boldsymbol{U}_{1}^{(I+1)},\boldsymbol{U}_{2}^{(I+1)},\boldsymbol{U}_{3}^{(I+1)})$. Since the PGD is an iterative optimization algorithm for finding a local minimum, large step sizes may overstep the local minima. Therefore, the choice of the step size $\eta$ lies at the heart of the PGD as it should not exceed the inverse Lipschitz constant of
the gradient. {In practice, we employ a line search strategy to adaptively set $\eta$ during optimization.} However, the updated $\boldsymbol{U}_{1}^{(I+1)}$ might not lie in the column space of $\Phi(\boldsymbol{X}_0)$ as desired, and thus an extra projection operator is included at each iterative step, which projects $\boldsymbol{U}_{1}^{(I+1)}$ onto the restricted tensor space, denoted by $\boldsymbol{U}_{1, X}^{(I+1)}$. {For theoretical reasons in Section \ref{sec:Statistical-Guarantee}, the updated $\boldsymbol{U}_2^{(I+1)}$ and $\boldsymbol{U}_3^{(I+1)}$ are projected to the Stiefel manifold to enforce column orthogonality.} A summary of the full algorithm is provided in Algorithm \ref{alg:PGD}.

\begin{algorithm}[!ht]
 \caption{\label{alg:PGD} Projected gradient descent for minimizing (\ref{eq:loss_1})}
\textbf{Input:} The observed outcome tensor $\mathcal{Y}$, estimated
weight tensor $\widehat{\mathcal{W}}$, the observation pattern $\Omega$,
the basis functions $\Phi(\boldsymbol{X}_{0})$, the multi-linear rank
$(r_{1},r_{2},r_{3})$, maximal number of iterations $I_{\max}$,
and step-size $\eta$.

\textit{Initialization}: Use the standard projected higher-order SVD (HOSVD) for initialization,
\begin{align*}
\boldsymbol{U}_{k}^{(0)} & =\text{HOSVD}_{r_{k}}\{\mathcal{M}_{(k)}(\mathcal{Y}_{w})\},\quad\boldsymbol{U}_{1,X}^{(0)}=\boldsymbol{P}_{\Phi(\boldsymbol{X}_{0})}\boldsymbol{U}_{k}^{(0)},\\
\mathcal{G}^{(0)} & =\mathcal{Y}_{w}\times_{1}\boldsymbol{U}_{1,X}^{(0)}{}^{\intercal}\times_{2}\boldsymbol{U}_{2}^{(0)\intercal}\times_{3}\boldsymbol{U}_{3}^{(0)\intercal},
\end{align*}
where $\text{HOSVD}_{r}(\cdot)$ returns the orthonormal matrix comprised
the top $r$ left singular vectors of a matrix, $\mathcal{Y}_{w}=\mathcal{Y}\odot\Omega\odot\mathcal{\widehat{\mathcal{W}}}$, and $$\boldsymbol{P}_{\Phi(\boldsymbol{X}_{0})}=\Phi(\boldsymbol{X}_{0})\{\Phi(\boldsymbol{X}_{0})^{\intercal}\Phi(\boldsymbol{X}_{0})\}^{-1}\Phi(\boldsymbol{X}_{0})^{\intercal}$$ is the projection matrix onto the sieves spaces spanned by the basis
function $\Phi(\boldsymbol{X}_{0})$.

\For{$I=1,\cdots,I_{\max}$}
{\begin{align*}
\boldsymbol{U}_{1}^{(I)} & =\boldsymbol{U}_{1,X}^{(I-1)}-\eta\frac{\partial L(\mathcal{G}^{(I-1)},\boldsymbol{U}_{1,X}^{(I-1)},\boldsymbol{U}_{2}^{(I-1)},\boldsymbol{U}_{3}^{(I-1)})}{\partial\boldsymbol{U}_{1,X}},\quad\boldsymbol{U}_{1,X}^{(I)}=\boldsymbol{P}_{\Phi(\boldsymbol{X}_{0})}\boldsymbol{U}_{1}^{(I)},\\
\boldsymbol{U}_{k}^{(I)} & =\boldsymbol{U}_{k}^{(I-1)}-\eta\frac{\partial L(\mathcal{G}^{(I-1)},\boldsymbol{U}_{1,X}^{(I-1)},\boldsymbol{U}_{2}^{(I-1)},\boldsymbol{U}_{3}^{(I-1)})}{\partial\boldsymbol{U}_{k}},
\quad\boldsymbol{U}_{k}^{(I)}=\boldsymbol{P}_{\boldsymbol{U}_k}\boldsymbol{U}_{k}^{(I)},\quad k=2,3,\\
\mathcal{G}^{(I)} & =\mathcal{G}^{(I-1)}-\eta\frac{\partial L(\mathcal{G}^{(I-1)},\boldsymbol{U}_{1,X}^{(I-1)},\boldsymbol{U}_{2}^{(I-1)},\boldsymbol{U}_{3}^{(I-1)})}{\partial\mathcal{G}}.
\end{align*}
}
Estimated potential outcomes tensor $\mathcal{\widehat{Y}}=\mathcal{G}^{(I_{\max})}\times_{1}\boldsymbol{U}_{1,X}^{(I_{\max})}\times_{2}\boldsymbol{U}_{2}^{(I_{\max})}\times_{3}\boldsymbol{U}_{3}^{(I_{\max})}$
\end{algorithm}

Besides, it is recognized that the direct rank computation of $(r_{1},r_{2},r_{3})$
is a NP-hard problem \citep{kolda2009tensor}, and we propose to tune $(r_{1},r_{2},r_{3})$ based on minimizing the BIC criterion:
\begin{equation}
\begin{split}
    \text{BIC} & =\log\left\{ \|
    \sqrt{\widehat{\mathcal{W}}}
    \odot
    \boldsymbol{P}_{\Omega}(\mathcal{Y}-\mathcal{G}^{(I_{\max})}\times_{1}\boldsymbol{U}_{1,X}^{(I_{\max})}\times_{2}\boldsymbol{U}_{2}^{(I_{\max})}\times_{3}\boldsymbol{U}_{3}^{(I_{\max})})
    \|_{F}^{2}\right\} \\
 & +\frac{\log(NTK)}{(NTK)}\{r_{1}r_{2}r_{3}+(N-r_{1})r_{1}+(T-r_{2})r_{2}+(K-r_{3})r_{3}\},
\end{split}
\label{eq:BIC}
\end{equation}
where $r_{1}r_{2}r_{3}+(N-r_{1})r_{1}+(T-r_{2})r_{2}+(K-r_{3})r_{3}$
is the degrees of freedom for any $(r_{1},r_{2},r_{3})$ multi-linear rank tensor in $\mathbb{R}^{N\times T\times K}$ \citep[Proposition 1]{zhang2019cross}; see Section \ref{subsec:effect_hyper} of the Supplemental Material for more details.

\section{Statistical guarantee}\label{sec:Statistical-Guarantee}

We first consider the following conditions which can be helpful to obtain the upper bound
for the estimation tensor error in tensor completion.
\begin{assumption}\label{assump:weight_G}
\begin{enumerate}
    \item [(a)] The estimated weight tensor $\widehat{\mathcal{W}}$ satisfies $$\frac{\sup_{t=1,\cdots,T}|\widehat{\mathcal{W}}_{i,t,l} -\mathcal{W}_{i,t,l}|}{\sqrt{TN^{-1} \log(N+T+K)}}\rightarrow 0.$$
    \item [(b)] The number of the basis functions $d_{\Phi}$, and the residual component $\boldsymbol{\Gamma}$ satisfy
    $$
    \frac{\sqrt{Td_{\Phi}^{-\tau}}\vee \sqrt{NT}\|\boldsymbol{\Gamma}\|}{\sqrt{(N\vee T)\log(N+T+K)}}\rightarrow 0,
    $$
    where $\tau$ controls the smoothness of the unknown function $G^*(\boldsymbol{X}_0)$.
\end{enumerate}
\end{assumption}
Assumption \ref{assump:weight_G}(a) states that the estimation error of the weight tensor should be bounded. Take the logistic modeling in Remark \ref{rmk:estimating_PS} for an example. Let $\mathcal{J}_t$ denote the indices of true nonzero coefficients for the logistic model $e(H_{i,t};\alpha_t)$. If the parameter $\alpha_t$ can be estimated with root-$N$ consistency, then we have $\sup_{t=1,\ldots,T} |\mathcal{J}_t|$ $/{\sqrt{T \log(N+T+K)}} \rightarrow 0$, which essentially limits the number of true nonzero coefficients, $|\mathcal{J}_t|$, under correct model specification; see Lemma \ref{lem:oracle_property}. Assumption \ref{assump:weight_G}(b) requires that the covariates $\boldsymbol{X}_{0}$ retain significant explanatory power for the unknown function $G^{*}(\boldsymbol{X}_{0})$. This assumption is crucial for establishing the upper bound induced by the projection matrix $\boldsymbol{P}_{\Phi(\boldsymbol{X}_{0})}^\perp$; see Lemma \ref{lem:basis_functions}.

Now, we can study the upper bound for the estimation
error in tensor completion. Denote the weighted loss function (\ref{eq:loss_1}) concisely by $L(\mathcal{X};\widehat{\mathcal{W}})=\sum_{\Omega_{i,t,l}=1}\widehat{\mathcal{W}}_{i,t,l}(\mathcal{Y}_{i,t,l}-\mathcal{X}_{i,t,l})^{2}$. To proceed with our claims on the statistical properties, we request
the following assumptions on the true potential outcomes tensor $\mathcal{Y}^*$ to bypass the ill-concentrated tensors.
\begin{assumption}\label{asmp:incoherence}
Suppose $\mathcal{Y}^{*}=\mathcal{G}^{*}\times_{1}\boldsymbol{U}_{1}^{*}\times_{2}\boldsymbol{U}_{2}^{*}\times_{3}\boldsymbol{U}_{3}^{*}$
and $\boldsymbol{U}_{k}^{*},k=1,2,3$ have orthonormal columns, there
exist some constants $\mu_{0}$ and $L_{0}$ such that $\mathcal{Y}^{*}\in\mathfrak{C}(r_{1},r_{2},r_{3},\mu_{0},L_{0})$,
where
\begin{align*}
 & \mathfrak{C}(r_{1},r_{2},r_{3},\mu_{0},L_{0})\\
 & =\left\{ \mathcal{X}=\mathcal{G}\times_{1}\boldsymbol{U}_{1}\times_{2}\boldsymbol{U}_{2}\times_{3}\boldsymbol{U}_{3},\quad\mu(\mathcal{X})\leq\mu_{0},\quad\max_{k\in\{1,2,3\}}\|\mathcal{M}_{k}(\mathcal{G})\|\leq L_{0}\sqrt{\frac{NTK}{\mu_{0}^{3/2}(r_{1}r_{2}r_{3})^{1/2}}}\right\} ,
\end{align*}
where $\mu(\mathcal{X})=\max\{\mu(\boldsymbol{U}_{1}),\mu(\boldsymbol{U}_{2}),\mu(\boldsymbol{U}_{3})\}$, $\mu(\boldsymbol{U}_{1})=N\|\boldsymbol{U}_{1}\|_{2,\infty}^{2}/r_1$, $\mu(\boldsymbol{U}_{2})=T\|\boldsymbol{U}_{2}\|_{2,\infty}^{2}/r_2$, and $\mu(\boldsymbol{U}_{3})=K\|\boldsymbol{U}_{3}\|_{2,\infty}^{2}/r_3$.
\end{assumption}

By enforcing $\mathcal{Y}^{*}$ in the restricted tensor space $\mathfrak{C}(r_{1},r_{2},r_{3},\mu_{0},L_{0})$,
the loading matrices $\boldsymbol{U}_{k}$ should satisfy the incoherence
condition, which is commonly assumed in the matrix/tensor
completion literature \citep{candes2009exact,ma2017exploration,yuan2017incoherent,cao2020multisample,xia2021statistically,cai2021nonconvex}.
Intuitively, the incoherence condition indicates that each tensor
entry contains a similar amount of information so missing any
of them will not prevent us from being able to reconstruct the entire
tensor. In particular, small $\mu(\mathcal{X})$ implies that the energy of tensor $\mathcal{X}$ is balanced across all modes, which is essential for establishing the restricted strong convexity in Lemma \ref{lem:restricted_convexity}. In addition, we also require an upper bound on the spectral norm of
each matricization of the core tensor $\mathcal{G}$, which is useful
to diminish the estimation error lying in the orthogonal space of
the sieve basis functions $\Phi(\boldsymbol{X}_{0})$. Encouragingly, Algorithm \ref{alg:PGD} can easily incorporate the
conditions in Assumption \ref{asmp:incoherence} by including multiple
projection steps to ensure that $(\mathcal{G}^{(I)},\boldsymbol{U}_{1,X}^{(I)},\boldsymbol{U}_{2}^{(I)},\boldsymbol{U}_{3}^{(I)})$
lie in the restricted domain $\mathfrak{C}(r_{1},r_{2},r_{3},\mu_{0},L_{0})$
throughout the iterations.

By the definition of $\widehat{\mathcal{Y}}$ that minimizes the weighted loss function, we have $L(\widehat{\mathcal{Y}};\widehat{\mathcal{W}})\leq L(\mathcal{Y}^{*};\widehat{\mathcal{W}})$, where
\begin{align}
 & \sum_{\Omega_{i,t,l}=1}\widehat{\mathcal{W}}_{i,t,l}(\mathcal{Y}_{i,t,l}-\mathcal{\widehat{Y}}_{i,t,l})^{2}-\sum_{\Omega_{i,t,l}=1}\widehat{\mathcal{W}}_{i,t,l}(\mathcal{Y}_{i,t,l}-\mathcal{Y}_{i,t,l}^{*})^{2}\nonumber \\
= & \sum_{\Omega_{i,t,l}=1}\widehat{\mathcal{W}}_{i,t,l}(\mathcal{Y}_{i,t,l}-\mathcal{\widehat{Y}}_{i,t,l})^{2}-\sum_{\Omega_{i,t,l}=1}\mathcal{W}_{i,t,l}(\mathcal{Y}_{i,t,l}-\mathcal{\widehat{Y}}_{i,t,l})^{2}\label{eq:part1}\\
 & +\sum_{\Omega_{i,t,l}=1}\mathcal{W}_{i,t,l}(\mathcal{Y}_{i,t,l}-\mathcal{\widehat{Y}}_{i,t,l})^{2}-\sum_{\Omega_{i,t,l}=1}\mathcal{W}_{i,t,l}(\mathcal{Y}_{i,t,l}-\mathcal{Y}_{i,t,l}^{*})^{2}\label{eq:part2}\\
 & +\sum_{\Omega_{i,t,l}=1}\mathcal{W}_{i,t,l}(\mathcal{Y}_{i,t,l}-\mathcal{Y}_{i,t,l}^{*})^{2}-\sum_{\Omega_{i,t,l}=1}\widehat{\mathcal{W}}_{i,t,l}(\mathcal{Y}_{i,t,l}-\mathcal{Y}_{i,t,l}^{*})^{2}\label{eq:part3}\\
= & I_{1}+I_{2}+I_{3}\leq0,\nonumber 
\end{align}
where the terms $I_1$ and $I_3$ are bounded under Assumption \ref{assump:weight_G}(a)
by estimating
$\mathcal{W}_{i,t,l}$ with vanishing error. The boundary of (\ref{eq:part3}) is challenging and is proved in the Appendix with details. Thus, we establish the statistical
guarantee for our proposed tensor completion in Theorem \ref{thm:bound-tensor}. 
\begin{theorem}
\label{thm:bound-tensor} Under Assumptions \ref{assum:HR}-\ref{asmp:incoherence}, and other regularity conditions, if $\widehat{\mathcal{Y}}$ minimizes the objective function in (\ref{eq:loss_1}),
we have 
\begin{align*}
 & \frac{\|\mathcal{Y}^{*}-\widehat{\mathcal{Y}}\|_{F}^{2}}{NT}\leq\frac{CL_{0}^{2}}{p_{\min}}\frac{\log(N+T+K)}{N}\\
 & \vee\frac{C^{'}k^{2}(r_{1}\wedge d_{\Phi})r_{2}r_{3}}{\max\{(r_{1}\wedge d_{\Phi}),r_{2},r_{3}\}p_{\min}^{2}}\frac{(N\vee T)}{NT}\log(N+T+K)\left\{ \sigma^{2}\left(\frac{p_{\max}}{p_{\min}}\right)^{2}\vee L_{0}^{2}\right\},
\end{align*}
with probability greater than $1-7(N+T+K)^{-2}$ for some constants
$C$, $C'$, $C''$ and $C'''$, {where $p_{\min}=\delta^{k}\leq \mathbb{P}(A_{i,t-k+1:t}=a_{(t-k+1):t}\mid H_{i,t})\leq(1-\delta)^{k}=p_{\max}.$}
\end{theorem}
Theorem \ref{thm:bound-tensor} shows that the upper bound converges to zero as $N$ and $T$ grow if the estimation errors for $G^{*}(\boldsymbol{X}_{0})$ and $\mathcal{W}$ are
sufficiently small for well-conditioned $\mathcal{Y}^*-\widehat{\mathcal{Y}}$. Here, we provide an interpretation of each term in the right-hand
side of the tensor estimation error. First, one can show that a deterministic
upper bound for $\sum_{\Omega_{i,t,l}=1}\mathcal{W}_{i,t,l}(\mathcal{Y}_{i,t,l}^{*}-\widehat{\mathcal{Y}}_{i,t,l})^{2}$
under any realization of the observation pattern $\Omega$, or equivalently
the treatment mechanism, which is controlled by the spectral norm of the weighted error tensor $\mathcal{E}_{\Omega}^{w}$. Combining
it with the restricted strong convexity condition, it yields the \textit{second}
error term; however, to legitimately use the restricted strong convexity
condition, one requires a lower bound on $\|\mathcal{Y}^{*}-\widehat{\mathcal{Y}}\|_F^2$, which is not necessarily satisfied for every $\mathcal{Y}^{*}$ and $\widehat{\mathcal{Y}}\in\mathfrak{C}(r_{1},r_{2},r_{3},\mu_{0},L_{0})$
\citep{hamidi2019low}. In fact, the required lower bound plays a
similar role as the incoherence condition on $\mathcal{Y}^{*}-\widehat{\mathcal{Y}}$
\citep{xia2021statistically}. Hence, failing to attain the lower
bound will give us another upper bound for $\|\mathcal{Y}^{*}-\widehat{\mathcal{Y}}\|_{F}^{2}$,
which is the \textit{first} term on the right-hand side.

\begin{remark}[Comparison with matrix completion on panel data]
The proposed tensorized HRMSM can be used to draw causal inferences on panel data with staggered entries. Under the staggered adoption design, it is common to assume that the potential outcomes are only influenced by the contemporaneous treatment, which rules out the dynamic treatment effects and focuses on estimating two potential outcomes matrices $Y(0)$ and $Y(1)$. By setting $k=1$, our tensorized HRMSM reduces to this setup, and a close analog can be drawn upon the matrix completion approach in \cite{athey2021matrix}, which exploits the low rankness via the matrix nuclear norm regularization. In fact, the first and the second error terms in Theorem \ref{thm:bound-tensor} are comparable to the first two terms in Theorem 2, \cite{athey2021matrix}. However, \cite{athey2021matrix} impute the potential outcomes matrices $Y(0)$ and $Y(1)$ separately, whereas we stack these two matrices into a tensor and employ the Bernstein-type inequality to bound the tolerance quantity $\vartheta$. Consider a special case where $N\asymp T$, our tolerance term satisfies $\vartheta\asymp N$, which is of a smaller order than that in \cite{athey2021matrix}, i.e., $\vartheta\asymp N^{3/2}$. An empirical
experiment demonstrates that our method achieves stable numerical results regardless of the
time that subjects are first exposed to the treatment; see Section \ref{subsubsec:comparison_MC} in the Supplementary
Material for more details.
\label{rmk:ate}
\end{remark}

\begin{remark}[On Global vs. Algorithmic Consistency.]
{
Theorem \ref{thm:bound-tensor} establishes the consistency of the global minimizer $\hat{Y}$. However, due to the non-convex nature of the optimization problem, the proposed algorithm cannot guarantee convergence to the global optimum. In practice, good empirical performance is often observed as demonstrated in our numerical analysis. Similar gaps between theoretical and computational guarantees are common in the tensor and nonconvex optimization literature (e.g., \cite{wang2020learning,zhou2024broadcasted,zhou2013tensor}). A more rigorous analysis linking algorithmic convergence and statistical consistency is left for future work.}
\end{remark}

\section{Simulation studies}\label{sec:Simulations}

 We assess the effectiveness of the tensorized HRMSMs in recovering the potential outcomes tensor with fixed multi-linear
ranks under Gaussian noises through extensive Monte
Carlo simulations. Specifically, we compare the covariate-assisted tensorized HRMSM with the parametric HRMSM
in \citet{neugebauer2007causal}, the matrix factor model in \cite{fan2016projected} based on the unfolded tensor $\mathcal{M}_{(1)}(\mathcal{Y})$, the \textit{vanilla} tensorized HRMSM without incorporating any baseline covariates, {and CT, as empirical evidence indicates its superiority over other state-of-the-art methods \citep{melnychuk2022causal}.} For the HRMSM, we postulate time-specific linear models $\mu(a_{(t-k+1):t},X_{i,0};\eta_t)=(a_{(t-k+1):t},X_{i,0}^{\intercal})^{\intercal}\eta_t$; The matrix factor model and the \textit{vanilla} tensorized HRMSM are included to demonstrate the usefulness of the tensor representation and covariate assistance. For a fair comparison, we minimize an IPTW version of their target loss functions, where the weights adjust for the time-varying confounding.  The performance { of the first four methods} is evaluated by the normalized tensor mean squared error of the difference between the recovered
tensor $\widehat{\mathcal{Y}}$ and the ground truth $\mathcal{Y}^{*}$
as $\ell_{2}(\widehat{\mathcal{Y}})=\|\widehat{\mathcal{Y}}-\mathcal{Y}^{*}\|_{F}^{2}/\|\mathcal{Y}^{*}\|_{F}^{2}$; { for CT, however, the difference is computed based on 10 randomly sampled paths only, due to the high computational cost associated with its pathwise prediction mechanism.} All simulation results are based on $100$ data replications. 

In what follows, the tensor estimation error is investigated under a set of simulation designs. Specifically, Section \ref{subsec:effect-models} studies the impact of different model  formulations of 
$\{\mathcal{Y}_{i,t,l},e(H_{i,t})\}$; Section \ref{subsec:effect-semi} examines the effects of the covariate-orthogonal loadings $\boldsymbol{\Gamma}$. For now, we assume the true multi-linear ranks are known, and a BIC-based tuning procedure is shown to be effective at the hyperparameter tuning in Section \ref{subsec:effect_hyper} of the Appendix.

\subsection{Effect of outcome and propensity score models}\label{subsec:effect-models}

We assess the robustness of our proposed model under various outcome and propensity score models.
For starter, the baseline covariate $\boldsymbol{X}_{0}\in\mathbb{R}^{N\times d_{0}}$
and the time-dependent covariate {counterfactuals} $X_{i,t}^{a_{(t-2):t}}$ are simulated as $
{X}_{i,0}\overset{i.i.d.}{\sim}\mathcal{N}(0,1), X_{i,t}^{a_{(t-2):t}}=
\text{cum}(X_{i,0})+
\eta_{1}a_{i,t}+\eta_{2}a_{i,t-1}+\eta_{3}a_{i,t-2}$, where  $\text{cum}(X_{i,0})$ is the sum of all the entries of $X_{i,0}$, $d_{0}=20$, and $\eta_{m}=2^{-m}, m=1,2,3$. 
For $a_{i,(t-k+1):t}\in\mathbb{A}_k$, we consider two data-generating procedures for $\mathcal{Y}_{i,t,l}$, labeled by (M1) and (M2). In particular, (M1) is a simple linear model of $X_{i,0:t}$ and $a_{i,(t-k+1):t}$ for $\mathcal{Y}_{i,t,\{a_{i,(t-k+1):t}\}_{(10)}}$, whereas (M2) is a complex additive model, which includes the sieve basis functions of $X_{i,0}$ and several quadratic terms; see Section \ref{subsubsec:tensor_rep} of the Appendix for the detailed definitions of (M1) and (M2). Next, we consider two sequential treatment assignment mechanisms for $A_{i,t}$ depending on the time-varying
confounders $(X_{i,t},X_{i,t-1})$:
\begin{itemize}
\item [(A1)]
$
\mathbb{P}(A_{i,t}=1\mid H_{i,t})=e(H_{i,t})=\exp(X_{i,t}+X_{i,t-1})/\{1+\exp(X_{i,t}+X_{i,t-1})\},
$
\item [(A2)]
$
\mathbb{P}(A_{i,t}=1\mid H_{i,t})=e(H_{i,t})=\exp(2X_{i,t}+2X_{i,t-1})/\{1+\exp(2X_{i,t}+2X_{i,t-1})\},
$
\end{itemize}
where (A2) is prone to assign treatments on the
previously treated subjects, leading to a wider range of possible propensity scores and therefore a smaller $\delta$ than (A1). The penalized likelihood estimation in (\ref{eq:penalized_Q_PS}) with SCAD penalty is used for estimating the propensity score at each time point.

Figure \ref{fig:mse-case1to6-N} displays the normalized tensor error $\ell_2(\widehat{\mathcal{Y}})$
versus $N$ and $T$ for various estimators when $k=5$. We observe that
the benchmark HRMSM achieves the best performance when the generative model of $\mathcal{Y}_{i,t,l}$ is correctly specified by $\mu(a_{(t-k+1):t},X_{i,0};\eta_t)$. However, as several
quadratic terms are involved in (M2), the estimation accuracy of HRMSM
is negatively affected due to its incorrect parameterization, whereas our
flexible tensor representation is able to capture the complex
relationship between the potential outcomes and covariates, time, and treatments.

More importantly, we observe that the estimation error of the tensorized HRMSMs (both the \textit{vanilla} and covariate-assisted) are favorably affected by the sample size $N$, which aligns with our theoretical bounds in Theorem \ref{thm:bound-tensor}. In addition, the normalized error is not greatly positively impacted by the time points $T$, which can be attributed to the increasingly time-specific propensity score estimations when $T$ grows. However, the tensorized HRMSMs can still significantly improve the performance over the matricized minimization, especially for larger $T$. { By contrast, CT suffers from cumulative prediction bias, likely due to its autoregressive attention mechanism through which small local errors propagate and accumulate over time. Besides, it also subject to confounding bias as the produced time-varying representations are only balanced on current treatment assignment, which do not account for the lagged treatment effect.} As the latent factors of potential outcomes tensor over the first mode is explainable by the baseline covariate $\boldsymbol{X}_{0}$ in both (M1) and (M2), additional information can be harnessed by the covariate-assisted tensorized HRMSM to compensate the efficiency loss due to increased variability of IPTWs in (A2).

\subsection{Effect of covariate-independent component}\label{subsec:effect-semi}

In this subsection, we assess the effectiveness of our proposed framework
when the strength of the covariate-independent components varies. In particular, the sequential treatment assignment for $A_{i,t}$ follows (A1) and the generative model of $\mathcal{Y}_{i,t,l}$
is similar to (M2) except that the covariate-orthogonal
noises $\Gamma_{i,j}$ are added, where $\Gamma_{i,j}$ controls the amplitude of the covariate-orthogonal part.

\begin{figure}[!t]
\centering
\includegraphics[width=0.9\linewidth]{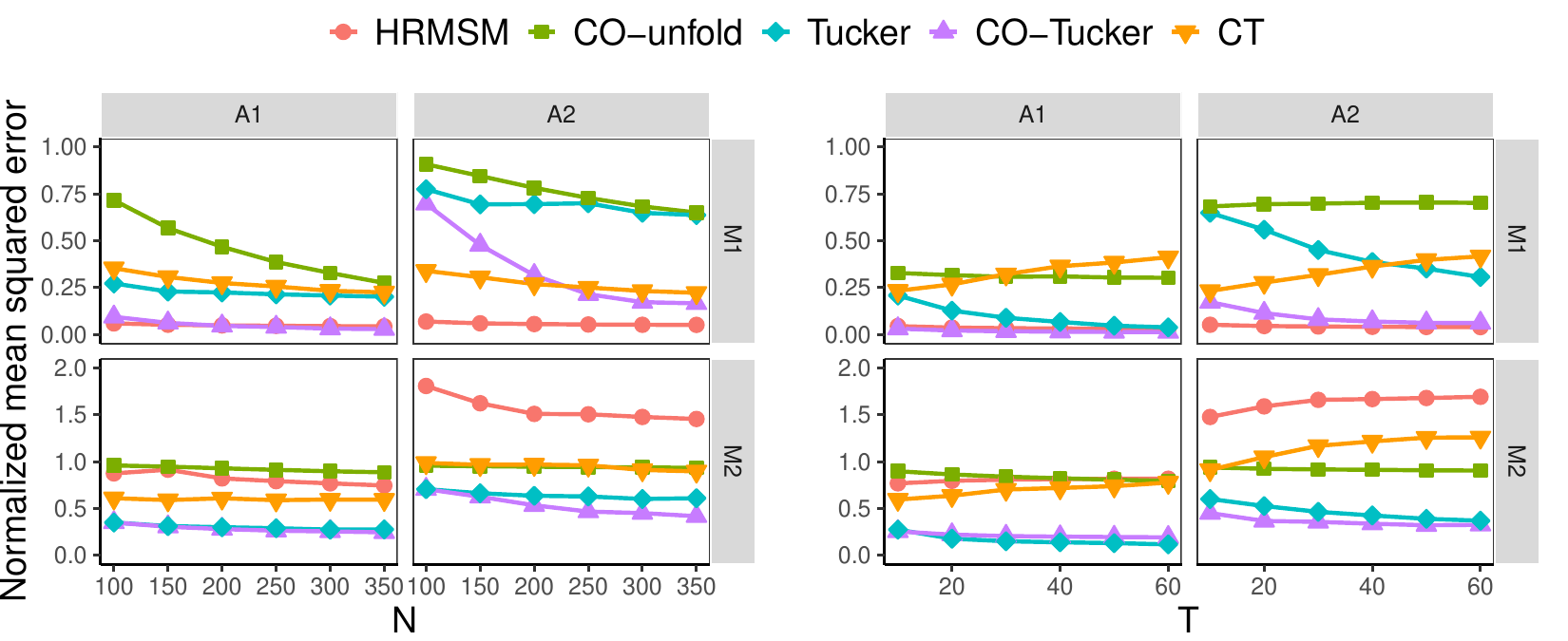}
\includegraphics[width=0.9\linewidth]{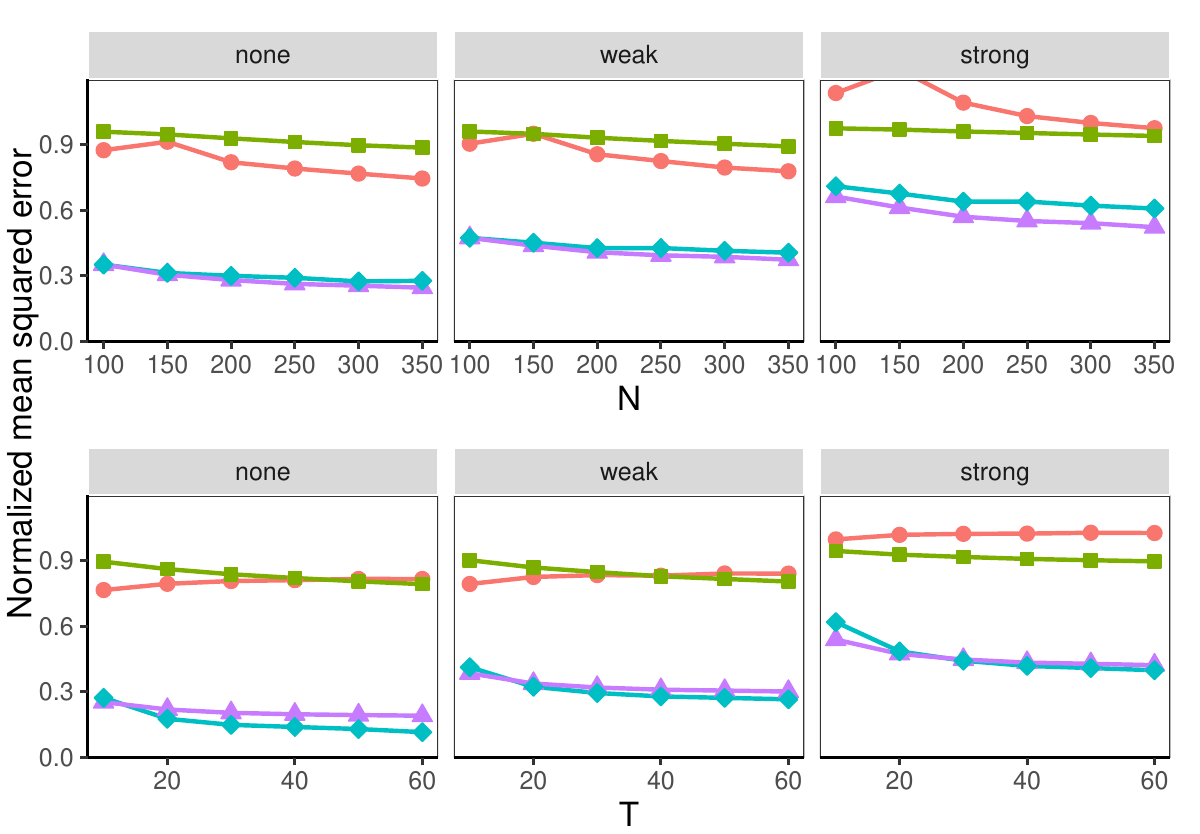}
\caption{\label{fig:mse-case1to6-N} Averaged normalized mean squared tensor
error when $T=10$, $N=100,150,\cdots,350$, and $N=300$, $T=10,20,\cdots,60$ under (Top panel) outcome models (M1) and (M2)
and treatment assignments (A1) and (A2), and (Bottom panel) none, weak and strong noises
due to the covariate-orthogonal factor ${\Gamma}_{i,j}$; \textsc{hrmsm} is the history-restricted marginal structural model, \textsc{co-unfold} is the matrix factor model based on unfolded tensor, \textsc{tucker} is the \textit{vanilla} tensorized HRMSM, \textsc{co-tucker} is the covariate-assisted tensorized HRMSM, { and {\sc ct} is the Causal Transformer.}}
\end{figure}

In this case, the potential outcome
tensor still admits the low-rank formulation except that the tensor factors of the first mode is changed to include the extra covariate-independent component $\Gamma_{i,j}$. Consider
three ways for generating the $\Gamma_{i,j}$:
\begin{enumerate}
    \item [1)] $\Gamma_{i,j}=0$, encoded by none covariate-orthogonal noises;
    \item [2)] $\Gamma_{i,j}\sim\mathcal{N}(0,5^{2})$, encoded by weak covariate-orthogonal noises; and
    \item [3)] $\Gamma_{i,j}\sim\mathcal{N}(0,20^{2})$, encoded by strong covariate-orthogonal noises. 
\end{enumerate}
Figure \ref{fig:mse-case1to6-N} displays the normalized mean squared error $\ell_{2}(\widehat{\mathcal{Y}})$ under varying strength of
${\Gamma}_{i,j}$ (none, weak or strong) versus $N$ and $T$. Similar to the phenomenons observed in Figure \ref{fig:mse-case1to6-N}, the covariate-assisted tensorized HRMSM can improve the
estimation accuracy upon the \textit{vanilla} tensorized HRMSM regardless of whether the magnitude of ${\Gamma}_{i,j}$ is strong or not. In particular, the merits of employing the covariate-assisted
tensorized HRMSM are more accentuated when the noises induced
by ${\Gamma}_{i,j}$ are amplified. The substantial performance improvement of the covariate-assisted tensorized HRMSM is largely attributed to the use of the auxiliary covariates, which aids
us in recovering a more accurate solution in the covariate-related
subspace.

\section{An application}\label{sec:Application}
We apply our proposed method to the Medical Information Mart for Intensive Care (MIMIC) III database. The MIMIC III database is an extensive, publicly available Intensive
Care Unit (ICU) database, which consists of de-identified health-related data associated with over $40000$ patients who stayed in critical care units of the Beth Israel Deaconess Medical Center between 2001 and 2012 \citep{goldberger2000physiobank,johnson2016mimic,johnson2016mimicb}. After removing the erroneous entries due to missingness and duplication, we extract $N=4006$ patients with trajectories up to $T=20$ timestamps
every four hours. Next, we consider the mechanical ventilation status as the treatment $A$ with $1$ as the presence of mechanical ventilation and $0$ otherwise. The outcome of interest $Y$ is the Sequential Organ Failure Assessment (SOFA) score, which measures the severity of organ dysfunction for a patient in ICU in terms of pulmonary, renal, hepatic, cardiovascular, hematologic, and neurologic systems. Each organ system gives a score ranging from $0$ to $4$, and the total
score is from $0$ to $24$ with $0$ as the lowest level of morbidity and $24$ as the highest level of morbidity \citep{vincent1996sofa}. Given the rapidly changing conditions of ICU patients, most of the effects of mechanical ventilation can be captured within a day; thus, we consider the treatment regimes for each patient as his/her previous $k=6$ binary treatment histories.

In that way, the potential outcomes tensor $\mathcal{Y}$ can be formed accordingly in the dimensions of $4006\times20\times2^{6}$.
Following the previous studies \citep{purushotham2018benchmarking},
we select the baseline covariates $\boldsymbol{X}_{0}\in\mathbb{R}^{N\times11}$
and the time-varying covariates $\boldsymbol{X}_{t}\in\mathbb{R}^{N\times7},t=1,\cdots,T$ 
as listed in Table \ref{tab:selected-covariate} for our subsequent analyses. We normalize each
covariate to have a unit $L_{2}$ norm at each timestamp, and construct
the basis functions $\Phi(\cdot)$ by the Legendre polynomial functions
up to order $10$.

Our purpose is to emphasize the flexibility of our low-rank tensor framework and highlight the advantages of including the baseline covariate in recovering the {counterfactuals}. Particularly, we randomly split the MIMIC III dataset into $V$ folds, denoted by $\mathcal{I}_{1},\cdots,\mathcal{I}_{V}$.
For each $\mathcal{I}_{v}$, the models are trained over the other folds $\mathcal{I}_{v}^{\complement}$, that is, the complement of $\mathcal{I}_{v}$ and are evaluated on $\mathcal{I}_{v}$ for testing.
The averaged training error and the testing error are  
\[
\ell_{2}(\widehat{\mathcal{Y}})_{\train}=\frac{1}{V}\sum_{v=1}^{V}\frac{\|\boldsymbol{P}_{\Omega_{\mathcal{I}_{v}^{\complement}}}(\widehat{\mathcal{Y}}_{\mathcal{I}_{v}^{\complement}}-\mathcal{Y}_{\mathcal{I}_{v}^{\complement}})\|_{F}}{\|\boldsymbol{P}_{\Omega_{\mathcal{I}_{v}^{\complement}}}(\mathcal{Y}_{\mathcal{I}_{v}^{\complement}})\|_{F}},\quad\ell_{2}(\widehat{\mathcal{Y}})_{\test}=\frac{1}{V}\sum_{v=1}^{V}\frac{\|\boldsymbol{P}_{\Omega_{\mathcal{I}_{v}}}(\widehat{\mathcal{Y}}_{\mathcal{I}_{v}}-\mathcal{Y}_{\mathcal{I}_{v}})\|_{F}}{\|\boldsymbol{P}_{\Omega_{\mathcal{I}_{v}}}(\mathcal{Y}_{\mathcal{I}_{v}})\|_{F}},
\]
where $\boldsymbol{P}_{\Omega_{\mathcal{I}_{v}^{\complement}}}(\cdot)$ and
$\boldsymbol{P}_{\Omega_{\mathcal{I}_{v}}}(\cdot)$ are the projections of
any tensor onto the observed entries within the training dataset $\mathcal{I}_{v}^{\complement}$
and the testing dataset $\mathcal{I}_{v}$; $\widehat{\mathcal{Y}}_{\mathcal{I}_{v}^{\complement}}$
and $\widehat{\mathcal{Y}}_{\mathcal{I}_{v}}$ are the estimated sub-tensor
for the sub-samples $\mathcal{I}_{v}^{\complement}$ and $\mathcal{I}_{v}$,
respectively. Under our framework in (\ref{eq:loss_1}), we can predict
the sub-tensor ${\mathcal{Y}}_{\mathcal{I}_{v}}$ using the held-out baseline covariate  $\boldsymbol{X}_{0,\mathcal{I}_{v}}$ by $\widehat{\mathcal{Y}}_{\mathcal{I}_{v}}=\widehat{\mathcal{G}}_{\mathcal{I}_{v}^{\complement}}\times_{1}\boldsymbol{\Phi}(\boldsymbol{X}_{0,\mathcal{I}_{v}})\widehat{\boldsymbol{B}}_{\mathcal{I}_{v}^{\complement}}\times_{2}\widehat{\boldsymbol{U}}_{2,\mathcal{I}_{v}^{\complement}}\times_{3}\widehat{\boldsymbol{U}}_{3,\mathcal{I}_{v}^{\complement}}$, where $\widehat{\boldsymbol{B}}_{\mathcal{I}_{v}^{\complement}}=\{\boldsymbol{\Phi}(\boldsymbol{X}_{0,\mathcal{I}_{v}^{\complement}})^{\intercal}\boldsymbol{\Phi}(\boldsymbol{X}_{0,\mathcal{I}_{v}^{\complement}})\}^{-1}\boldsymbol{\Phi}(\boldsymbol{X}_{0,\mathcal{I}_{v}^{\complement}})^{\intercal}\widehat{\boldsymbol{U}}_{1,X,\mathcal{I}_{v}^{\complement}}$
and $(\widehat{\mathcal{G}}_{\mathcal{I}_{v}^{\complement}},\widehat{\boldsymbol{U}}_{1,X,\mathcal{I}_{v}^{\complement}},\widehat{\boldsymbol{U}}_{2,\mathcal{I}_{v}^{\complement}},\widehat{\boldsymbol{U}}_{3,\mathcal{I}_{v}^{\complement}})$
are the minimizer of the loss function (\ref{eq:loss_1}) over the
sub-sample $\mathcal{I}_{v}^{\complement}$. For the other methods, the
simple kernel smoothing over the first mode is used with $\boldsymbol{X}_{0,\mathcal{I}_{v}}$
and $\boldsymbol{X}_{0,\mathcal{I}_{v}^{\complement}}$ for obtaining
the prediction. In specific,
the kernel weight matrix $\boldsymbol{H}_{v}\in\mathbb{R}^{|\mathcal{I}_{v}|\times|\mathcal{I}_{v}^{\complement}|}$
is 
\[
H_{v,i,j}=\frac{\kappa(\|X_{j,0,\mathcal{I}_{v}^{\complement}}-X_{i,0,\mathcal{I}_{v}}\|^{2})}{\sum_{j=1}^{|\mathcal{I}_{v}^{\complement}|}\kappa(\|X_{j,0,\mathcal{I}_{v}^{\complement}}-X_{i,0,\mathcal{I}_{v}}\|^{2})},
\]
where $\kappa(\cdot,\cdot)$ is a user-specified kernel distance function
and the Gaussian kernel distance is chosen with the kernel variance
parameter $\sigma_{\kappa}=1$. Then, we can predict the sub-tensor
${\mathcal{Y}}_{\mathcal{I}_{v}}$ using $\boldsymbol{X}_{0,\mathcal{I}_{v}}$ based on kernel smoothing as
$\widehat{\mathcal{Y}}_{\mathcal{I}_{v}}=\widehat{\mathcal{Y}}_{\mathcal{I}_{v}^{\complement}}\times_{1}\boldsymbol{H}_v$, which is the best feasible linear predictor under the tensor factor model \citep{chen2020semiparametric}.

\begin{table}[!htbp]
\caption{\label{tab:results_real} Averaged normalized mean squared training
$\ell_{2}(\widehat{\mathcal{Y}})_{\train}$ $(\times10^{-2})$, testing
errors $\ell_{2}(\widehat{\mathcal{Y}})_{\test}$ $(\times10^{-2})$ over cross-validation folds $V=2,3,5,10$ }
\vspace{0.5em}
\centering
{\tabcolsep=0pt
\begin{tabular*}{\textwidth}{@{\extracolsep{\fill}}lcccccc@{\extracolsep{\fill}}}
\hline 
 & $V$ & HRMSM & CO-unfold & Tucker & CO-Tucker\tabularnewline
\hline 
\hline 
$\ell_{2}(\widehat{\mathcal{Y}})_{\train}$ & 2 & 39.02&62.74&33.91&45.91\\
 & 3 & 38.96&59.44&31.62&45.79\\
 & 5 & 38.98&58.00&30.57&45.74\\
 & 10 & 39.01&57.266&29.95&45.70\\
\hline 
$\ell_{2}(\widehat{\mathcal{Y}})_{\test}$ & 2 & 56.63&60.70&51.11&46.08\\
 & 3 & 160.0&123.58&95.19&45.93\\
 & 5 & 397.64&322.83&258.08&45.83\\
 & 10 & 1001.1&838.66&686.03&45.79\\
\hline 
\end{tabular*}
}
\end{table}

Table \ref{tab:results_real} presents the training and testing errors averaged over the cross-validation folds with varying $V$. We find that tensorized HRMSMs (both the \textit{vanilla} and covariate-assisted) achieve comparable performance as the traditional HRMSM over the training sets. However, the covariate-assisted tensorized HRMSM outperforms when evaluated on the testing sets by leveraging the baseline covariates in completing the missing entries, whereas the \textit{vanilla} tensorized HRMSM suffers the issue of overfitting as it does not constrain the low-rank factors to be in a certain functional space spanned by the sieve basis.

In addition, we report the estimated ATE $\widehat{\tau}_{\ate}$ for the full-day presence of mechanical ventilation on the SOFA score. Based on the recovered potential outcome tensor $\widehat{\mathcal{Y}}$, $\widehat{\tau}_{\ate}$ over the training set $\mathcal{I}_v^{\complement}$ and testing set $\mathcal{I}_v$ can be computed easily by  
\begin{equation}
    \widehat{\tau}_{\ate, \rm train} =
    \sum_{i\in \mathcal{I}_{v}^{\complement}, t}\frac{\widehat{\mathcal{Y}}_{i,t,(\mathbf{1}_6)_{(10)}}  
    - 
    \widehat{\mathcal{Y}}_{i,t,(\mathbf{0}_6)_{(10)}}}{|\mathcal{I}_{v}^{\complement}|T},
    \quad 
    \widehat{\tau}_{\ate, \rm test} =
    \sum_{i\in \mathcal{I}_{v},t}\frac{\widehat{\mathcal{Y}}_{i,t,(\mathbf{1}_6)_{(10)}}  
    - 
    \widehat{\mathcal{Y}}_{i,t,(\mathbf{0}_6)_{(10)}}}{|\mathcal{I}_{v}|T}.
    \label{eq:est_ATT}
\end{equation}

\begin{table}[!htbp]
\caption{\label{tab:results_real_ATT} Averaged ATE on the training sets 
$\widehat{\tau}_{\ate,\train}$ and the testing sets $\widehat{\tau}_{\ate,\test}$ over cross-validation folds $V=2,3,5,10$ }
\vspace{0.5em}
\centering
{
\tabcolsep=0pt
\begin{tabular*}{\textwidth}{@{\extracolsep{\fill}}lcccccc@{\extracolsep{\fill}}}
\hline 
 & $V$ & HRMSM & CO-unfold & Tucker & CO-Tucker\tabularnewline
\hline 
\hline 
$\widehat{\tau}_{\ate,\rm train}$ & 2 & 1.12&3.03&2.05&0.39\\
 & 3 & 1.12&3.01&2.04&0.37\\
 & 5 &1.11&2.78&2.02&0.36\\
 & 10 & 1.11&2.56&2.00&0.37\\
\hline 
$\widehat{\tau}_{\ate,\rm test}$ & 2 & 1.53&4.12&2.92&0.39\\
 & 3 &3.05&8.20&5.82&0.37\\
 & 5 & 6.07&15.2&11.5&0.36\\
 & 10 & 13.6&31.5&25.5&0.37\\
\hline 
\end{tabular*}
}
\end{table}

Table \ref{tab:results_real_ATT} reports the estimated ATE, averaged over the folds with various $V$. As expected, the covariate-assisted tensorized HRMSM yields a stable estimation of the ATE when evaluated on the sets $\mathcal{I}_v^{\complement}$ and $\mathcal{I}_v$ across all the number of folds, similar to the phenomenons observed from Table \ref{tab:results_real}. However, other methods are subject to the over-fitting issue, indicated by the unstable predictions when the number of folds increases (i.e., more training samples and fewer testing samples). Therefore, it demonstrates the potential benefit of using the flexible tensor representation in conjunction with the sieve approximation for robust prediction. According to our analysis, mechanical ventilation may increase the SOFA score of the treated patients and is therefore harmful to critically ill patients in ICU, which agrees with the findings in \cite{mutlu2000complications} that mechanical ventilation can bring substantial risks to ICU patients and result in various complications, such as ventilator-associated pneumonia \citep{american2005guidelines,kalil2016management}.

\section{Discussion}\label{sec:discussion}
In this paper, we explore the flexibility introduced by the low-rank tensorized HRMSMs to estimate the unobserved potential outcome via the IPTW, which extends the history-restricted marginal structural models and exploits the full structure of the potential outcomes tensor for completion. Moreover, we adopt the semi-parametric framework to include the baseline covariates in the tensor completion while suppressing the impact of the covariate-independent components by the projected gradient descent algorithm.

The tensor completion framework can be readily adapted for multi-level or cluster treatments by expanding the treatment mode of the tensor to include potential outcomes under treatments of varying levels or structures. Additionally, leveraging the generalized propensity score \citep{yang2016propensity} and cluster calibration \citep{yang2018propensity} facilitates confounding adjustment through weighting. Most existing causal models use a discrete-time setup, which requires all subjects to be followed at the same pre-fixed time points but can be restrictive in many practical situations. In practice, the number of observations and observation times differ from subject to subject \citep{yang2022semiparametric}, yielding irregularly-spaced observation times. The proposed tensor completion framework can adapt to such situations by refining and expanding the time mode of the tensor. We can further include additional weighting to adjust for informative observation times. The tensor completion procedure remains the same. Extending the proposed framework for general outcomes such as survival outcomes \citep{yang2018modeling,yang2020semiparametric} is an important future work. 
Instead of assuming a low-rank structure on the data tensor, one can impose low-rankness on the model parameter tensor from canonical exponential distributions \citep{mao2024mixed}. Another intriguing direction is to de-confound the sequential treatments along the lines of \cite{bica2020time}, as the key assumption underpinning our framework is the unmeasured confounding condition, whose plausibility relies on collecting a rich set of auxiliary covariates from the observational studies. Lastly, it would be interesting to develop an inferential framework for the tensorized HRMSMs since it is hardly explored in the current tensor completion literature \citep{xia2022inference}. {A promising approach would be to integrate our method as the outcome model within a doubly robust framework \citep{kennedy2020optimal}. The statistical inference is valid under the upper bound for the estimation errors established in this paper, which can be a valuable tool for decision-making.} 

\bibliographystyle{apacite}
\bibliography{ref}

@article{zhou2013tensor,
  title={Tensor regression with applications in neuroimaging data analysis},
  author={Zhou, Hua and Li, Lexin and Zhu, Hongtu},
  journal={Journal of the American Statistical Association},
  volume={108},
  number={502},
  pages={540--552},
  year={2013},
  publisher={Taylor \& Francis}
}

@article{zhou2024broadcasted,
  title={Broadcasted nonparametric tensor regression},
  author={Zhou, Ya and Wong, Raymond KW and He, Kejun},
  journal={Journal of the Royal Statistical Society Series B: Statistical Methodology},
  volume={86},
  number={5},
  pages={1197--1220},
  year={2024},
  publisher={Oxford University Press UK}
}

@inproceedings{melnychuk2022causal,
  title={Causal transformer for estimating counterfactual outcomes},
  author={Melnychuk, Valentyn and Frauen, Dennis and Feuerriegel, Stefan},
  booktitle={International conference on machine learning},
  pages={15293--15329},
  year={2022},
  organization={PMLR}
}

@article{mutlu2000complications,
  title={Complications of mechanical ventilation.},
  author={Mutlu, GM and Factor, P},
  journal={Respiratory Care Clinics of North America},
  volume={6},
  pages={213--252},
  year={2000}
}

@article{american2005guidelines,
  title={Guidelines for the management of adults with hospital-acquired, ventilator-associated, and healthcare-associated pneumonia},
  author={American Thoracic Society and Infectious Diseases Society of America and others},
  journal={American Journal of Respiratory and Critical Care Medicine},
  volume={171},
  pages={388--416},
  year={2005},
  publisher={American Thoracic Society}
}

@article{kalil2016management,
  title={Management of adults with hospital-acquired and ventilator-associated pneumonia: 2016 clinical practice guidelines by the Infectious Diseases Society of America and the American Thoracic Society},
  author={Kalil, Andre C and Metersky, Mark L and Klompas, Michael and Muscedere, John and Sweeney, Daniel A and Palmer, Lucy B and Napolitano, Lena M and O'Grady, Naomi P and Bartlett, John G and Carratal{\`a}, Jordi and others},
  journal={Clinical Infectious Diseases},
  volume={63},
  pages={61--111},
  year={2016},
  publisher={Oxford University Press}
}

@article{zhou2021partially,
  title={Partially observed dynamic tensor response regression},
  author={Zhou, Jie and Sun, Will Wei and Zhang, Jingfei and Li, Lexin},
  journal={Journal of the American Statistical Association},
  pages={1--25},
  year={2021},
  publisher={Taylor \& Francis}
}

@article{yang2020semiparametric,
  title={Semiparametric estimation of structural failure time models in continuous-time processes},
  author={Yang, Shu and Pieper, Karen and Cools, Frank},
  journal={Biometrika},
  volume={107},
  pages={123--136},
  year={2020},
  publisher={Oxford University Press}
}

@article{lee2021improving,
  title={Improving trial generalizability using observational studies},
  author={Lee, Dasom and Yang, Shu and Dong, Lin and Wang, Xiaofei and Zeng, Donglin and Cai, Jianwen},
  journal={Biometrics},
  year={2021},
  pages={doi.org/10.1111/biom.13609},
  publisher={Wiley Online Library}
}

@article{yang2022semiparametric,
  title={Semiparametric estimation of structural nested mean models with irregularly spaced longitudinal observations},
  author={Yang, Shu},
  journal={Biometrics},
  volume={78},
  pages={937--949},
  year={2022},
  publisher={Wiley Online Library}
}

@article{yang2016propensity,
  title={Propensity score matching and subclassification in observational studies with multi-level treatments},
  author={Yang, Shu and Imbens, Guido W and Cui, Zhanglin and Faries, Douglas E and Kadziola, Zbigniew},
  journal={Biometrics},
  volume={72},
  pages={1055--1065},
  year={2016},
  publisher={Wiley Online Library}
}

@article{kennedy2020optimal,
  title={Towards optimal doubly robust estimation of heterogeneous causal effects},
  author={Kennedy, Edward H and others},
  journal={Electronic Journal of Statistics},
  volume={17},
  year={2023}
}

@article{zhen2024nonnegative,
  title={Nonnegative tensor completion for dynamic counterfactual prediction on covid-19 pandemic},
  author={Zhen, Yaoming and Wang, Junhui},
  journal={The Annals of Applied Statistics},
  volume={18},
  number={1},
  pages={224--245},
  year={2024},
  publisher={Institute of Mathematical Statistics}
}

@article{mandal2019weighted,
  title={Weighted Tensor Completion for Time-Series Causal Inference},
  author={Mandal, Debmalya and Parkes, David},
  journal={arXiv preprint arXiv:1902.04646},
  year={2019}
}

@article{chen2020semiparametric,
  title={Semiparametric tensor factor analysis by iteratively projected svd},
  author={Chen, Elynn Y and Xia, Dong and Cai, Chencheng and Fan, Jianqing},
  journal={arXiv preprint arXiv:2007.02404},
  year={2020}
}

@article{yu2006double,
  title={Double robust estimation in longitudinal marginal structural models},
  author={Yu, Zhuo and van der Laan, Mark},
  journal={Journal of Statistical Planning and Inference},
  volume={136},
  pages={1061--1089},
  year={2006},
  publisher={Elsevier}
}

@article{candes2009exact,
  title={Exact matrix completion via convex optimization},
  author={Cand{\`e}s, Emmanuel J and Recht, Benjamin},
  journal={Foundations of Computational Mathematics},
  volume={9},
  pages={717--772},
  year={2009},
  publisher={Springer}
}

@article{yang2018propensity,
  title={Propensity score weighting for causal inference with clustered data},
  author={Yang, Shu},
  journal={Journal of Causal Inference},
  volume={6},
  pages={20170027},
  year={2018},
  publisher={De Gruyter}
}

@article{mao2024mixed,
  title={Mixed Matrix Completion in Complex Survey Sampling under Heterogeneous Missingness},
  author={Mao, Xiaojun and Wang, Hengfang and Wang, Zhonglei and Yang, Shu},
  journal={Journal of Computational and Graphical Statistics},
  pages={1320--1328},
  year={2024},
  publisher={Taylor \& Francis}
}

@article{yang2018modeling,
  title={Modeling survival distribution as a function of time to treatment discontinuation: A dynamic treatment regime approach},
  author={Yang, Shu and Tsiatis, Anastasios A and Blazing, Michael},
  journal={Biometrics},
  volume={74},
  pages={900--909},
  year={2018},
  publisher={Wiley Online Library}
}

@article{heron2010ecological,
  title={Ecological momentary interventions: incorporating mobile technology into psychosocial and health behaviour treatments},
  author={Heron, Kristin E and Smyth, Joshua M},
  journal={British Journal of Health Psychology},
  volume={15},
  pages={1--39},
  year={2010},
  publisher={Wiley Online Library}
}

@article{semerci2014tensor,
  title={Tensor-based formulation and nuclear norm regularization for multienergy computed tomography},
  author={Semerci, Oguz and Hao, Ning and Kilmer, Misha E and Miller, Eric L},
  journal={IEEE Transactions on Image Processing},
  volume={23},
  pages={1678--1693},
  year={2014},
  publisher={IEEE}
}

@article{xie2016topicsketch,
  title={Topicsketch: Real-time bursty topic detection from twitter},
  author={Xie, Wei and Zhu, Feida and Jiang, Jing and Lim, Ee-Peng and Wang, Ke},
  journal={IEEE Transactions on Knowledge and Data Engineering},
  volume={28},
  pages={2216--2229},
  year={2016},
  publisher={IEEE}
}

@article{xu2013block,
  title={A block coordinate descent method for regularized multiconvex optimization with applications to nonnegative tensor factorization and completion},
  author={Xu, Yangyang and Yin, Wotao},
  journal={SIAM Journal on Imaging Sciences},
  volume={6},
  pages={1758--1789},
  year={2013},
  publisher={SIAM}
}

@article{tomioka2010estimation,
  title={Estimation of low-rank tensors via convex optimization},
  author={Tomioka, Ryota and Hayashi, Kohei and Kashima, Hisashi},
  journal={arXiv preprint arXiv:1010.0789},
  year={2010}
}

@article{gandy2011tensor,
  title={Tensor completion and low-n-rank tensor recovery via convex optimization},
  author={Gandy, Silvia and Recht, Benjamin and Yamada, Isao},
  journal={Inverse Problems},
  volume={27},
  pages={025010},
  year={2011},
  publisher={IOP Publishing}
}

@article{liu2012tensor,
  title={Tensor completion for estimating missing values in visual data},
  author={Liu, Ji and Musialski, Przemyslaw and Wonka, Peter and Ye, Jieping},
  journal={IEEE transactions on pattern analysis and machine intelligence},
  volume={35},
  pages={208--220},
  year={2012},
  publisher={IEEE}
}

@article{yuan2016tensor,
  title={On tensor completion via nuclear norm minimization},
  author={Yuan, Ming and Zhang, Cun-Hui},
  journal={Foundations of Computational Mathematics},
  volume={16},
  pages={1031--1068},
  year={2016},
  publisher={Springer}
}

@article{yuan2017incoherent,
  title={Incoherent tensor norms and their applications in higher order tensor completion},
  author={Yuan, Ming and Zhang, Cun-Hui},
  journal={IEEE Transactions on Information Theory},
  volume={63},
  pages={6753--6766},
  year={2017},
  publisher={IEEE}
}

@article{xia2017polynomial,
  title={On polynomial time methods for exact low rank tensor completion},
  author={Xia, Dong and Yuan, Ming},
  journal={arXiv preprint arXiv:1702.06980},
  year={2017}
}

@inproceedings{jain2014provable,
  title={Provable tensor factorization with missing data},
  author={Jain, Prateek and Oh, Sewoong},
  booktitle={Advances in Neural Information Processing Systems},
  pages={1431--1439},
  year={2014}
}

@article{barak2016noisy,
  title={Noisy tensor completion via the sum-of-squares hierarchy},
  author={Barak, Boaz and Moitra, Ankur},
  journal={Journal of Machine Learning Research},
  booktitle={Conference on Learning Theory},
  volume={49},
  pages={1--29},
  year={2016},
  organization={PMLR}
}

@article{cai2021nonconvex,
  title={Nonconvex low-rank tensor completion from noisy data},
  author={Cai, Changxiao and Li, Gen and Poor, H Vincent and Chen, Yuxin},
  journal={Operations Research},
  year={2021},
  volume={70},
  pages={1--19},
  publisher={INFORMS}
}

@article{xia2021statistically,
  title={Statistically optimal and computationally efficient low rank tensor completion from noisy entries},
  author={Xia, Dong and Yuan, Ming and Zhang, Cun-Hui},
  journal={The Annals of Statistics},
  volume={49},
  pages={76--99},
  year={2021},
  publisher={Institute of Mathematical Statistics}
}

@article{neugebauer2007causal,
  title={Causal inference in longitudinal studies with history-restricted marginal structural models},
  author={Neugebauer, Romain and van der Laan, Mark J and Joffe, Marshall M and Tager, Ira B},
  journal={Electronic Journal of Statistics},
  volume={1},
  pages={119--154},
  year={2007},
  publisher={NIH Public Access}
}

@article{feldman2004administration,
  title={Administration of parenteral iron and mortality among hemodialysis patients},
  author={Feldman, Harold I and Joffe, Marshall and Robinson, Bruce and Knauss, Jill and Cizman, Borut and Guo, Wensheng and Franklin-Becker, Eunice and Faich, Gerald},
  journal={Journal of the American Society of Nephrology},
  volume={15},
  pages={1623--1632},
  year={2004},
  publisher={Am Soc Nephrol}
}

@incollection{robins2000marginal,
  title={Marginal structural models versus structural nested models as tools for causal inference},
  author={Robins, James M},
  booktitle={Statistical Models in Epidemiology, the Environment, and Clinical Trials},
  volume = {116},
  pages={95--133},
  year={2000},
  publisher={Springer}
}

@article{robins2000marginal2,
  title={Marginal structural models and causal inference in epidemiology},
  author={Robins, James M and Hernan, Miguel Angel and Brumback, Babette},
  journal={Epidemiology},
  volume={11},
  pages={550--560},
  year={2000},
  publisher={Lww}
}

@book{van2003unified,
  title={Unified methods for censored longitudinal data and causality},
  author={Van der Laan, Mark J and Laan, MJ and Robins, James M},
  year={2003},
  publisher={Springer Science \& Business Media}
}

@article{robins1986new,
  title={A new approach to causal inference in mortality studies with a sustained exposure period—application to control of the healthy worker survivor effect},
  author={Robins, James},
  journal={Mathematical Modelling},
  volume={7},
  pages={1393--1512},
  year={1986},
  publisher={Elsevier}
}

@article{mortimer2005application,
  title={An application of model-fitting procedures for marginal structural models},
  author={Mortimer, Kathleen M and Neugebauer, Romain and Van Der Laan, Mark and Tager, Ira B},
  journal={American Journal of Epidemiology},
  volume={162},
  pages={382--388},
  year={2005},
  publisher={Oxford University Press}
}

@article{rosenbaum1983central,
  title={The central role of the propensity score in observational studies for causal effects},
  author={Rosenbaum, Paul R and Rubin, Donald B},
  journal={Biometrika},
  volume={70},
  pages={41--55},
  year={1983},
  publisher={Oxford University Press}
}

@article{kolda2009tensor,
  title={Tensor decompositions and applications},
  author={Kolda, Tamara G and Bader, Brett W},
  journal={SIAM Review},
  volume={51},
  pages={455--500},
  year={2009},
  publisher={SIAM}
}

@article{athey2021matrix,
  title={Matrix completion methods for causal panel data models},
  author={Athey, Susan and Bayati, Mohsen and Doudchenko, Nikolay and Imbens, Guido and Khosravi, Khashayar},
  journal={Journal of the American Statistical Association},
  pages={1--15},
  volume={116},
  year={2021},
  publisher={Taylor \& Francis}
}

@article{abadie2010synthetic,
  title={Synthetic control methods for comparative case studies: Estimating the effect of California’s tobacco control program},
  author={Abadie, Alberto and Diamond, Alexis and Hainmueller, Jens},
  journal={Journal of the American statistical Association},
  volume={105},
  pages={493--505},
  year={2010},
  publisher={Taylor \& Francis}
}

@inproceedings{zhou2017tensor,
  title={Tensor completion with side information: A riemannian manifold approach.},
  author={Zhou, Tengfei and Qian, Hui and Shen, Zebang and Zhang, Chao and Xu, Congfu},
  booktitle={Proceedings of the Twenty-Sixth International Joint Conference on Artificial Intelligence},
  pages={3539--3545},
  year={2017}
}

@article{acar2011all,
  title={All-at-once optimization for coupled matrix and tensor factorizations},
  author={Acar, Evrim and Kolda, Tamara G and Dunlavy, Daniel M},
  journal={arXiv preprint arXiv:1105.3422},
  year={2011}
}

@article{van2005history,
  title={History-adjusted marginal structural models and statically-optimal dynamic treatment regimens},
  author={van der Laan, Mark J and Petersen, Maya L and Joffe, Marshall M},
  journal={The International Journal of Biostatistics},
  volume={1},
  year={2005},
  publisher={De Gruyter}
}

@article{klopp2014noisy,
  title={Noisy low-rank matrix completion with general sampling distribution},
  author={Klopp, Olga},
  journal={Bernoulli},
  volume={20},
  pages={282--303},
  year={2014},
  publisher={Bernoulli Society for Mathematical Statistics and Probability}
}

@article{tucker1966some,
  title={Some mathematical notes on three-mode factor analysis},
  author={Tucker, Ledyard R},
  journal={Psychometrika},
  volume={31},
  pages={279--311},
  year={1966},
  publisher={Springer}
}

@article{massart2000constants,
  title={About the constants in Talagrand's concentration inequalities for empirical processes},
  author={Massart, Pascal},
  journal={The Annals of Probability},
  volume={28},
  pages={863--884},
  year={2000},
  publisher={Institute of Mathematical Statistics}
}

@article{luo2020tensor,
  title={Tensor Bernstein concentration inequalities with an application to sample estimators for high-order moments},
  author={Luo, Ziyan and Qi, Liqun and Toint, Philippe L},
  journal={Frontiers of Mathematics in China},
  volume={15},
  pages={367--384},
  year={2020},
  publisher={Springer}
}

@article{fan2001variable,
  title={Variable selection via nonconcave penalized likelihood and its oracle properties},
  author={Fan, Jianqing and Li, Runze},
  journal={Journal of the American statistical Association},
  volume={96},
  pages={1348--1360},
  year={2001},
  publisher={Taylor \& Francis}
}

@article{chen2007large,
  title={Large sample sieve estimation of semi-nonparametric models},
  author={Chen, Xiaohong},
  journal={Handbook of Econometrics},
  volume={6},
  pages={5549--5632},
  year={2007},
  publisher={Elsevier}
}

@article{chang2022convenient,
  title={Convenient tail bounds for sums of random tensors},
  author={Chang, Shih Yu and Lin, Wen-Wei},
  journal={Taiwanese Journal of Mathematics},
  volume={26},
  pages={571--606},
  year={2022},
  publisher={Mathematical Society of the Republic of China}
}

@article{negahban2012restricted,
  title={Restricted strong convexity and weighted matrix completion: Optimal bounds with noise},
  author={Negahban, Sahand and Wainwright, Martin J},
  journal={The Journal of Machine Learning Research},
  volume={13},
  pages={1665--1697},
  year={2012},
  publisher={JMLR. org}
}

@article{hamidi2019low,
  title={On low-rank trace regression under general sampling distribution},
  author={Hamidi, Nima and Bayati, Mohsen},
  journal={arXiv preprint arXiv:1904.08576},
  year={2019}
}

@article{kong2018new,
  title={New estimations on the upper bounds for the nuclear norm of a tensor},
  author={Kong, Xu and Li, Jicheng and Wang, Xiaolong},
  journal={Journal of Inequalities and Applications},
  volume={2018},
  pages={1--17},
  year={2018},
  publisher={SpringerOpen}
}

@article{hu2015relations,
  title={Relations of the nuclear norm of a tensor and its matrix flattenings},
  author={Hu, Shenglong},
  journal={Linear Algebra and its Applications},
  volume={478},
  pages={188--199},
  year={2015},
  publisher={Elsevier}
}

@article{lim2013blind,
  title={Blind multilinear identification},
  author={Lim, Lek-Heng and Comon, Pierre},
  journal={IEEE Transactions on Information Theory},
  volume={60},
  pages={1260--1280},
  year={2013},
  publisher={IEEE}
}

@article{derksen2016nuclear,
  title={On the nuclear norm and the singular value decomposition of tensors},
  author={Derksen, Harm},
  journal={Foundations of Computational Mathematics},
  volume={16},
  pages={779--811},
  year={2016},
  publisher={Springer}
}

@article{fan2016projected,
  title={Projected principal component analysis in factor models},
  author={Fan, Jianqing and Liao, Yuan and Wang, Weichen},
  journal={The Annals of Statistics},
  volume={44},
  pages={219--254},
  year={2016},
  publisher={NIH Public Access}
}

@article{landsberg2012tensors,
  title={Tensors: geometry and applications},
  author={Landsberg, Joseph M},
  journal={Representation Theory},
  volume={381},
  pages={3},
  year={2012}
}

@book{hogben2013handbook,
  title={Handbook of linear algebra},
  author={Hogben, Leslie},
  year={2013},
  publisher={CRC press}
}

@article{zhang2019cross,
  title={Cross: Efficient low-rank tensor completion},
  author={Zhang, Anru},
  journal={The Annals of Statistics},
  volume={47},
  pages={936--964},
  year={2019},
  publisher={Institute of Mathematical Statistics}
}

@article{bai2012statistical,
  title={Statistical analysis of factor models of high dimension},
  author={Bai, Jushan and Li, Kunpeng},
  journal={The Annals of Statistics},
  volume={40},
  pages={436--465},
  year={2012},
  publisher={Institute of Mathematical Statistics}
}

@article{yang2020doubly,
  title={Doubly robust inference when combining probability and non-probability samples with high dimensional data},
  author={Yang, Shu and Kim, Jae Kwang and Song, Rui},
  journal={Journal of the Royal Statistical Society: Series B (Statistical Methodology)},
  volume={82},
  pages={445--465},
  year={2020},
  publisher={Wiley Online Library}
}

@book{vershynin2018high,
  title={High-dimensional probability: An introduction with applications in data science},
  author={Vershynin, Roman},
  volume={47},
  year={2018},
  publisher={Cambridge university press}
}

@article{ma2017exploration,
  title={Exploration of large networks via fast and universal latent space model fitting},
  author={Ma, Zhuang and Ma, Zongming},
  journal={arXiv preprint arXiv:1705.02372},
  year={2017}
}

@article{cao2020multisample,
  title={Multisample estimation of bacterial composition matrices in metagenomics data},
  author={Cao, Yuanpei and Zhang, Anru and Li, Hongzhe},
  journal={Biometrika},
  volume={107},
  pages={75--92},
  year={2020},
  publisher={Oxford University Press}
}

@article{gigli2013log,
  title={From log Sobolev to Talagrand: a quick proof},
  author={Gigli, Nicola and Ledoux, Michel},
  journal={Discrete and Continuous Dynamical Systems-Series A},
  pages={dcds--2013},
  year={2013}
}

@misc{vincent1996sofa,
  title={The SOFA (Sepsis-related Organ Failure Assessment) score to describe organ dysfunction/failure},
  author={Vincent, J-L and Moreno, Rui and Takala, Jukka and Willatts, Sheila and De Mendon{\c{c}}a, Arnaldo and Bruining, Hajo and Reinhart, CK and Suter, PeterM and Thijs, Lambertius G},
  year={1996},
  publisher={Springer-Verlag}
}

@article{johnson2016mimic,
  title={MIMIC-III clinical database (version 1.4)},
  author={Johnson, Alistair and Pollard, Tom and Mark, Roger},
  journal={PhysioNet},
  year={2016}
}

@article{johnson2016mimicb,
  title={MIMIC-III, a freely accessible critical care database},
  author={Johnson, Alistair EW and Pollard, Tom J and Shen, Lu and Lehman, Li-wei H and Feng, Mengling and Ghassemi, Mohammad and Moody, Benjamin and Szolovits, Peter and Anthony Celi, Leo and Mark, Roger G},
  journal={Scientific Data},
  volume={3},
  pages={1--9},
  year={2016},
  publisher={Nature Publishing Group}
}

@article{goldberger2000physiobank,
  title={PhysioBank, PhysioToolkit, and PhysioNet: components of a new research resource for complex physiologic signals},
  author={Goldberger, Ary L and Amaral, Luis AN and Glass, Leon and Hausdorff, Jeffrey M and Ivanov, Plamen Ch and Mark, Roger G and Mietus, Joseph E and Moody, George B and Peng, Chung-Kang and Stanley, H Eugene},
  journal={Circulation},
  volume={101},
  pages={215--220},
  year={2000},
  publisher={Am Heart Assoc}
}

@article{purushotham2018benchmarking,
  title={Benchmarking deep learning models on large healthcare datasets},
  author={Purushotham, Sanjay and Meng, Chuizheng and Che, Zhengping and Liu, Yan},
  journal={Journal of Biomedical Informatics},
  volume={83},
  pages={112--134},
  year={2018},
  publisher={Elsevier}
}

@article{ning2020robust,
  title={Robust estimation of causal effects via a high-dimensional covariate balancing propensity score},
  author={Ning, Yang and Sida, Peng and Imai, Kosuke},
  journal={Biometrika},
  volume={107},
  pages={533--554},
  year={2020},
  publisher={Oxford University Press}
}

@article{xia2022inference,
  title={Inference for low-rank tensors—no need to debias},
  author={Xia, Dong and Zhang, Anru R and Zhou, Yuchen},
  journal={The Annals of Statistics},
  volume={50},
  pages={1220--1245},
  year={2022},
  publisher={Institute of Mathematical Statistics}
}

@inproceedings{bica2020time,
  title={Time series deconfounder: Estimating treatment effects over time in the presence of hidden confounders},
  author={Bica, Ioana and Alaa, Ahmed and Van Der Schaar, Mihaela},
  booktitle={International Conference on Machine Learning},
  pages={884--895},
  year={2020},
  organization={PMLR}
}

@article{tropp2015introduction,
  title={An introduction to matrix concentration inequalities},
  author={Tropp, Joel A and others},
  journal={Foundations and Trends{\textregistered} in Machine Learning},
  volume={8},
  pages={1--230},
  year={2015},
  publisher={Now Publishers, Inc.}
}

@article{nguyen2015tensor,
  title={Tensor sparsification via a bound on the spectral norm of random tensors},
  author={Nguyen, Nam H and Drineas, Petros and Tran, Trac D},
  journal={Information and Inference: A Journal of the IMA},
  volume={4},
  pages={195--229},
  year={2015}
}

@article{wang2020learning,
  title={Learning from binary multiway data: Probabilistic tensor decomposition and its statistical optimality},
  author={Wang, Miaoyan and Li, Lexin},
  journal={Journal of Machine Learning Research},
  volume={21},
  number={154},
  year={2020}
}

@article{agarwal2022synthetic,
  title={Synthetic blip effects: Generalizing synthetic controls for the dynamic treatment regime},
  author={Agarwal, Anish and Syrgkanis, Vasilis},
  journal={arXiv preprint arXiv:2210.11003},
  year={2022}
}

@inproceedings{agarwal2023causal,
  title={Causal matrix completion},
  author={Agarwal, Anish and Dahleh, Munther and Shah, Devavrat and Shen, Dennis},
  booktitle={The Thirty Sixth Annual Conference on Learning Theory},
  pages={3821--3826},
  year={2023},
  organization={PMLR}
}

@article{ibriga2022covariate,
  title={Covariate-assisted sparse tensor completion},
  author={Ibriga, Hilda S and Sun, Will Wei},
  journal={Journal of the American Statistical Association},
  pages={1--15},
  year={2022},
  publisher={Taylor \& Francis}
}

@article{belloni2017program,
  title={Program evaluation and causal inference with high-dimensional data},
  author={Belloni, Alexandre and Chernozhukov, Victor and Fernandez-Val, Ivan and Hansen, Christian},
  journal={Econometrica},
  volume={85},
  pages={233--298},
  year={2017},
  publisher={Wiley Online Library}
}

@article{robins1994correcting,
  title={Correcting for non-compliance in randomized trials using structural nested mean models},
  author={Robins, James M},
  journal={Communications in Statistics-Theory and methods},
  volume={23},
  pages={2379--2412},
  year={1994},
  publisher={Taylor \& Francis}
}

@article{li2020g,
  title={G-Net: a deep learning approach to G-computation for counterfactual outcome prediction under dynamic treatment regimes},
  author={Li, Rui and Shahn, Zach and Li, Jun and Lu, Mingyu and Chakraborty, Prithwish and Sow, Daby and Ghalwash, Mohamed and Lehman, Li-wei H},
  journal={Machine Learning for Health},
  year={2021}
}
\newpage

\appendix

\global\long\def\theequation{A\arabic{equation}}%
\setcounter{equation}{0}

\global\long\def\thesection{A\arabic{section}}%
\setcounter{section}{0}

\global\long\def\thetable{A\arabic{table}}%
\setcounter{table}{0}

\global\long\def\thelemma{A\arabic{lemma}}%

\global\long\def\thetheorem{A\arabic{theorem}}%
\setcounter{theorem}{0}

\global\long\def\thecondition{A\arabic{condition}}%
\global\long\def\theremark{A\arabic{remark}}%
\global\long\def\thestep{A\arabic{step}}%

\global\long\def\theassumption{A\arabic{assumption}}%
\setcounter{assumption}{0}

\global\long\def\thefigure{A\arabic{figure}}%
\setcounter{figure}{0}

\global\long\def\theproposition{A\arabic{proposition}}%

\setcounter{algocf}{0} 
\makeatletter
\renewcommand{\thealgocf}{A\@arabic\c@algocf}
\makeatother

\section{Additional illustrations}\label{sec:additional_illustration}
\subsection{Details for simulation}\label{subsubsec:tensor_rep}
\subsubsection{Tensor representation for the outcomes}
Let the time-varying covariate $X_{i,t}^{a_{i,(t-2):t}}=\mathbf{1}_{d_{0}}^{\intercal} X_{i,0}+\eta_1 a_{i,t} + \eta_2 a_{i,t-1} + \eta_3 a_{i,t-2}$, the generative models for $\mathcal{Y}_{i,t,l}=\mathcal{Y}_{i,t,\{a_{i,(t-k+1):t}\}_{(10)}}$ with $l=\{a_{i,(t-k+1):t}\}_{(10)}$ under (M1) and (M2) will be
\begin{enumerate}
\item [(M1)] $\mathcal{Y}_{i,t,\{a_{i,(t-k+1):t}\}_{(10)}}=4\cdot\mathbf{1}_{d_{0}}^{\intercal}X_{i,0}+\beta_{1}X_{i,t}^{a_{(t-2):t}}+\beta_{2}X_{i,t-1}^{a_{(t-3):(t-1)}}+\gamma_{1}a_{i,t}+\gamma_{2}a_{i,t-1}+\mathcal{E}_{i,t,l}.$
\item [(M2)]
\begin{align*}
 & \mathcal{Y}_{i,t,\{a_{i,(t-k+1):t}\}_{(10)}}=4\sum_{j=1}^{J^{*}}d_{0}^{-1}\mathbf{1}_{d_{0}}^{\intercal}\Upsilon_{j}(X_{i,0})+4\cdot\mathbf{1}_{d_{0}}^{\intercal}X_{i,0}\cdot2^{-t}\\
 & +3\cdot\mathbf{1}_{k}^{\intercal}a_{(t-k+1):t}\cdot\mathbf{1}_{d_{0}}^{\intercal}X_{i,0}+2a_{i,t}\cdot\mathbf{1}_{d_{0}}^{\intercal}X_{i,0}+a_{i,t-1}\cdot\mathbf{1}_{d_{0}}^{\intercal}X_{i,0}\\
 & +\mathbf{1}_{k}^{\intercal}a_{(t-k+2):t}\cdot\mathbf{1}_{d_{0}}^{\intercal}X_{i,0}\cdot X_{i,t}^{a_{(t-2):t}}+\mathbf{1}_{k}^{\intercal}a_{(t-k+1):(t-1)}\cdot\mathbf{1}_{d_{0}}^{\intercal}X_{i,0}\cdot X_{i,t-1}^{a_{(t-3):(t-1)}}\\
 & +5\beta_{1}X_{i,t}^{a_{(t-2):t}}\cdot\mathbf{1}_{d_{0}}^{\intercal}X_{i,0}+3\beta_{2}X_{i,t-1}^{a_{(t-3):(t-1)}}\cdot\mathbf{1}_{d_{0}}^{\intercal}X_{i,0}\\
 & +\beta_{1}X_{i,t}^{a_{(t-2):t}}+\beta_{2}X_{i,t-1}^{a_{(t-3):(t-1)}}+\gamma_{1}a_{i,t}+\gamma_{2}a_{i,(t-1)}+\mathcal{E}_{i,t,l},
\end{align*}
\end{enumerate}
where $\mathcal{E}_{i,t,l}$ is generated with independent standard
normal distribution $\mathcal{N}(0,1)$, $\beta_{1}=\beta_{2}=1$,
$\gamma_{1}=\gamma_{2}=3$, $\Upsilon_{j}(\cdot)$ is the $j$-th
Legendre polynomial function defined by 
\[
\Upsilon_{0}(X)=1,\quad\Upsilon_{1}(X)=X,\quad\Upsilon_{j}(X)=\frac{(2j-1)\cdot\Upsilon_{j-1}(X)X-(j-1)\Upsilon_{j-1}(X)}{j},
\]
and $J^{*}=2$ is the highest order of the Legendre polynomial function. Note that our low-rank model formulation in (\ref{eq:Tucker-Y}) encompasses
the generative models of $\mathcal{Y}_{i,t,l}$ in both (M1) and (M2) with the potential outcomes tensor defined as $\mathcal{Y}_{i,t,l}=\mathcal{G}\times_{1}\boldsymbol{U}_{1,i}^{\intercal}\times_{2}\boldsymbol{U}_{2,t}^{\intercal}\times_{1}\boldsymbol{U}_{3,l}^{\intercal}+\mathcal{E}_{i,t,l}$ with $\mathcal{G}$ representing the loading coefficients for the associated factors $(\boldsymbol{U}_{1,i},\boldsymbol{U}_{2,i},\boldsymbol{U}_{3,i })$. For (M1), we have
\begin{align*}
& \boldsymbol{U}_{1,i}=\{1,\mathbf{1}_{d_{0}}^{\intercal}X_{i,0}\}^{\intercal}\in\mathbb{R}^{2\times 1},\quad\boldsymbol{U}_{2,t}=(1)\in\mathbb{R}^{1\times 1},\quad \boldsymbol{U}_{3,l}=(1,l_{(b),1},\cdots,l_{(b),k})\in\mathbb{R}^{(k+1)\times 1},
\end{align*}
 For (M2), let the basis functions $\Phi(\boldsymbol{X}_{0})=\{1,\Upsilon_{1}(\boldsymbol{X}_{0}),\Upsilon_{2}(\boldsymbol{X}_{0})\}\in\mathbb{R}^{N\times(2d_{0}+1)}$, we have
\begin{align*}
 & \boldsymbol{U}_{1,i}=\{1,\sum_{j=1}^{J^{*}}\mathbf{1}_{d_{0}}^{\intercal}\Upsilon_{j}(X_{i,0}),\mathbf{1}_{d_{0}}^{\intercal}X_{i,0}, (\mathbf{1}_{d_{0}}^{\intercal}X_{i,0})^2\}^{\intercal}\in\mathbb{R}^{4\times 1},\quad\boldsymbol{U}_{2,t}=(1,2^{-t})\in\mathbb{R}^{2\times 1},\\
 & \boldsymbol{U}_{3,l}=(1,l_{(b),1},\cdots,l_{(b),k},\mathbf{1}_{k}^{\intercal}l_{(b)},u_{3}^{l})\in\mathbb{R}^{(k+3)\times 1},\\
 & u_{3}^{l}=
 \eta_{1}a_{i,t}\mathbf{1}_{k-1}^{\intercal}a_{(t-k+2):t}+\eta_{2}a_{i,t-1}\mathbf{1}_{k-1}^{\intercal}a_{(t-k+2):t}+\eta_{3}a_{i,t-2}\mathbf{1}_{k-1}^{\intercal}a_{(t-k+2):t}\\
 & +\eta_{1}a_{i,t-1}\mathbf{1}_{k-1}^{\intercal}a_{(t-k+1):(t-1)}+\eta_{2}a_{i,t-2}\mathbf{1}_{k-1}^{\intercal}a_{(t-k+1):(t-1)}+\eta_{3}a_{i,t-3}\mathbf{1}_{k-1}^{\intercal}a_{(t-k+1):(t-1)},
\end{align*}
where the factor $u_3^l$ is induced by $X_{i,t}^{a_{(t-2):t}} \mathbf{1}_{k-1}^{\intercal}a_{(t-k+2):t}  + X_{i,t-1}^{a_{(t-3):(t-1)}}\mathbf{1}_{k-1}^{\intercal}a_{(t-k+1):(t-1)}$. For (M2) with covariate-orthogonal component $\boldsymbol{\Gamma}$, the potential outcome tensor admits the same low-rank model formulation as (M2) with the tensor factors of mode-1 modified to $\boldsymbol{U}_{1,i}=\{1,\sum_{j=1}^{J^{*}}\mathbf{1}_{d_{0}}^{\intercal}\Upsilon_{j}(X_{i,0})+\sum_{j=1}^{J^{*}}\Gamma_{i,j},\mathbf{1}_{d_{0}}^{\intercal}X_{i,0},(\mathbf{1}_{d_{0}}^{\intercal}X_{i,0})^2\}\in\mathbb{R}^{4}$, where  $\mathcal{Y}_{i,t,\{a_{i,(t-k+1):t}\}_{(10)}}$  is defined by
\begin{align*}
 & \mathcal{Y}_{i,t,\{a_{i,(t-k+1):t}\}_{(10)}}=4\sum_{j=1}^{J^{*}}\{d_{0}^{-1}\mathbf{1}_{d_{0}}^{\intercal}\Upsilon_{j}(X_{i,0})+\Gamma_{i,j}\}+4\cdot\mathbf{1}_{d_{0}}^{\intercal}X_{i,0}\cdot2^{-t}\\
 & +3\cdot\mathbf{1}_{k}^{\intercal}a_{(t-k+1):t}\cdot\mathbf{1}_{d_{0}}^{\intercal}X_{i,0}+2a_{i,t}\cdot\mathbf{1}_{d_{0}}^{\intercal}X_{i,0}+a_{i,t-1}\cdot\mathbf{1}_{d_{0}}^{\intercal}X_{i,0}\\
 & +\mathbf{1}_{k}^{\intercal}a_{(t-k+1):t}\cdot\mathbf{1}_{d_{0}}^{\intercal}X_{i,0}\cdot X_{i,t}^{a_{(t-2):t}}+\mathbf{1}_{k}^{\intercal}a_{(t-k+1):t}\cdot\mathbf{1}_{d_{0}}^{\intercal}X_{i,0}\cdot X_{i,t-1}^{a_{(t-3):(t-1)}}\\
 & +5\beta_{1}X_{i,t}^{a_{(t-2):t}}\cdot\mathbf{1}_{d_{0}}^{\intercal}X_{i,0}+3\beta_{2}X_{i,t-1}^{a_{(t-3):(t-1)}}\cdot\mathbf{1}_{d_{0}}^{\intercal}X_{i,0}\\
 & +\beta_{1}X_{i,t}^{a_{(t-2):t}}+\beta_{2}X_{i,t-1}^{a_{(t-3):(t-1)}}+\gamma_{1}a_{i,t}+\gamma_{2}a_{i,t-1}+\mathcal{E}_{i,t,l}.
\end{align*}

\subsubsection{Hyperparameter tuning}\label{subsec:effect_hyper}

In this subsection, we examine the effect of the tuning process for the hyperparameters. First, we generate $(A_{i,t},\mathcal{Y}_{i,t,l})$ following
the same data generating procedures as (A1) and (M2) with fixed multi-linear ranks and covariate-orthogonal noises ${\Gamma}_{i,j}$ (none or strong) as defined in Section \ref{subsec:effect-semi}. Instead of perceiving the multi-linear ranks as known in Sections \ref{subsec:effect-models} and \ref{subsec:effect-semi}, we use the BIC criterion in (\ref{eq:BIC}) to sequentially tune these hyperparameters. In particular, we first randomly draw the parameter candidates $r_1^{\dagger}$ and $r_2^{\dagger}$ from their tuning ranges, and tune the remaining rank $r_3$ via (\ref{eq:BIC}) with fixed $r_1^{\dagger}$ and $r_2^{\dagger}$. Next, given the tuned rank $r_3$, we tune the other two ranks in a similar vein. We set the tuning range for rank $r_1$ to be $\{1,3,5,7\}$, $r_2$ to be $\{2,4,6\}$ and $r_3$ to be $\{6,8,10\}$. 

\begin{figure}[!htbp]
\centering
\includegraphics[width=.9\linewidth]{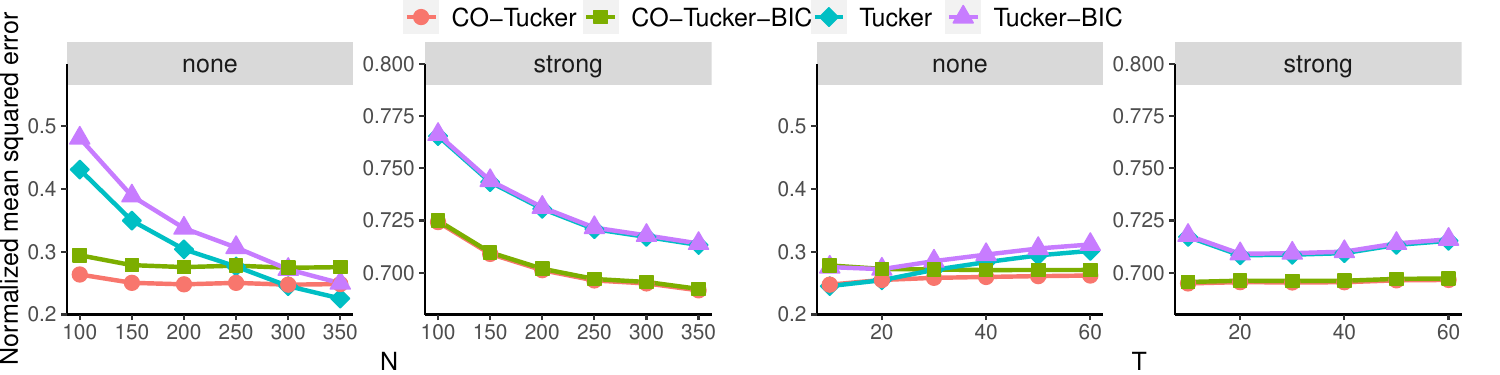}
\caption{\label{fig:mse-N-BIC} Averaged normalized mean squared tensor
error, when $T=10$, $N=100,150,\cdots,350$ (left), and $N=300$, $T=10,20,\cdots,60$ (right) under none, and strong noises due to the covariate-orthogonal component ${\Gamma}_{i,j}$; \textsc{tucker} and \textsc{co-tucker} are the \textit{vanilla} and the covariate-assisted tensorized HRMSMs with known multi-linear ranks $(r_1,r_2,r_3)$, \textsc{co-tucker-bic} and \textsc{tucker-bic} are the estimators with adaptively tuned multi-linear ranks $(r_1,r_2,r_3)$.}
\end{figure}

Figures \ref{fig:mse-N-BIC} presents the normalized mean squared error $\ell_{2}(\widehat{\mathcal{Y}})$ when the data-adaptive tuning process is adopted. It can be seen that the BIC-based tuning procedure is subject to a minor quality drop in the absence of covariate-independent components ${\Gamma}_{i,j}$ when compared to the oracle estimator with known multi-linear ranks. But when the covariate-orthogonal noise ${\Gamma}_{i,j}$ is strong, BIC-tuned estimators achieve comparable normalized mean squared
errors $\ell_{2}(\widehat{\mathcal{Y}})$ as the oracle estimators over the varying amounts of $N$ and $T$. These findings highlight the practicality of the BIC-based tuning scheme in real-world setups, where covariate-orthogonal noises are ubiquitous and inevitable. 

\subsubsection{Additional simulation studies for ATE estimation\label{subsubsec:sim_add}}
Considering the estimation error of the average treatment effect $\tau_{\text{ATE}}^{l,l'}$ defined in Section \ref{subsubsec:low-rank}, the Frobenius norm error in Theorem \ref{thm:bound-tensor} ensures the convergence of the ATE estimation: 
    $$
    |\tau_{\ate}^{l,l'} - \widehat{\tau}_{\ate}^{l,l'}| = 
    (\mathcal{Y}^* - \widehat{\mathcal{Y}}) \times_1 \frac{\mathbf{1}_N}{N}
    \times_2
    \frac{\mathbf{1}_T}{T}
    \times_3 \{e_l(K) - e_{l'}(K)\}
    \lesssim 
    \frac{\|\mathcal{Y}^* - \widehat{\mathcal{Y}}\|}{\sqrt{NT}},
    $$
where the right-hand side converges to zero as $N$ and $T$ grow. To provide empirical evidence, we present the estimated ATE comparing the fully active regime $\mathbf{1}_k$ and the baseline regime $\mathbf{0}_k$ using the same data-generating procedures outlined in Section \ref{sec:Simulations}. 

\begin{figure}[htbp]
    \centering
    \includegraphics[width=.9\linewidth]{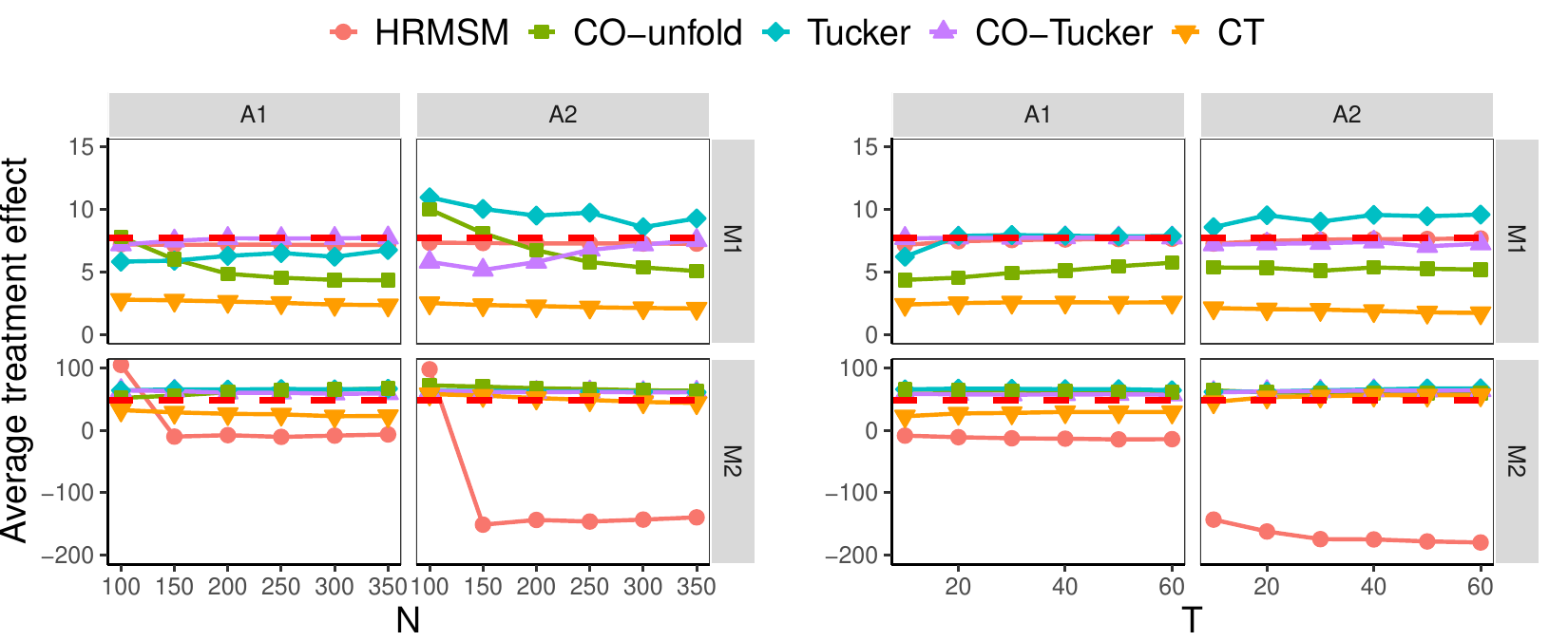}
    \caption{Estimations of the averaged treatment effect $\tau_{\ate}^{K,0}$ when $T=10$, $N=100,150,\cdots,350$ (left), and $N=300$, $T=10,20,\cdots,60$ (right) under outcome models (M1) and (M2)
and treatment assignments (A1) and (A2); \textsc{hrmsm} is the history-restricted marginal structural model, \textsc{co-unfold} is the matrix factor model based on unfolded tensor, \textsc{tucker} is the \textit{vanilla} tensorized HRMSM, \textsc{co-tucker} is the covariate-assisted tensorized HRMSM{ , and {\sc ct} is the Causal Transformer}; the red solid line stands for the true average treatment effect.}
    \label{fig:ATE}
\end{figure}

Figure \ref{fig:ATE} illustrates that HRMSM can identify the ATE in the presence of time-varying confounding when the outcome follows a simple linear model (M1). However, when the outcome model follows the complex additive model (M2), HRMSM exhibits substantial bias in ATE estimation. { In addition, the CT exhibits a non-negligible bias across all settings, which does not diminish as $N$ or $T$ increases. This appears to stem both from the accumulation of autoregressive prediction errors and its inability to update time-varying balanced representations under counterfactual interventions. } In contrast, our tensorized HRMSMs demonstrate robustness and satisfactory performance regardless of the complexity of the outcome model. In cases where propensity scores present increased variability, as in (A2), the covariate-assisted tensorized HRMSM is preferable as it leverages additional information from the baseline covariates to impute the missing potential outcomes.

\subsection{Details for application}
\subsubsection{Comparisons with matrix completion}\label{subsubsec:comparison_MC}

In this subsection, we numerically compare our proposed method with the matrix completion method via nuclear norm minimization (MC-NNM) \citep{athey2021matrix} and the synthetic nearest neighbors (SNN) algorithm \citep{agarwal2023causal}. In theory, both methods require that the expected number of control units be large enough to recover the post-treatment entries. They also implicitly assume that the potential outcomes are indexed only by the contemporaneous treatment for that unit and not by its past treatment. Hence, we set the length of the treatment regime $k$ to be $1$. The outcome of interest here is the potential outcomes matrix $Y(0)$, and the goal is to impute the missing entries of $Y(0)$ under staggered adoption. 

\begin{figure}[!htbp]
    \centering
    \includegraphics[width=.65\linewidth]{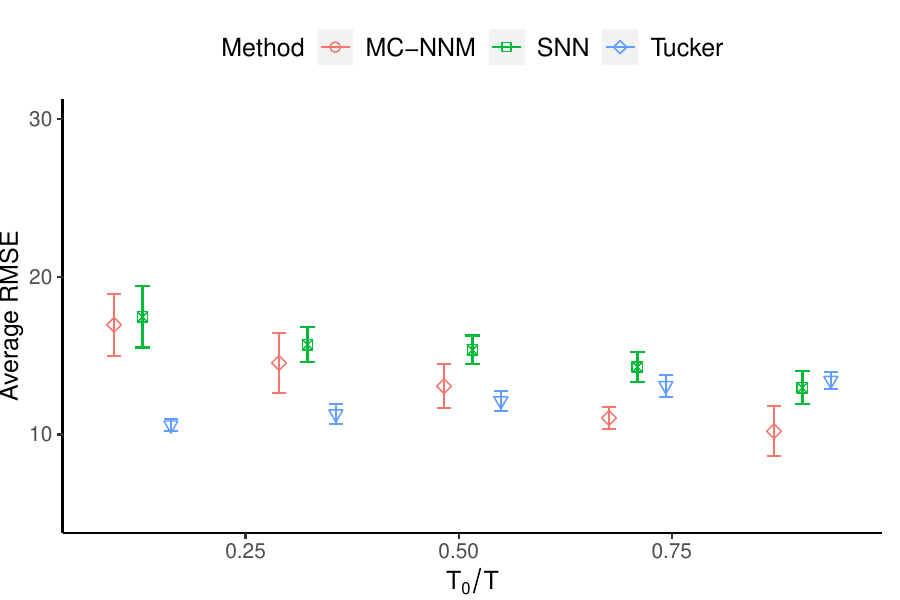}
    \caption{Averaged root mean squared error for different ratios $T_0/T$; \textsc{mc-nnm} is the matrix completion via nuclear norm minimization \citep{athey2021matrix}, \textsc{snn} is the matrix completion via synthetic nearest neighbors \citep{agarwal2023causal}, \textsc{tucker} is the \textit{vanilla} tensorized HRMSM with $k=1$.}
    \label{fig:cal_RMSE}
\end{figure}

We consider the California smoking data studied in \cite{abadie2010synthetic} with $N=39$ states over $T=31$ years, including $38$ controls units and one treated unit (i.e., California) which will be removed from our experiment. Next, we randomly selected $N_t=35$ control units and consider them to be treated at some point after $T_0$ years. The random selection procedure is repeated $10$ times and Figure \ref{fig:cal_RMSE} displays the averaged root mean squared error for the post-treatment entries against different ratios $T_0/T$. Our method utilizes the post-treatment outcomes by storing them in $Y(1)$ and stacks $Y(1)$ with the observed $Y(0)$ to form the observed outcomes tensor. By leveraging additional information from $Y(1)$, it achieves desirable performances in predicting the post-treatment outcomes across all the scenarios even when $T_0$ is small (i.e., the data presents a large proportion of missingness in the control group). Therefore, our method achieves stable numerical results regardless of the time that subjects are first
exposed to the treatment.

\subsubsection{MIMIC III dataset}
Table \ref{tab:selected-covariate} lists the selected baseline covariates $\boldsymbol{X}_{0}\in\mathbb{R}^{N\times11}$
and the time-varying covariates $\boldsymbol{X}_{t}\in\mathbb{R}^{N\times7},t=1,\cdots, T$ for the Medical Information Mart for Intensive Care (MIMIC) database.
\begin{table}[!htbp]
\centering
\caption{\label{tab:selected-covariate} Selected baseline and time-varying
covariates for the MIMIC III database with units in parentheses for
continuous variables and number of levels in parentheses for categorical
variables. A detailed summary of the covariates is given in \citet{johnson2016mimic}.}
\vspace{0.5cm}
\begin{tabular}{lll}
\hline 
 & Covariate & Type\tabularnewline
\hline 
Baseline  & gender (2, 0=female, 1=male) & Categorical\tabularnewline
 & age (years) & \multirow{10}{*}{Continuous}\tabularnewline
 & admission weight (kg) & \tabularnewline
 & admission temperature (Celsius) & \tabularnewline
 & glucose amount (mg/dL) & \tabularnewline
 & blood urea nitrogen amount (mg/dL) & \tabularnewline
 & creatinine amount (mg/dL) & \tabularnewline
 & white blood cell count (E9/L) & \tabularnewline
 & Glasgow Coma Score (0-15) & \tabularnewline
 & sodium amount (mEq/L) & \tabularnewline
 & total input amount (mL) & \tabularnewline
 & SOFA score at baseline (0-24) \\
\hline 
Time-varying & heart rate (bpm) & \multirow{7}{*}{Continuous}\tabularnewline
 & respiratory rate (cpm) & \tabularnewline
 & systolic blood pressure (mmHg) & \tabularnewline
 & diastolic blood pressure (mmHg) & \tabularnewline
 & oxygen saturation (\%) & \tabularnewline
 & PaO2 / FiO2 ratio (\%) & \tabularnewline
 & dose of vasopressin administration (unit)  & \tabularnewline
\hline 
\end{tabular}
\end{table}

\section{Proofs}\label{sec:additional_proof}
\subsection{Proof of Theorem \ref{thm:bound-tensor}\label{subsubsec:proof_thm1}}

\subsubsection{Upper bounds of (\ref{eq:part1}) and (\ref{eq:part3}) under penalized logistic modeling}

Under Assumption \ref{assum:SRA}, $\mathcal{W}_{i,t,l}=\mathbb{P}(A_{i,(t-k+1):t}=l_{(b)}\mid H_{i,t})^{-1}$,
and we assume that the propensity score at time $t$ follows a logistic model with covariate $H_{i,t}$. The estimation error of $\widehat{\mathcal{W}}_{i,t,l}$ can be bounded by
\begin{align*}
|\widehat{\mathcal{W}}_{i,t,l}-\mathcal{W}_{i,t,l}| & =\left|\prod_{j=t-k+1}^{t}\text{expit}(\widehat{\alpha}^{\intercal} H_{i,j})-\prod_{j=t-k+1}^{t}\text{expit}(\alpha^{*\intercal}H_{i,j})\right|\\
 & \leq C_{0}\sup_{j=t-k+1,\cdots,t}|\text{expit}(\widehat{\alpha}^{\intercal}H_{i,j})-\text{expit}(\alpha^{*\intercal}H_{i,j})|.
\end{align*}
Further, we have
\begin{align*}
 & \mathbb{P}(\sup_{t=1,\cdots,T}|\text{expit}(\widehat{\alpha}^{\intercal}H_{i,t})-\text{expit}(\alpha^{*\intercal}H_{i,t})|\geq\nu)\\
\leq & \mathbb{P}(\sup_{t=1,\cdots,T}\{\text{expit}(\tilde{\alpha}^{*\intercal}H_{i,t})+o_{\pr}(1)\}H_{i,t}^{\intercal}(\widehat{\alpha}_{t}-\alpha_{t}^{*})\geq\nu)\\
\leq & \sum_{t=1}^{T}\mathbb{P}\left\{C_{0}(1-\delta) \sum_{j\in\mathcal{J}_t}N^{1/2}(\widehat{\alpha}_{t,j}-\alpha_{t,j}^{*})\geq N^{1/2}\nu\right\}\\
\leq & 2\sum_{t=1}^T\exp\left\{ -\frac{cN\nu^{2}}{|\mathcal{J}_{t}|^2C_{0}^{2}(1-\delta)^{2}}\right\} ,
\end{align*}
where $\tilde{\alpha}^{*}$ lies between $\widehat{\alpha}_{t}$ and
$\alpha_{t}^{*}$, the second inequality uses Assumptions \ref{assmp:on_W(x)}
(a) and \ref{assum:positivity}, and the third inequality uses the concentration inequality for sub-Gaussian where $N^{1/2}(\widehat{\alpha}_{t,j}-\alpha_{t,j}^{*}),j\in\mathcal{J}_{t}$
is sub-Gaussian and $\widehat{\alpha}_{t,j}\rightarrow0,j\in\mathcal{J}_{t}^{\complement}$
by Lemma \ref{lem:oracle_property}. Therefore, we have 
\begin{align}
\sup_{t=1,\cdots,T}|\widehat{\mathcal{W}}_{i,t,l}-\mathcal{W}_{i,t,l}|&\leq C_{0}\sup_{t=1,\cdots,T}|\text{expit}(\widehat{\alpha}^{\intercal}H_{i,t})-\text{expit}(\alpha^{*\intercal}H_{i,t})|\nonumber\\
    &\leq C\max_{t=1,\cdots,T}|\mathcal{J}_t|\log^{1/2}(N+T+K)N^{-1/2}\label{eq:bound_W},
\end{align}
for some constant $C$ with a probability greater than $1-(N+T+K)^{-2}$ for any $\alpha\geq1$. The remaining terms require to bound the first-
and second-moment of a sub-Gaussian error, that is, $\||\mathcal{Y}_{i,t,l}-\mathcal{Y}_{i,t,l}^{*}|^{2}\|_{\psi_{1}}=\|\mathcal{Y}_{i,t,l}-\mathcal{Y}_{i,t,l}^{*}\|_{\psi_{2}}^{2}=\sigma^{2}$
\citep{vershynin2018high},
\[
\mathbb{P}\left\{\sum_{\Omega_{i,t,l}=1}(\mathcal{Y}_{i,t,l}-\mathcal{Y}_{i,t,l}^{*})^{2}\geq t\right\}\leq2\exp\left\{ -c\min\left(\frac{t^{2}}{NT\sigma^{4}},\frac{t}{\sigma^{2}}\right)\right\} =2\exp\left\{ -\frac{ct^{2}}{NT\sigma^{4}}\right\} .
\]
By matching the tail probabilities under Assumption \ref{asmp:incoherence}, we have 
\begin{align*}
\sum_{\Omega_{i,t,l}=1}(\mathcal{Y}_{i,t,l}-\mathcal{Y}_{i,t,l}^{*})^{2} & \leq C\sigma^{2}\sqrt{NT\log(N+T+K)},
\end{align*}
with probability higher than $1-(N+T+K)^{-2}$. Combine with the bound (\ref{eq:bound_W}), we have
\begin{align*}
 & \sum_{\Omega_{i,t,l}=1}(\widehat{\mathcal{W}}_{i,t,l}-\mathcal{W}_{i,t,l})(\mathcal{Y}_{i,t,l}-\mathcal{\widehat{Y}}_{i,t,l})^{2}\\
 & \leq C^{2}\max_{t=1,\cdots,T}|\mathcal{J}_t|N^{-1/2}\log^{1/2}(N+T+K)\left\{ \sigma^{2}\sqrt{NT\log(N+T+K)}+
 \|\mathcal{Y}^* - \widehat{\mathcal{Y}}\|_F^2
 \right\} 
\end{align*}
with probability greater than $1-2(N+T+K)^{-2}$ since $\sum_{\Omega_{i,t,l}=1}(\mathcal{Y}_{i,t,l}-\mathcal{\widehat{Y}}_{i,t,l})^{2}\leq 2\{\sum_{\Omega_{i,t,l}=1}(\mathcal{Y}_{i,t,l}-\mathcal{Y}^*_{i,t,l}) + \sum_{\Omega_{i,t,l}=1}(\mathcal{Y}^*_{i,t,l}-\mathcal{\widehat{Y}}_{i,t,l})\}$. Therefore,
\[
|I_{1}|+|I_{3}|\leq C_{1}\max_{t=1,\cdots,T}|\mathcal{J}_{t}|N^{-1/2}\log^{1/2}(N+T+K)\left\{ \sigma^{2}\sqrt{NT\log(N+T+K)}+
\|\mathcal{Y}^* - \widehat{\mathcal{Y}}\|_F^2
\right\} ,
\]
for some constant $C_{1}$.

\subsubsection{Upper bound of (\ref{eq:part2})}
Denote $L_{N,T}(\mathcal{Y};\mathcal{W})=\sum_{i,t}\mathcal{W}_{i,t,l}\{Y_{i,t}-\mathcal{Y}(i,t,l)\}^{2}/(NT), L_{N,T}(\mathcal{Y})=\E\{L_{N,T}(\mathcal{Y};\mathcal{W})\}$. Consider the restricted tensor space, which requires an upper bound
on the incoherence of the loading matrices and the spectral norm of
each matricization of the core tensor $\mathcal{G}$. Let
\begin{align*}
 & \mu(\boldsymbol{U}_{1})=\frac{N}{r_{1}}\|\boldsymbol{U}_{1}\|_{2,\infty}^{2},\mu(\boldsymbol{U}_{2})=\frac{T}{r_{2}}\|\boldsymbol{U}_{2}\|_{2,\infty}^{2},\mu(\boldsymbol{U}_{3})=\frac{K}{r_{3}}\|\boldsymbol{U}_{3}\|_{2,\infty}^{2},\\
 & \mu(\mathcal{X})=\max\{\mu(\boldsymbol{U}_{1}),\mu(\boldsymbol{U}_{2}),\mu(\boldsymbol{U}_{3})\},
\end{align*}
and the restricted tensor space $\mathfrak{C}(r_{1},r_{2},r_{3},\mu_{0},L_{0})$
is defined as
\begin{align*}
 & \mathfrak{C}(r_{1},r_{2},r_{3},\mu_{0},L_{0})\\
 & =\left\{ \mathcal{X}=\mathcal{G}\times_{1}\boldsymbol{U}_{1}\times_{2}\boldsymbol{U}_{2}\times_{3}\boldsymbol{U}_{3},\quad\mu(\mathcal{X})\leq\mu_{0},\quad\max_{k\in\{1,2,3\}}\|\mathcal{M}_{k}(\mathcal{G})\|\leq L_{0}\sqrt{\frac{NTK}{\mu_{0}^{3/2}(r_{1}r_{2}r_{3})^{1/2}}}\right\} .
\end{align*}
Now, using the definition of $\widehat{\mathcal{Y}}$, i.e., $L_{N,T}(\widehat{\mathcal{Y}};\widehat{\mathcal{W}})\leq L_{N,T}(\mathcal{Y}^{*};\mathcal{W}),$
we have 
\begin{align}
&\frac{1}{NT}\sum_{i,t}\mathcal{W}_{i,t,l}(\mathcal{Y}_{i,t,l}^{*}-\widehat{\mathcal{Y}}_{i,t,l})^{2}  \leq-\frac{2}{NT}\sum_{i,t}\mathcal{W}_{i,t,l}\mathcal{E}_{i,t,l}(\mathcal{Y}_{i,t,l}^{*}-\widehat{\mathcal{Y}}_{i,t,l})-\frac{I_{1}+I_{3}}{NT}\nonumber\\
 & \leq-\frac{2}{NT}\langle\mathcal{E}_{\Omega}^{w},\mathcal{Y}^{*}-\widehat{\mathcal{Y}}\rangle-\frac{I_{1}+I_{3}}{NT}\nonumber\\
 & \leq-\frac{2}{NT}\langle\mathcal{E}_{\Omega}^{w},(\mathcal{Y}^{*}-\widehat{\mathcal{Y}})\times_{1}\boldsymbol{P}_{\Phi(\boldsymbol{X})}+(\mathcal{Y}^{*}-\widehat{\mathcal{Y}})\times_{1}\boldsymbol{P}_{\Phi(\boldsymbol{X})}^{\perp}\rangle\nonumber\\
 & +\frac{1}{NT}(\|I_{1}\|+\|I_{3}\|)\nonumber\\
 & \leq\frac{2}{NT}\|\mathcal{E}_{\Omega}^{w}\|\left\{ \|(\mathcal{Y}^{*}-\widehat{\mathcal{Y}})\times_{1}\boldsymbol{P}_{\Phi(\boldsymbol{X})}\|_{*}+\|(\mathcal{Y}^{*}-\widehat{\mathcal{Y}})\times_{1}\boldsymbol{P}_{\Phi(\boldsymbol{X})}^{\perp}\|_{*}\right\} \label{eq:part2_supple} \\
 & +\frac{1}{NT}(\|I_{1}\|+\|I_{3}\|),\nonumber
\end{align}
where $\mathcal{E}_{\Omega}^{w}=\sum_{(i,t)\in\mathcal{O}}\mathcal{W}_{i,t,l}\mathcal{E}_{i,t,l}\cdot e_{i}(N)\otimes e_{j}(T)\otimes e_{l}(K)$
whose entries are all zero except its $(i,t,l)$-th entry and the fourth inequality is derived by Lemma \ref{lem:duality_tensor_nuclear}. The boundary of (\ref{eq:part2_supple}) is challenging and can be solved in the following steps.

\paragraph{Step 1} Obtain an upper bound for $\|\mathcal{E}_{\Omega}^{w}\|$. Previous literature mainly focuses on independent sampling \citep{negahban2012restricted,klopp2014noisy,hamidi2019low},
but it is not applicable in our setup since the entries in $\Omega$ tie to the received treatment regimes and are correlated by nature. To curb the impact of dependence between $\Omega_{i,t,l}$ and $\Omega_{i,t',l'}$ for any $t'\neq t$ and $l'\neq l$, an additional assumption is needed.
\begin{assumption}\label{assum:decay-covariance}
 For any $t'\geq t+k$ and some constant $c_0>0$, the treatment regimes $A_{i,(t-k+1):t}$ and $A_{i,(t'-k+1):t'}$ satisfy 
 $$
 1-c_0\leq \frac{\mathbb{P}(A_{i,(t'-k+1):t'}=a_{(t-k+1):t}\mid \mathbb{V}_{i,t},A_{i,(t-k+1):t})}{\mathbb{P}(A_{i,(t'-k+1):t'})=a_{(t'-k+1):t'})}\leq 1+c_0,
 $$
 where $a_{(t-k+1):t}$, $a_{(t'-k+1):t'}\in\mathbb{A}_k$. 
\end{assumption}
Assumption \ref{assum:decay-covariance} implies that two treatment regimes assigned at time points far apart are almost independent \citep{mandal2019weighted}. Under Assumption \ref{assum:decay-covariance}, we consider separating the sum of tensors $\mathcal{E}_{\Omega}^w$ by computing
$t$ modulo $k$ 
\[
\mathcal{E}_{\Omega}^{[h]}=\sum_{i=1}^N\sum_{\mod(t,k)=h}\mathcal{W}_{i,t,l}\mathcal{E}_{i,t,(A_{i,(t-k+1):t})_{(10)}}\cdot e_{i}(N)\otimes e_{t}(T)\otimes e_{(A_{i,(t-k+1):t})_{(10)}}(K),
\]
for $h=0,\cdots,k-1$, where $\mod(t,k)$ returns the remainder of $t$ divided by $k$, and $\mathcal{E}_{\Omega}^{[h]}$ is the sum of several
(nearly) independent random tensors under Assumption \ref{assum:decay-covariance}. Therefore, it reduces the likelihood where the mass of observation pattern $\Omega$ is concentrated on one or two particular slices. For example, if all the patients have always been treated during the study
period, the observation pattern $\Omega$ will not be well-conditioned
since its one-values shall concentrate on one particular slice of
$\Omega$, causing its spectral norm to be abnormally large. Together with Assumption \ref{assum:decay-covariance}, they collectively entail a similar sparse structure on $\Omega$ as implied by the uniform sampling design in \cite{klopp2014noisy}, \cite{cai2021nonconvex}, and \cite{xia2021statistically}. Therefore, the spectral norm of $\mathcal{E}_{\Omega}^w$ can be well-controlled by the Bernstein-type inequality in Lemma \ref{lem:(Tensor-Bernstein-Inequality)}. Similar results are presented in \cite{nguyen2015tensor}.
\begin{lemma}
\label{lem:(Tensor-Bernstein-Inequality)}(Tensor Bernstein Inequality
\citep[Theorem 4.3,][]{luo2020tensor}) Let $Z_{1},\cdots,Z_{N}$
be independent tensor in $\mathbb{R}^{d\times d\times d}$, such that
$\E(Z_{i})=0$ and $\|Z_{i}\|\leq D_{Z}$ for all $i\in[N]$. Let
$\sigma_{Z}$ be such that 
\[
\sigma_{Z}^{2}\geq\max\left\{ \left\|\E\left(\sum_{i=1}^{N}Z_{i}\overline{\square}\sum_{i=1}^{N}Z_{i}\right)\right\|,
\left\|\E\left(\sum_{i=1}^{N}Z_{i}\underline{\square}\sum_{i=1}^{N}Z_{i}\right)\right\|\right\} .
\]
Then for any $\alpha\geq0$ 
\[
\mathbb{P}(\|\sum_{i=1}^{N}Z_{i}\|^{\overline{\square}}\geq\alpha)\leq d^{2}\exp\left\{ \frac{-\alpha^{2}}{2\sigma_{Z}^{2}+(2D_{Z}\alpha)/3}\right\},
\]
where $\overline{\square}$ and $\underline{\square}$ are two generalized Einstein products of tensors, which are defined in Section \ref{proof_Tensor-Bernstein-Inequality}.
\end{lemma}
Define the sequence of independent random tensors as $\mathcal{E}_{\Omega_{1}}^{w},\cdots,\mathcal{E}_{\Omega_{N}}^{w}$.
For subject $i\in[N]$, define 
\begin{align*}
\mathcal{E}_{\Omega_{i}}^{w} & =\sum_{t\in\mathcal{O}_{i}}\mathcal{W}_{i,t,l}\mathcal{E}_{i,t,l}\cdot e_{i}(N)\otimes e_{j}(T)\otimes e_{l}(K)\\
 & =\sum_{t=1}^{T}\mathcal{W}_{i,t,l}\mathcal{E}_{i,t,(A_{i,(t-k+1):t})_{(10)}}\cdot e_{i}(N)\otimes e_{t}(T)\otimes e_{(A_{i,(t-k+1):t})_{(10)}}(K),
\end{align*}
where $\mathcal{O}_{i}=\{t:\Omega_{i,t,l}=1,t=1,\cdots,T\}$ and $\mathcal{E}_{\Omega}^{w}=\sum_{i=1}^{N}\mathcal{E}_{\Omega_{i}}^{w}$,
and $\E\{\mathcal{E}_{\Omega_{i}}^{w}\}=0$. By the dilation arguments
in \citet{chang2022convenient}, we define an operator $D(\cdot)$
that dilate the non-square tensor to a square tensor. For example,
$D(\mathcal{E}_{\Omega}^{w})\in\mathbb{R}^{(N+T+K)\times(N+T+K)\times(N+T+K)}$
and $\|D(\mathcal{E}_{\Omega}^{w})\|=\|\mathcal{E}_{\Omega}^{w}\|$
follows by the definition of norm. To resolve the dependence of $(A_{i,(t-k+1):t})_{(10)}$
with $i$ and $t$, the summation of $D(\mathcal{E}_{\Omega_{i,t}}^{w,[h]})$
over $t$ modulo $k$ and define 
\begin{align*}
    D(\mathcal{E}_{\Omega_{i}}^{w,[h]})
    &=\sum_{\mod(t,k)=h}D(\mathcal{E}_{\Omega_{i,t}}^{w,[h]})\\
    &=\sum_{\mod(t,k)=h}\mathcal{W}_{i,t,l}\mathcal{E}_{i,t,(A_{i,(t-k+1):t})_{(10)}}\cdot e_{i}(N)\otimes e_{t}(T)\otimes e_{(A_{i,(t-k+1):t})_{(10)}}(K).
\end{align*}
By Lemma \ref{lem:(Tensor-Bernstein-Inequality)} and union bound, we have 
$$
\mathbb{P}(\max_{i}\|D(\mathcal{E}_{\Omega_{i}}^{w,[h]})\|\geq \alpha)\leq 
N(N+T+K)^2\exp\left\{-\frac{\alpha^2}{2p_{\min}^{-2}k^{-1}T\|D(\mathcal{E}_{\Omega_{i,t}}^{w,[h]})\|_{\psi}^{2}+(2D_Z^{[h]}\alpha)/3}\right\},
$$
where
$\|D(\mathcal{E}_{\Omega_{i,t}}^{w,[h]})\|_{\psi}^{2}=\sigma^{2}$ and $D_Z^{[h]}\leq p_{\min}^{-1}\sigma^2 \log(N+T+K)$ with probability greater than $1-(N+T+K)^{-2}$. Hence, it implies that
\[
\max_{i}\|D(\mathcal{E}_{\Omega_{i}}^{w,[h]})\|=\max_{i}\left\|\sum_{\mod(t,k)=h}D(\mathcal{E}_{\Omega_{i,t}}^{w,[h]})\right\|\leq C\sigma p_{\min}^{-1}k^{-1/2}T^{1/2}\log(N+T+K)\cdot \log(N),
\]
with probability greater $1-(N+T+K)^{-2}$. For the upper bound of the variance, the iterated expectation technique
can be used and we have
\begin{align*}
 \|\sum_{i=1}^{N}\E\{\mathcal{E}_{\Omega_{i}}^{w,[h]}\underline{\square}\mathcal{E}_{\Omega_{i}}^{w,[h]}\}\|
 &=
  \left\|\sum_{i=1}^{N}\E\left\{
  \sum_{\mod(t,k)=\mod(t',k)=h} \mathcal{E}_{\Omega_{i,t}}^{w,[h]}\underline{\square}\mathcal{E}_{\Omega_{i,t'}}^{w,[h]}\right\}\right\|\\
&=
 \left\|\sum_{i=1}^{N}\E\left\{
  \sum_{\mod(t,k)=h}  \mathcal{E}_{\Omega_{i,t}}^{w,[h]}\underline{\square}\mathcal{E}_{\Omega_{i,t}}^{w,[h]} + 
   \sum_{t\neq t'}  \mathcal{E}_{\Omega_{i,t}}^{w,[h]}\underline{\square}\mathcal{E}_{\Omega_{i,t'}}^{w,[h]}
  \right\}\right\|\\
&= \left\|\sum_{i=1}^{N} \sum_{\mod(t,k)=h} \E\left\{
 \mathcal{E}_{\Omega_{i,t}}^{w,[h]}\underline{\square}\mathcal{E}_{\Omega_{i,t}}^{w,[h]}
  \right\}\right\|\\
&\leq \max_{t,l}\sum_{i}\mathcal{W}_{i,t,l}^{2}\sigma^{2}
 \lesssim Np_{\min}^{-2}\sigma^{2}\log(N+T+K),
\end{align*}
and
\begin{align*}
 \left\|\sum_{i=1}^{N}\E\{\mathcal{E}_{\Omega_{i}}^{w,[h]}\overline{\square}\mathcal{E}_{\Omega_{i}}^{w,[h]}\}\right\|
 &=\left\|\sum_{i=1}^{N}\E\left\{
  \sum_{\mod(t,k)=\mod(t',k)=h}  \mathcal{E}_{\Omega_{i,t}}^{w,[h]}\overline{\square}\mathcal{E}_{\Omega_{i,t'}}^{w,[h]}\right\}\right\| \\
  &\leq\max_{i}\sum_{t,l}\mathcal{W}_{i,t,l}^{2}\sigma^{2}
 \lesssim Tp_{\min}^{-2}\sigma^{2}\log(N+T+K),
\end{align*}
with probability greater than $1-(N+T+K)^{-2}$, where $\E\{\mathcal{E}_{\Omega_{i,t}}^{w,[h]}\underline{\square}\mathcal{E}_{\Omega_{i,t'}}^{w,[h]}\}=0$ for $t\neq t'$ since $\E\{\mathcal{E}_{i,t,(a_{i,(t-k+1):t})_{(10)}}\mid a_{i,(t-k+1):t},\mathcal{E}_{i,t',(a_{i,(t-k+1):t})_{(10)}}\}=0$ under Assumption \ref{assum:decay-covariance}. 

Let $D_{Z}=\sigma p_{\min}^{-1}k^{-1/2}T^{1/2}\log(N+T+K)\log(N)$
and $\sigma_{Z}^{2}=p_{\min}^{-2}(N\vee T)\sigma^{2}\log(N+T+K)$,
we obtain the upper bound for $\|\mathcal{E}_{\Omega}^{w}\|$ by applying
Lemma \ref{lem:(Tensor-Bernstein-Inequality)} as 
\begin{align*}
&\mathbb{P}(\|\sum_{i=1}^{N}D(\mathcal{E}_{\Omega_{i}}^{w,[h]})\|^{\overline{\square}}\geq\alpha)  \leq(N+T+K)^{2}\exp\left\{ -\frac{\alpha^{2}}{2\sigma_{Z}^{2}+(2D_{Z}\alpha)/3}\right\} \\
 & \leq(N+T+K)^{2}\\
 & \times\exp\left[-\frac{3}{4}\min\left\{ \frac{\alpha^{2}}{p_{\min}^{-2}(N\vee T)\sigma^{2}\log(N+T+K)},\frac{\alpha}{\sigma p_{\min}^{-1}k^{-1/2}T^{1/2}\log(N+T+K)\log(N)}\right\} \right],
\end{align*}
which holds with probability greater $1-2(N+T+K)^{-2}$. By union
bound over $h$, we have with probability greater than $1-3(N+T+K)^{-2}$
\begin{align*}&\|\sum_{i=1}^{N}\mathcal{E}_{\Omega_{i}}^{w}\|=\|\sum_{i=1}^{N}D(\mathcal{E}_{\Omega_{i}}^{w})\|\leq\|\sum_{i=1}^{N}D(\mathcal{E}_{\Omega_{i}}^{w})\|^{\overline{\square}}\leq\sum_{h=1}^{k}\|\sum_{i=1}^{N}D(\mathcal{E}_{\Omega_{i}}^{w,[h]})\|^{\overline{\square}}\\
&\leq C_{3}k\sigma p_{\min}^{-1}\sqrt{(N\vee T)}\log(N+T+K),
\end{align*}
for a constant $C_{3}$, where the first inequality is shown in \citet{luo2020tensor}.

\paragraph{Step 2} We aim to bound the term $\|\mathcal{Y}^{*}-\widehat{\mathcal{Y}}\|_{*}$. First, we can decompose it into two parts: $\mathcal{Y}^{*}-\widehat{\mathcal{Y}}=(\mathcal{Y}^{*}-\widehat{\mathcal{Y}})\times_{1}\boldsymbol{P}_{\Phi(\boldsymbol{X}_{0})}+(\mathcal{Y}^{*}-\widehat{\mathcal{Y}})\times_{1}\boldsymbol{P}_{\Phi(\boldsymbol{X}_{0})}^{\perp}$,
where $\boldsymbol{P}_{\Phi(\boldsymbol{X}_{0})}$ is the projection
matrix onto the column spaces of $\Phi(\boldsymbol{X}_{0})$, and
$\boldsymbol{P}_{\Phi(\boldsymbol{X}_{0})}^{\perp}$ is the projection matrix
onto the orthogonal space of $\Phi(\boldsymbol{X}_{0})$. For the first term, we have 
\begin{align*}
\|(\mathcal{Y}^{*}-\widehat{\mathcal{Y}})\times_{1}\boldsymbol{P}_{\Phi(\boldsymbol{X})}\|_{*} & =\sup_{\|\mathcal{Z}\|\leq1}\langle(\mathcal{Y}^{*}-\widehat{\mathcal{Y}})\times_{1}\boldsymbol{P}_{\Phi(\boldsymbol{X})},\mathcal{Z}\rangle\\
 & =\sup_{\|\mathcal{Z}\|\leq1}\langle(\mathcal{Y}^{*}-\widehat{\mathcal{Y}})\times_{1}\Phi(\boldsymbol{X})^{\intercal},\mathcal{Z}\times_{1}\Phi(\boldsymbol{X})^{\intercal}\rangle\\
 & \leq\sup_{\|\mathcal{Z}\|\leq1}\langle(\mathcal{Y}^{*}-\widehat{\mathcal{Y}})\times_{1}\Phi(\boldsymbol{X})^{\intercal},\mathcal{Z}\rangle\\
 & \leq\|(\mathcal{Y}^{*}-\widehat{\mathcal{Y}})\times_{1}\Phi(\boldsymbol{X})^{\intercal}\|_{*}\\
 & \leq2\sqrt{\frac{(r_{1}\wedge d_{\Phi})r_{2}r_{3}}{\max\{(r_{1}\wedge d_{\Phi}),r_{2},r_{3}\}}}\|(\mathcal{Y}^{*}-\widehat{\mathcal{Y}})\times_{1}\Phi(\boldsymbol{X})^{\intercal}\|_{F}\\
 & \leq2\sqrt{\frac{(r_{1}\wedge d_{\Phi})r_{2}r_{3}}{\max\{(r_{1}\wedge d_{\Phi}),r_{2},r_{3}\}}}\|(\mathcal{Y}^{*}-\widehat{\mathcal{Y}})\|_{F},
\end{align*}
where the last two inequalities are based on Lemma \ref{lem:tensor_nuclear_norm},
which are adapted from Theorem 1.2 in \citet{kong2018new} and Lemma
5.1 in \citet{hu2015relations}. For the second term, we have
\begin{align}
\|(\mathcal{Y}^{*}-\widehat{\mathcal{Y}})\times_{1}\boldsymbol{P}_{\Phi(\boldsymbol{X})}^{\perp}\|_{*} & =\|\mathcal{Y}^{*}\times_{1}\boldsymbol{P}_{\Phi(\boldsymbol{X})}^{\perp}\|_{*}\nonumber \\
 & =\|\mathcal{G}^{*}\times_{1}\{\boldsymbol{P}_{\Phi(\boldsymbol{X})}^{\perp}\cdot(\boldsymbol{R}+\boldsymbol{\Gamma})\}\times_{2}\boldsymbol{U}_{2}^{*}\times_{3}\boldsymbol{U}_{3}^{*}\|_{*}\nonumber \\
 & =\sup_{\|\mathcal{Z}\|\leq1}\langle\mathcal{G}^{*}\times_{1}\{\boldsymbol{P}_{\Phi(\boldsymbol{X})}^{\perp}\cdot(\boldsymbol{R}+\boldsymbol{\Gamma})\}\times_{2}\boldsymbol{U}_{2}^{*}\times_{3}\boldsymbol{U}_{3}^{*},\mathcal{Z}\rangle\nonumber \\
 & =\sup_{\|\mathcal{Z}\|\leq1}\langle\{\boldsymbol{P}_{\Phi(\boldsymbol{X})}^{\perp}\cdot(\boldsymbol{R}+\boldsymbol{\Gamma})\}\mathcal{M}_{1}(\mathcal{G}^{*})(\boldsymbol{U}_{3}^{*}\otimes\boldsymbol{U}_{2}^{*})^{\intercal},\mathcal{M}_{1}(\mathcal{Z})\rangle\nonumber \\
 & =\sup_{\|\mathcal{Z}\|\leq1}\langle\mathcal{M}_{1}(\mathcal{G}^{*}),\{(\boldsymbol{R}^{\intercal}+\boldsymbol{\Gamma}^{\intercal})\boldsymbol{P}_{\Phi(\boldsymbol{X})}^{\perp}\}\mathcal{M}_{1}(\mathcal{Z})(\boldsymbol{U}_{3}^{*}\otimes\boldsymbol{U}_{2}^{*})\rangle\nonumber \\
 & =\sup_{\|\mathcal{Z}\|\leq1}\langle\mathcal{G}^{*},\mathcal{Z}\times_{1}\{(\boldsymbol{R}^{\intercal}+\boldsymbol{\Gamma}^{\intercal})\boldsymbol{P}_{\Phi(\boldsymbol{X})}^{\perp}\}\times_{2}\boldsymbol{U}_{2}^{*\intercal}\times_{3}\boldsymbol{U}_{3}^{*\intercal}\rangle\nonumber \\
 & \leq\sup_{\|\mathcal{Z}\|\leq1}\langle\mathcal{G}^{*},\mathcal{Z}\rangle\|\boldsymbol{P}_{\Phi(\boldsymbol{X})}^{\perp}(\boldsymbol{R}+\boldsymbol{\Gamma})\|\|\boldsymbol{U}_{2}^{*}\|\|\boldsymbol{U}_{3}^{*}\|\nonumber \\
 & \leq\|\mathcal{G}^{*}\|_{*}\{\|\boldsymbol{R}\|_{F}+\|\boldsymbol{\Gamma}\|\}\nonumber \\
 & \leq C_{2}r_{3}\sqrt{r_{1}\wedge r_{2}}\max_{k=1,2,3}\|\mathcal{M}_{k}(\mathcal{G}^{*})\|\{\sqrt{Nr_{1}}\cdot d_{\Phi}^{-\tau/2}\vee\|\boldsymbol{\Gamma}\|\}\label{eq:upper_bound_NTN}\\
 & \leq C_{2}r_{3}\sqrt{r_{1}\wedge r_{2}}L_{0}\sqrt{\frac{NTK}{\mu_{0}^{3}r_{1}r_{2}r_{3}}}\{\sqrt{Nr_{1}}\cdot d_{\Phi}^{-\tau/2}\vee\|\boldsymbol{\Gamma}\|\},\nonumber 
\end{align}
where the tensor nuclear norm is the dual norm to the tensor spectral
norm \citep{lim2013blind,derksen2016nuclear}, (\ref{eq:upper_bound_NTN})
is based on Lemma \ref{lem:G_nuclear_norm}, $\boldsymbol{P}_{\Phi(\boldsymbol{X})}\Phi(\boldsymbol{X}_{0})=\Phi(\boldsymbol{X}_{0})$,
$\boldsymbol{P}_{\Phi(\boldsymbol{X})}^{\perp}\boldsymbol{\Gamma}=\boldsymbol{\Gamma}$,
$\boldsymbol{P}_{\Phi(\boldsymbol{X})}\boldsymbol{\Gamma}=0$ by definition,
and $\|\boldsymbol{R}\|_{F}=O(\sqrt{Nr_{1}}\cdot d_{\Phi}^{-\tau/2})$
under Assumption \ref{assmp:on_G(x)}. Thus, we are able to prove that:
\begin{align*}
 & \frac{1}{NT}\sum_{i,t}\mathcal{W}_{i,t,l}(\mathcal{Y}_{i,t,l}^{*}-\widehat{\mathcal{Y}}_{i,t,l})^{2}\\
 & \leq\frac{2}{NT}\|\mathcal{E}_{\Omega}^{w}\|\left\{ \|(\mathcal{Y}^{*}-\widehat{\mathcal{Y}})\times_{1}\boldsymbol{P}_{\Phi(\boldsymbol{X})}\|_{*}+\|(\mathcal{Y}^{*}-\widehat{\mathcal{Y}})\times_{1}\boldsymbol{P}_{\Phi(\boldsymbol{X})}^{\perp}\|_{*}\right\} \\
 & +\frac{1}{NT}(\|I_{1}\|+\|I_{3}\|)\\
 & \leq\frac{4C_{3}k\sigma}{NT}p_{\min}^{-1}\sqrt{(N\vee T)}\log(N+T+K) \cdot \sqrt{\frac{(r_{1}\wedge d_{\Phi})r_{2}r_{3}}{\max\{(r_{1}\wedge d_{\Phi}),r_{2},r_{3}\}}} \|(\mathcal{Y}^{*}-\widehat{\mathcal{Y}})\|_{F}\\
 & + \frac{2C_{2}k\sigma}{NT}p_{\min}^{-1}\sqrt{(N\vee T)}\log(N+T+K) 
\cdot r_{3}\sqrt{r_{1}\wedge r_{2}}L_{0}\sqrt{\frac{NTK}{\mu_{0}^{3}r_{1}r_{2}r_{3}}}\left\{\sqrt{Nr_{1}}\cdot d_{\Phi}^{-\tau/2}\vee\|\boldsymbol{\Gamma}\|\right\}\\ 
 & +\frac{1}{NT}(\|I_{1}\|+\|I_{3}\|).
\end{align*}
Notice that if $\|\Delta\|_{F}^{2}\leq4L_{0}^{2}T\log(N+T+K)/p_{\min}$,
then the bound in Theorem holds. Hence, our focus now is the error
bound conditional on the event where $\|\Delta\|_{F}^{2}\geq4L_{0}^{2}T\log(N+T+K)/p_{\min}$.

\subsubsection{Restricted strong convexity}\label{subsubsec:rsc}

We now aim to prove the restricted strong convexity of the loss function,
that is $\|\mathcal{Y}^{*}-\widehat{\mathcal{Y}}\|_{F}^{2}$ is larger
than a constant fraction of $\|\boldsymbol{P}_{\Omega}(\mathcal{Y}^{*}-\widehat{\mathcal{Y}})\|_{F}^{2}$
up to an additive term. A summary of the proof adapted to our setting.
For starter, we can provide an entry-wise upper bound for any $\mathcal{X}\in\mathfrak{C}(r_{1},r_{2},r_{3},\mu_{0},L_{0})$.
Let the maximum entry of $\mathcal{X}$ be $\|\mathcal{X}\|_{\max}$
and
\begin{align*}
\|\mathcal{X}\|_{\max} & =\max_{i,t,l}|\langle\mathcal{X},e_{i}\otimes e_{t}\otimes e_{l}\rangle|\\
&=\max_{i,t,l}|\langle\mathcal{G}\times_1\boldsymbol{U}_{1}\times_2\boldsymbol{U}_{2}\times_3 \boldsymbol{U}_{3},
e_{i}\otimes e_{t}\otimes e_{l}\rangle|\\
 & =\max_{i,t,l}|\langle\mathcal{G},(\boldsymbol{U}_{1}^{\intercal}e_{i})\otimes(\boldsymbol{U}_{2}^{\intercal}e_{t})\otimes(\boldsymbol{U}_{3}^{\intercal}e_{l})\rangle|\\
 & \leq\max_{k\in\{1,2,3\}}\|\mathcal{M}_k(\mathcal{G})\|\max_{i}\|\boldsymbol{U}_{1}^{\intercal}e_{i}\|_{F}\max_{t}\|\boldsymbol{U}_{2}^{\intercal}e_{t}\|_{F}\max_{l}\|\boldsymbol{U}_{3}^{\intercal}e_{l}\|_{F}\\
 & \leq\max_{k\in\{1,2,3\}}\|\mathcal{M}_k(\mathcal{G})\|\|\boldsymbol{U}_{1}\|_{2,\infty}\|\boldsymbol{U}_{2}\|_{2,\infty}\|\boldsymbol{U}_{3}\|_{2,\infty}\\
 & \leq
 L_0\sqrt{\frac{NTK}{\mu_0^{3/2}(r_1r_2r_3)^{1/2}}}
 \cdot
 \mu_{0}^{3/2}(r_{1}r_{2}r_{3})^{1/2}(NTK)^{-1/2}\leq L_0,
\end{align*}
which implies the entry-wise upper bound. Next, Lemma \ref{lem:restricted_convexity} proves the restricted strong convexity by
the Massart’s inequality \citep{massart2000constants} and peeling
arguments in Lemma 14, \citet{klopp2014noisy}.
\begin{lemma}
\label{lem:restricted_convexity} Let $\Delta=\mathcal{Y}^{*}-\widehat{\mathcal{Y}}$
and $\|\Delta\|_{L^{2}(\Omega)}=\sqrt{\mathbb{E}_{\mathcal{A}}\left(\|P_{\Omega}(\Delta)\|_{F}^{2}\right)}$,
if $\Delta\in\mathfrak{C}(r_{1},r_{2},r_{3},\mu_{0},L_{0})$ and $\|\Delta\|_{F}^{2}\geq L_{0}^{2}\theta/p_{\min}$
for $\theta=C'T\log(N+T+K)$, we have for small enough constant $C_{4}$
(e.g., $C_{4}=0.005$),
\[
\mathbb{P}_{\mathcal{A}}\left\{ \frac{p_{\min}}{2}\|\Delta\|_{F}^{2}\geq\|P_{\Omega}(\Delta)\|_{F}^{2}+2\|\Delta\|_{\max}^{2}\vartheta\right\} \leq2\exp\left(-\frac{C_{4}\theta}{T}\right)
\]
where $p_{\min}=\delta^{k}\leq \mathbb{P}(A_{i,t-k+1:t}=a_{(t-k+1):t}\mid H_{i,t})\leq(1-\delta)^{k}=p_{\max}$
by Lemma \ref{lem:bound-P(A)}, 
\[
\vartheta=C''\frac{(r_{1}\wedge d_{\Phi})r_{2}r_{3}}{\max\{(r_{1}\wedge d_{\Phi}),r_{2},r_{3}\}p_{\min}}k^2(N\vee T)\log(N+T+K),
\]
and $\mathbb{P}_{\mathcal{A}},\E_{\mathcal{A}}$ are taken with respect to
the marginal distribution of the treatment assignments. 
\end{lemma}
Combing Lemma \ref{lem:restricted_convexity} with the previous steps and the upper bound on $I_{1}$
and $I_{3}$, we have for $\mathcal{Y}^{*},\widehat{\mathcal{Y}}\in\mathfrak{C}(r_{1},r_{2},r_{3},\mu_{0},L_{0})$
and $\|\Delta\|_{F}^{2}\geq L_{0}^{2}\theta/p_{\min}$,
\begin{align}
 & \frac{p_{\min}}{2p_{\max}}\|\mathcal{Y}^{*}-\widehat{\mathcal{Y}}\|_{F}^{2}\leq\frac{p_{\min}}{4p_{\max}}\|\mathcal{Y}^{*}-\widehat{\mathcal{Y}}\|_{F}^{2} \nonumber\\
 & +C_{1}\max_{t=1,\cdots,T}|\mathcal{J}_{t}|\log^{1/2}(N+T+K)N^{-1/2} \|\mathcal{Y}^* - \widehat{\mathcal{Y}}\|_F^2 \label{eq:term_negligible}\\
 & +16C_{2}^{2}\frac{(r_{1}\wedge d_{\Phi})r_{2}r_{3}}{\max\{(r_{1}\wedge d_{\Phi}),r_{2},r_{3}\}}\sigma^{2}p_{\min}^{-2}k^2(N\vee T)\log(N+T+K)p_{\min}^{-1}p_{\max}\nonumber\\
 & +2C_{2}C_{3} p_{\min}^{-1}k
 \{\log(N+T+K)\}^{1/2}
 \sigma L_{0}\sqrt{\frac{r_3(r_1\wedge r_2)(N\vee T)NTK}{\mu^{3}r_{1}r_{2}}}\{\sqrt{Nr_{1}}\cdot d_{\Phi}^{-\tau/2}\vee\|\boldsymbol{\Gamma}\|\}\label{eq:term_bound1}\\
 & +8C_{2}C_{6}L_{0}^{2}\frac{(r_{1}\wedge d_{\Phi})r_{2}r_{3}}{\max\{(r_{1}\wedge d_{\Phi}),r_{2},r_{3}\}p_{\min}p_{\max}}k^2(N\vee T)\log(N+T+K)\nonumber\\
 & +C_{1}\max_{t=1,\cdots,T}|\mathcal{J}_{t}|\log^{1/2}(N+T+K)N^{-1/2}\sigma^{2}\sqrt{NT\log(N+T+K)} \label{eq:term_bound2},
\end{align}
with probability greater than $1-7(N+T+K)^{-2}$. Under Assumption \ref{assump:weight_G}, (\ref{eq:term_bound1}) and (\ref{eq:term_bound2}) are negligible compared to the rest terms. Therefore, it completes the
proof of Theorem \ref{thm:bound-tensor} by rearranging the terms as follows:
\begin{align*}
& \frac{\|\mathcal{Y}^*-\widehat{\mathcal{Y}}\|_F^2}{N T} \leq \frac{C L_0^2}{p_{\min }} \frac{\log (N+T+K)}{N} \\
& \vee \frac{C^{\prime} k^2\left(r_1 \wedge d_{\Phi}\right) r_2 r_3}{\max \left\{\left(r_1 \wedge d_{\Phi}\right), r_2, r_3\right\} p_{\min }^2} \frac{(N \vee T)}{N T} \log (N+T+K)\left\{\sigma^2\left(\frac{p_{\max }}{p_{\min }}\right)^2 \vee L_0^2\right\},
\end{align*}
with probability greater than $1-7(N+T+K)^{-2} $ for some constants $C$ and $C'$, where (\ref{eq:term_negligible}) is negligible to $\|\mathcal{Y}^*-\widehat{\mathcal{Y}}\|_F^2$ under Assumption \ref{assmp:on_W(x)}(e).

\subsection{Additional Lemmas}
In this section, we include technical proof for all the lemmas.
\subsection{Proof of Lemma \ref{lem:oracle_property}}\label{subsec:prove_oracle}
\begin{lemma}
\label{lem:oracle_property} Under Assumption \ref{assum:HR}-\ref{assum:positivity} and regularity conditions in Assumption \ref{assmp:on_W(x)} of the Appendix, let $p_{\lambda_\alpha}(\cdot)$ be the SCAD penalty function 
$$
p_{\lambda_{\alpha}}(|\alpha_{1,t}|)=\begin{cases}
\lambda_{\alpha}|\alpha_{1,t}| & \text{if }|\alpha_{1,t}|<\lambda_{\alpha}\\
\frac{\epsilon\lambda_{\alpha}|\alpha_{1,t}|-\alpha_{1,t}^{2}-\lambda_{\alpha}^{2}}{\epsilon-1} & \text{if }\lambda_{\alpha}<|\alpha_{1,t}|\leq\epsilon\lambda_{\alpha}\\
\frac{\lambda_{\alpha}^{2}(\epsilon+1)}{2} & \text{if }|\alpha_{1,t}|>\epsilon\lambda_{\alpha}
\end{cases},
$$
with $\epsilon=3.7$ as suggested in \citet{fan2001variable}, there exists an approximate
penalized minimizer $\widehat{\alpha}_{t}$ to (\ref{eq:penalized_Q_PS})
at each time point, such that
$$
 \widehat{\alpha}_{t,\mathcal{J}_{t}}-\alpha_{t,\mathcal{J}_{t}}^{*}=O_{\mathbb{P}}(\sqrt{|\mathcal{J}_{t}|/N}),\quad \mathbb{P}(\widehat{\alpha}_{t,\mathcal{J}_{t}^{\complement}}=0)\rightarrow1 \text{ as }N\rightarrow\infty,\label{eq:selection-consistency}
$$
where $\alpha_{t,\mathcal{J}_t}$ is a sub-vector of $\alpha_{t}$ formed
by its elements whose indexes are in $\mathcal{J}_t$, and $\mathcal{J}_{t}^{\complement}$
is the complement of $\mathcal{J}_{t}$.
\end{lemma}
We request the following regularity conditions to hold.

\begin{assumption}\label{assmp:on_W(x)}

(a) $\|X_{i,t}\|$ are bounded uniformly by $C_{0}$ for $i=1,\cdots,N$
and $t=1,\cdots,T$.

(b) Let $\alpha_{t}^{*}$ lies in the interior of the compact ball
and
\[
Z_{i,t}=\left\{ \frac{\exp(\alpha_{0,t}^{*}+\alpha_{1,t}^{*}H_{i,t})}{1+\exp(\alpha_{0,t}^{*}+\alpha_{1,t}^{*}H_{i,t})}-A_{i,t}\right\} .
\]
Assume $\E(Z_{i,t})=0$ for $i=1,\cdots,N,t=1,\cdots,T$ and there
exists constant $C_{1}$ such that $\E(|Z_{i,t}|^{2+\delta})\leq C_{1}$
for some $\delta>0$.

(c) Let $\Sigma_{Z_{i,t}}=\E\{\partial Z_{i,t}/\partial\alpha_{t}^{\intercal}\cdot(1,H_{i,t})\}$,
that is
\[
\Sigma_{Z_{i,t}}=\E\left\{ (1,H_{i,t})^{\intercal}\frac{\exp(\alpha_{0,t}^{*}+\alpha_{1,t}^{*}H_{i,t})}{\{1+\exp(\alpha_{0,t}^{*}+\alpha_{1,t}^{*}H_{i,t})\}^{2}}(1,H_{i,t})\right\} .
\]
Assume $\|\Sigma_{Z_{i,t}}\|$ is bounded and satisfied the restricted
eigenvalue condition, that is, for any $b$
\[
C_{2}\|b\|_{2}^{2}\leq b^{\intercal}\Sigma_{Z_{i,t}}b\leq C_{3}\|b\|_{2}^{2},
\]
for some constants $C_{2}$ and $C_{3}$.

(d) For any $t=1,\cdots,T$, there exist constants $C_{4}$ and $C_{5}$
such that 
\[
C_{4}<\sigma_{\min}\left(\frac{1}{N}\sum_{i=1}^{N}H_{i,t}H_{i,t}^{\intercal}\right)<\sigma_{\max}\left(\frac{1}{N}\sum_{i=1}^{N}H_{i,t}H_{i,t}^{\intercal}\right)<C_{5},
\]
where $\sigma_{\min}(\cdot)$ and $\sigma_{\max}(\cdot)$ return the
minimum and the maximum eigenvalue of a symmetric matrix, respectively. 

(e) $|\mathcal{J}_{t}|=O(1)$ for any $t=1,\cdots,T$. The tuning
parameter satisfies that $\lambda_{\alpha}\rightarrow0$ and $\sqrt{N}\lambda_{\alpha}\rightarrow\infty$
as $N\rightarrow\infty$.

\end{assumption}

Assumptions \ref{assmp:on_W(x)} (a)-(d) are typical regularity conditions
in the penalization literature \citep{fan2001variable,yang2020doubly}.
Assumption \ref{assmp:on_W(x)}(e) specifies the restrictions on
the dimension of the true nonzero coefficients and places an asymptotic nature for the tuning parameter. Strictly speaking, a weaker sparsity assumption can be imposed on the number of nonzero coefficients for the propensity score model. Assume that $\max_{t=1,\cdots,T}|\mathcal{J}_t|\log^{1/2}(N+T+k)N^{-1/2} = o(1)$, our non-asymptotic analysis in Theorem \ref{thm:bound-tensor} remains valid as shown in Section \ref{subsubsec:rsc}. Similar conditions can be found in \cite{belloni2017program} and \cite{ning2020robust}.

\begin{proof}[Proof of Lemma \ref{lem:oracle_property}]
   For each time point $t$, we aim to prove that there exists a large
constant $\tau$ for any given $\varepsilon>0$ such that 
\begin{equation}
\mathbb{P}\left\{ \sup_{\alpha_{t}\in\partial\mathfrak{M}_{\tau}}Q(\alpha_{t})>Q(\alpha_{t}^{*})\right\} \geq1-\varepsilon,\quad\mathfrak{M}_{\tau}=\{\alpha_{t}:\|\alpha_{t}-\alpha_{t}^{*}\|\leq\tau\sqrt{|\mathcal{J}_{t}|/N}\},\label{eq:existence_min}
\end{equation}
and $\partial\mathfrak{M}_{\tau}$ means that $\alpha_{t}$ attains
the boundary of the restricted set, i.e., $\|\alpha_{t}-\alpha_{t}^{*}\|=\tau\sqrt{|\mathcal{J}_{t}|/N}$.
(\ref{eq:existence_min}) implies that there exists a local minimum
$\widehat{\alpha}_{t}$ in the ball $\mathfrak{M}_{\tau}$ as continuous
function must have a local minimum within a compact subset, i.e.,
$\mathfrak{M}_{r}$. Using $p_{\lambda_{\alpha}}(|\alpha_{1,t,j}|)=0,j\in\mathcal{J}_{t}^{\complement}$,
we have
\begin{align}
&Q(\alpha_{t})-Q(\alpha_{t}^{*})  =L(\alpha_{t})-L(\alpha_{t}^{*})+p_{\lambda_{\alpha}}(|\alpha_{1,t}|)-p_{\lambda_{\alpha}}(|\alpha_{1,t}^{*}|)\nonumber \\
 & \geq\frac{\partial L(\alpha_{t}^{*})}{\partial\alpha_{t}^{\intercal}}(\alpha_{t}-\alpha_{t}^{*})+(\alpha_{t}-\alpha_{t}^{*})^{\intercal}\frac{\partial L(\widehat{\alpha}_{t}^{*})}{\partial\alpha_{t}\partial\alpha_{t}^{\intercal}}(\alpha_{t}-\alpha_{t}^{*})+\sum_{j\in\mathcal{J}_{t}}\left\{ p_{\lambda_{\alpha}}(|\alpha_{1,t,j}|)-p_{\lambda_{\alpha}}(|\alpha_{1,t,j}^{*}|)\right\} \nonumber \\
 & =\mathcal{T}_{1}+\mathcal{T}_{2}+\mathcal{T}_{3},\label{eq:error_1_2_3}
\end{align}
where $\widehat{\alpha}_{t}^{*}$ is between $\alpha_{t}$ and $\alpha_{t}^{*}$.
Considering the first term $\mathcal{T}_{1}$, by Cauchy-Schwarz inequality,
we have 
\[
|\mathcal{T}_{1}|\leq\tau\sqrt{|\mathcal{J}_{t}|/N}\left\Vert \frac{\partial L(\alpha_{t}^{*})}{\partial\alpha_{t}^{\intercal}}\right\Vert \leq\tau\sqrt{C_{1}C_{5}}|\mathcal{J}_{t}|/N,
\]
where 
\begin{align*}
\E\left\{ \left\Vert \frac{\partial L(\alpha_{t}^{*})}{\partial\alpha_{t}^{\intercal}}\right\Vert ^{2}\right\}  & =\E\left\{ \left\Vert \frac{1}{N}\sum_{i=1}^{N}\left\{ \frac{\exp(\alpha_{0,t}+\alpha_{1,t}H_{i,t})}{1+\exp(\alpha_{0,t}+\alpha_{1,t}H_{i,t})}-A_{i,t}\right\} (1,H_{i,t})\right\Vert ^{2}\right\} \\
 & =\E\left\{ \left\Vert \frac{1}{N}\sum_{i=1}^{N}Z_{i,t}(1,H_{i,t})\right\Vert ^{2}\right\} \\
 & =\frac{1}{N^{2}}\text{trace}\left\{ \sum_{i=1}^{N}\E\left(Z_{i,t}^{2}\right)H_{i,t}H_{i,t}^{\intercal}\right\} \\
 & \leq\frac{C_{1}}{N^{2}}|\mathcal{J}_{t}|\sigma_{\max}\left(\sum_{i=1}^{N}H_{i,t}H_{i,t}^{\intercal}\right) \leq C_{1}C_{5}|\mathcal{J}_{t}|/N,
\end{align*}
under Assumptions \ref{assmp:on_W(x)} (b) and (d). Considering $\mathcal{T}_{2}$,
we have 
\begin{align*}
\mathcal{T}_{2} & =(\alpha_{t}-\alpha_{t}^{*})^{\intercal}\frac{\partial L(\widehat{\alpha}_{t}^{*})}{\partial\alpha_{t}\partial\alpha_{t}^{\intercal}}(\alpha_{t}-\alpha_{t}^{*})\\
 & =(\alpha_{t}-\alpha_{t}^{*})^{\intercal}\Sigma_{Z_{i,t}}(\alpha_{t}-\alpha_{t}^{*})\\
 & +(\alpha_{t}-\alpha_{t}^{*})^{\intercal}\left\{ \frac{\partial L(\widehat{\alpha}_{t}^{*})}{\partial\alpha_{t}\partial\alpha_{t}^{\intercal}}-\frac{\partial L(\alpha_{t}^{*})}{\partial\alpha_{t}\partial\alpha_{t}^{\intercal}}\right\} (\alpha_{t}-\alpha_{t}^{*})\\
 & =\mathcal{T}_{21}+\mathcal{T}_{22}.
\end{align*}
For $\mathcal{T}_{21}$, 
\begin{align*}
\mathcal{T}_{21} & =(\alpha_{t}-\alpha_{t}^{*})^{\intercal}\Sigma_{Z_{i,t}}(\alpha_{t}-\alpha_{t}^{*})\\
 & \leq C_{3}\|\alpha_{t}-\alpha_{t}^{*}\|^{2}\leq C_{3}\tau^{2}|\mathcal{J}_{t}|/N,
\end{align*}
under Assumptions \ref{assmp:on_W(x)} (c). For $\mathcal{T}_{22}$,
\begin{align*}
\mathcal{T}_{22} & =(\alpha_{t}-\alpha_{t}^{*})^{\intercal}\left\{ \frac{\partial L(\widehat{\alpha}_{t}^{*})}{\partial\alpha_{t}\partial\alpha_{t}^{\intercal}}-\frac{\partial L(\alpha_{t}^{*})}{\partial\alpha_{t}\partial\alpha_{t}^{\intercal}}\right\} (\alpha_{t}-\alpha_{t}^{*})\\
 & =B\|\alpha_{t}-\alpha_{t}^{*}\|^{2}\|\widehat{\alpha}_{t}^{*}-\alpha_{t}^{*}\|\cdot\sigma_{\max}\left(\frac{1}{N}\sum_{i=1}^{N}H_{i,t,\mathcal{J}_{t}}H_{i,t,\mathcal{J}_{t}}^{\intercal}\right)\\
 & =B\cdot(\tau\sqrt{|\mathcal{J}_{t}|/N})^{3}=o(|\mathcal{J}_{t}|/N),
\end{align*}
where 
\begin{align*}
 & \frac{\partial L(\widehat{\alpha}_{t}^{*})}{\partial\alpha_{t}\partial\alpha_{t}^{\intercal}}-\frac{\partial L(\alpha_{t}^{*})}{\partial\alpha_{t}\partial\alpha_{t}^{\intercal}}\\
 & =\frac{1}{N}\sum_{i=1}^{N}\left\{ \frac{\exp(\widehat{\alpha}_{0,t}^{*}+\widehat{\alpha}_{1,t}^{*}H_{i,t})}{\{1+\exp(\widehat{\alpha}_{0,t}^{*}+\widehat{\alpha}_{1,t}^{*}H_{i,t})\}^{2}}-\frac{\exp(\alpha_{0,t}^{*}+\alpha_{1,t}^{*}H_{i,t})}{\{1+\exp(\alpha_{0,t}^{*}+\alpha_{1,t}^{*}H_{i,t})\}^{2}}\right\} (1,H_{i,t})(1,H_{i,t})^{\intercal}\\
 & =\frac{1}{N}\sum_{i=1}^{N}B\cdot(\widehat{\alpha}_{t}^{*}-\alpha_{t}^{*})(1,H_{i,t})(1,H_{i,t})^{\intercal},
\end{align*}
where 
\[
B=\sup_{\alpha_{t}\in\mathfrak{M}_{\tau}}\left\{ \frac{\exp(\widehat{\alpha}_{0,t}^{*}+\widehat{\alpha}_{1,t}^{*}H_{i,t})}{\{1+\exp(\widehat{\alpha}_{0,t}^{*}+\widehat{\alpha}_{1,t}^{*}H_{i,t})\}^{2}}-\frac{\exp(\alpha_{0,t}^{*}+\alpha_{1,t}^{*}H_{i,t})}{\{1+\exp(\alpha_{0,t}^{*}+\alpha_{1,t}^{*}H_{i,t})\}^{2}}\right\} \|H_{i,t}\|_{\max},
\]
which is bounded by Assumptions \ref{assmp:on_W(x)} (a). Considering
the third term $\mathcal{T}_{3}$ 
\begin{align*}
\mathcal{T}_{3} & =\sum_{j\in\mathcal{J}_{t}}\left\{ p_{\lambda_{\alpha}}(|\alpha_{1,t,j}|)-p_{\lambda_{\alpha}}(|\alpha_{1,t,j}^{*}|)\right\} \\
 & =\sum_{j\in\mathcal{J}_{t}}p'_{\lambda_{\alpha}}(|\alpha_{1,t,j}^{*}|)\text{sgn}(\alpha_{1,t,j}^{*})(\alpha_{1,t,j}-\alpha_{1,t,j}^{*})\\
 & +\sum_{j\in\mathcal{J}_{t}}p''_{\lambda_{\alpha}}(|\widehat{\alpha}_{1,t,j}^{*}|)(\alpha_{1,t,j}-\alpha_{1,t,j}^{*})^{2}\\
 & =\sqrt{|\mathcal{J}_{t}|}\max\left\{ |p'_{\lambda_{\alpha}}(|\alpha_{1,t,j}^{*}|)|:j\in\mathcal{J}_{t}\right\} \cdot\tau\sqrt{|\mathcal{J}_{t}|/N}\\
 & +\max\left\{ |p''_{\lambda_{\alpha}}(|\alpha_{1,t,j}^{*}|)|:j\in\mathcal{J}_{t}\right\} \cdot\tau^{2}|\mathcal{J}_{t}|/N,
\end{align*}
where $\widehat{\alpha}_{1,t,j}^{*}$ is between $\alpha_{1,t,j}$ and
$\alpha_{1,t,j}^{*}$. If $p_{\lambda_{\alpha}}(x)$ be the SCAD penalty
function, we have 
\begin{align*}
p'_{\lambda}(|\theta|) & =\lambda\left\{ I(|\theta|<\lambda)+\frac{(a\lambda-|\theta|)_{+}}{(a-1)\lambda}I(|\theta|\geq\lambda)\right\} ,\\
p''_{\lambda}(|\theta|) & =\frac{-1}{(a-1)\lambda}I(\lambda\leq|\theta|\leq a\lambda).
\end{align*}
Therefore, under Assumption \ref{assmp:on_W(x)}(e), $\mathcal{T}_{3}$
is dominated by $\mathcal{T}_{2}$ with a proper choice of $\lambda_{\alpha}$.
In addition, by choosing a sufficiently large $\tau$, $\mathcal{T}_{21}$
dominates the right-hand side of (\ref{eq:error_1_2_3}) in the ball
$\mathfrak{M}_{\tau}$. Since $\mathcal{T}_{21}$ is positive for
sufficiently large $N$, (\ref{eq:existence_min}) holds and hence
$\widehat{\alpha}_{t}-\alpha_{t}^{*}=O(\sqrt{|\mathcal{J}_{t}|/N})$. 

For proving (\ref{eq:selection-consistency}), it is sufficient to show that for
any $j\in\mathcal{J}_{t}^{\complement}$, we have
\begin{align*}
\frac{\partial Q(\alpha_{t})}{\partial\alpha_{t,j}} & <0,\quad\alpha_{t,j}<0,\\
\frac{\partial Q(\alpha_{t})}{\partial\alpha_{t,j}} & >0,\quad\alpha_{t,j}>0,
\end{align*}
where $\alpha_{t,j}\in\mathfrak{M}_{\tau}$. To show this, by Taylor's
expansion, we have
\begin{align*}
\frac{\partial Q(\alpha_{t})}{\partial\alpha_{t,j}} & =\frac{\partial L(\alpha_{t})}{\partial\alpha_{t,j}}+p'_{\lambda_{\alpha}}(|\alpha_{t,j}|)\text{sgn}(\alpha_{t,j})\\
 & =\frac{\partial L(\alpha_{t}^{*})}{\partial\alpha_{t,j}}+\sum_{l}\frac{\partial L(\alpha_{t}^{*})}{\partial\alpha_{t,j}\partial\alpha_{t,l}}(\alpha_{t,l}-\alpha_{t,l}^{*})\\
 & +\sum_{l,k}\frac{\partial L(\widehat{\alpha}_{t}^{*})}{\partial\alpha_{t,j}\partial\alpha_{t,l}\partial\alpha_{t,k}}\times(\alpha_{t,l}-\alpha_{t,l}^{*})(\alpha_{t,k}-\alpha_{t,k}^{*})+p'_{\lambda_{\alpha}}(|\alpha_{t,j}|)\text{sgn}(\alpha_{t,j})
\end{align*}
where $\widehat{\alpha}_{t}^{*}$ lies between $\alpha_{t}$ and $\alpha_{t}^{*}$.
Under Assumptions \ref{assmp:on_W(x)} (a), (b) and (c),
\[
\frac{\partial L(\alpha_{t}^{*})}{\partial\alpha_{t,j}}=O(N^{-1/2}),\quad\frac{\partial L(\alpha_{t}^{*})}{\partial\alpha_{t,j}\partial\alpha_{t,l}}=\E\left\{ \frac{\partial L(\alpha_{t}^{*})}{\partial\alpha_{t,j}\partial\alpha_{t,l}}\right\} +o(1).
\]
As $\alpha_{t}\in\mathfrak{M}_{r}$, we have 
\[
\frac{\partial Q(\alpha_{t})}{\partial\alpha_{t,j}}=\lambda_{\alpha}\{\lambda_{\alpha}^{-1}p'_{\lambda_{\alpha}}(|\alpha_{t,j}|)\text{sgn}(\alpha_{t,j})+O(N^{-1/2}/\lambda_{\alpha})\}.
\]
where $\lambda_{\alpha}^{-1}p'_{\lambda_{\alpha}}(|\alpha_{t,j}|)>0$
under the SCAD penalty and $N^{-1/2}/\lambda_{\alpha}\rightarrow0$
by Assumption \ref{assmp:on_W(x)}(e). Therefore, the sign of derivative
is completely determined by $\alpha_{t,j}$ and hence completes the proof. 
\end{proof}

\subsection{Proof of Lemma \ref{lem:basis_functions}}
\begin{lemma}
\label{lem:basis_functions} Under Assumptions \ref{assum:HR}-\ref{assum:positivity} and regularity conditions in Assumption \ref{assmp:on_G(x)} of the Appendix, we have for $r=1,\cdots r_{1}$
\[
\sup_{i=1,\cdots,N}|G_{r}^{*}(X_{0,i})-\sum_{j=1}^{d_{\Phi}}b_{j,r}\Phi_{j}(X_{0,i})|^{2}=O(d_{\Phi}^{-\tau}),
\]
where $\tau=2(q+\beta)\geq4$ for $(q,\beta)$ defined in (\ref{eq:smoothness}),
and the sieve coefficients $b_{j,r}$ satisfy that $\max_{j,r}b_{j,r}^{2}<\infty$.
\end{lemma}
\begin{assumption}\label{assmp:on_G(x)} 
(a) $\boldsymbol{U}_{1}^{*\intercal}\boldsymbol{U}_{1}^{*}=I_{r_{1}}=G^{*}(\boldsymbol{X}_{0})^{\intercal}G^{*}(\boldsymbol{X}_{0})+\boldsymbol{\Gamma}^{\intercal}\boldsymbol{\Gamma}$
and $G^{*}(\boldsymbol{X}_{0})^{\intercal}\boldsymbol{\Gamma}=0$.

(b) For the basis functions, there exist constants $C_{1}, C_{2}>0$
such that
\[
C_{0}<\frac{1}{N}\sigma_{\min}\left\{ \Phi(\boldsymbol{X}_{0})^{\intercal}\Phi(\boldsymbol{X}_{0})\right\} <\frac{1}{N}\sigma_{\max}\left\{ \Phi(\boldsymbol{X}_{0})^{\intercal}\Phi(\boldsymbol{X}_{0})\right\} <C_{1},
\]
and $\max_{i=1,\cdots,N}\|\Phi(X_{i,0})\|<C_{2}$.

(c) Let $G^{*}(X_{i,0})=\{G_{1}^{*}(X_{i,0}),\cdots,G_{r_{1}}^{*}(X_{i,0})\}^{\intercal}$,
the functions $G_{r}^{*}(X),r=1,\cdots,r_{1}$ belong to a H{\"o}lder
class $\mathfrak{G}$ defined by
\begin{equation}
\mathfrak{G}=\{G\in\mathfrak{G}^{q}:\sup_{\boldsymbol{u},\boldsymbol{v}}|G^{(q)}(\boldsymbol{u})-G^{(q)}(\boldsymbol{v})|\leq C_{3}\|\boldsymbol{u}-\boldsymbol{v}\|_{2}^{\beta}\},\label{eq:smoothness}
\end{equation}
for some positive number $C_{3}$. Here, $\mathfrak{G}^{q}$ is the
space of all $q$-times continuously differentiable real-value functions,
and $G^{(q)}$ is the $q$-th derivative of the function $G$. 

\end{assumption}
Assumption (a) is the identification assumption used in factor analysis
\citep{bai2012statistical} and implies that the residual loading
component $\boldsymbol{\Gamma}$ is bounded. Assumption (b) ensures
that $\Phi(\boldsymbol{X}_{0})^{\intercal}\Phi(\boldsymbol{X}_{0})$
is well-conditioned and this condition can be achieved by common basis functions, such as polynomial basis functions or B-splines under proper
normalization. Assumption (c) imposes mild smoothness on the unknown
function $G^{*}(\boldsymbol{X}_{0})$, which is critical for sieve
approximation \citep{chen2007large}.

\begin{proof}[Proof of Lemma \ref{lem:basis_functions}]
   Under Assumptions \ref{assmp:on_G(x)}, the sieve approximation
is well-controlled and its accuracy is guaranteed; see \citet{chen2007large}
for more details. 
\end{proof}
\subsubsection{Proof of Lemma \ref{lem:(Tensor-Bernstein-Inequality)}}\label{proof_Tensor-Bernstein-Inequality}
\begin{proof}
    Let $\mathcal{A}$ and $\mathcal{B}$ be two real tensors in $\mathbb{R}^{I_1\times\cdots\times I_M}$. Define two generalized Einstein products $\mathcal{A}\overline{\square}\mathcal{B}$ and $\mathcal{A}\underline{\square}\mathcal{B}$ be
    $$
(\mathcal{A} \bar{\square} \mathcal{B})_{i_1,\cdots ,i_m ,j_1, \cdots ,j_m} =\sum_{k_1, \ldots, k_{M-m}} a_{i_1 \cdots i_m k_1 \cdots k_{M-m}} b_{j_1 \cdots j_m k_1 \cdots k_{M-m}}
$$
and
$$
(\mathcal{A} \underline{\square}  \mathcal{B})_{k_1, \cdots, k_{N-m}, k_1^{\prime}, \cdots, k_{N-m}^{\prime}}  =\sum_{i_1, \ldots, i_m} a_{i_1, \cdots, i_m, k_1, \cdots, k_{N-m}} b_{i_1, \cdots, i_m, k_1^{\prime} ,\cdots ,k_{M-m}^{\prime}},
$$
where $m=[M/2]$, $\mathcal{A} \overline{\square} \mathcal{B}\in \mathbb{R}^{I_1 \times \cdots \times I_m \times I_1 \times \cdots \times I_m}$ and $\mathcal{A} \underline{\square}  \mathcal{B}\in \mathbb{R}^{I_{m+1} \times \cdots \times I_M \times I_{m+1} \times \cdots \times I_M}$. Similarly, we define the spectral norm of $\mathcal{A}$ by
$\|\mathcal{A}\|^{\overline{\square}}=\sqrt{\|\mathcal{A}{\overline{\square}}\mathcal{A}\|^{\square}}$, and it has been shown by Lemma 4.2 and 4.3 in \cite{luo2020tensor} that $\|\mathcal{A}\|^{\overline{\square}} = \|D(\mathcal{A})\|^{\overline{\square}}= \|D(\mathcal{A})\|^{{\square}}$, where $D(\cdot)$ is the dilation operator. 

By applying Theorem 6.1.1 in \cite{tropp2015introduction} to $\overline{f}(\mathcal{A})$, where $\overline{f}(\cdot)$ is a bijective linear transformation:
$$
\overline{f}: \mathbb{R}^{I_1 \times \cdots \times I_M} \rightarrow \mathbb{R}^{\prod_{k=1}^m I_k \times \prod_{k=1}^{M-m} I_{k+m}}
$$
such that for any tensor $\mathcal{A} \in \mathbb{R}^{I_1 \times \cdots \times I_M}$,
$$
(\bar{f}(\mathcal{A}))_{i j}=a_{i_1, \cdots, i_m, j_1, \cdots, j_{M-m}},
$$
where
$$
i=i_1+\sum_{k=2}^m\left(\left(i_k-1\right) \prod_{l=1}^{k-1} I_l\right), \quad j=j_1+\sum_{k=2}^{M-m}\left(\left(j_k-1\right) \prod_{l=1}^{k-1} I_{l+m}\right),
$$
the results of Lemma \ref{lem:(Tensor-Bernstein-Inequality)} follow. 
\end{proof}

\subsubsection{Proof of Lemma \ref{lem:restricted_convexity}}
\begin{proof}
We aim to prove that
\[
\sum_{\Omega_{i,t,l}=1}\Delta_{i,t,l}^{2}\geq\frac{p_{\min}}{2}\sum_{i,t,l}\Delta_{i,t,l}^{2}-2\|\Delta\|_{\max}^{2}\vartheta
\]
with high probability. Let for any $\Delta\in\mathfrak{C}(r_{1},r_{2},r_{3},\mu_{0},L_{0})$:
\[
\|\Delta\|_{L^{2}(\Omega)}^{2}\geq p_{\min}\|\Delta\|_{F}^{2}\geq L_{0}^{2}\theta,\quad\|\Delta\|_{\max}\leq2L_{0},
\]
where the first inequality is shown in Lemma \ref{lem:bound-P(A)}
and the second inequality by definition. By Lemma \ref{lem:Delta-Inequality},
we have for small enough $C_{4}$ (e.g., $C_{4}=0.005$)
\begin{align*}
 & \mathbb{P}_{\mathcal{A}}\left\{ \frac{p_{\min}}{2}\sum_{i,t,l}(\mathcal{Y}_{i,t,l}^{*}-\widehat{\mathcal{Y}}_{i,t,l})^{2}\geq\sum_{\Omega_{i,t,l}=1}(\mathcal{Y}_{i,t,l}^{*}-\widehat{\mathcal{Y}}_{i,t,l})^{2}+2\|\Delta\|_{\max}^{2}\vartheta\right\} \\
 & =\mathbb{P}_{\mathcal{A}}\left\{ \frac{p_{\min}}{2}\|\Delta\|_{F}^{2}\geq\|P_{\Omega}(\Delta)\|_{F}^{2}+2\|\Delta\|_{\max}^{2}\vartheta\right\} \\
 & \leq \mathbb{P}_{\mathcal{A}}\left\{ \frac{1}{2}\|\frac{\Delta}{L_{0}}\|_{L^{2}(\Omega)}^{2}\geq\|P_{\Omega}\frac{\Delta}{L_{0}}\|_{F}^{2}+2\|\frac{\Delta}{L_0}\|_{\max}^{2}\vartheta\right\} \\
 & \leq2\exp\left(-\frac{C_{4}\theta}{T}\right)\overset{\theta=C'T\log(N+T+K)}{\leq}\frac{1}{(N+T+K)^{2}},
\end{align*}
which completes the proof.
\end{proof}
\subsubsection{Proof of Lemma \ref{lem:nuclear_norm}}
\begin{lemma}
\label{lem:nuclear_norm} Let $\mathcal{X}\in\mathbb{R}^{p_{1}\times p_{2}\times p_{3}}$,
we have 
\[
\|\mathcal{X}\|_{*}=\left\{ \min\sum_{i=1}^{r}|\lambda_{i}|:\mathcal{X}=\sum_{i=1}^{r}\lambda_{i}\boldsymbol{a}_{i}\otimes\boldsymbol{b}_{i}\otimes\boldsymbol{c}_{i},\|\boldsymbol{a}_{i}\|=\|\boldsymbol{b}_{i}\|=\|\boldsymbol{c}_{i}\|=1,i=1,\cdots,r\right\} .
\]
\end{lemma}
\begin{proof}
see \citep{hogben2013handbook} for proofs. Note that the nuclear
matrix norm has a similar characterization as well.
\end{proof}

\subsubsection{Proof of Lemma \ref{lem:duality_tensor_nuclear}}
\begin{lemma}    \label{lem:duality_tensor_nuclear}
Let $\mathcal{X}\in\mathbb{R}^{p_{1}\times p_{2}\times p_{3}}$ and $\mathcal{Y}\in\mathbb{R}^{p_{1}\times p_{2}\times p_{3}}$, then, $|\langle\mathcal{X},\mathcal{Y}\rangle|\leq \|\mathcal{X}\|\|\mathcal{Y}\|_*$.
\end{lemma}
\begin{proof}
By Lemma \ref{lem:nuclear_norm}, we know that 
$
\mathcal{Y}=\sum_{i=1}^r 
\lambda_i \boldsymbol{a}_{i}\otimes\boldsymbol{b}_{i}\otimes\boldsymbol{c}_{i}
$ with $\|\boldsymbol{a}_{i}\|=\|\boldsymbol{b}_{i}\|=\|\boldsymbol{c}_{i}\|=1$ and $\|\mathcal{Y}\|_*=\sum_{i=1}^r \lambda_i$. Therefore, we have
\begin{align*}
|\langle\mathcal{X},\mathcal{Y}\rangle|
=|
\langle\mathcal{X},
\sum_{i=1}^r 
\lambda_i \boldsymbol{a}_{i}\otimes\boldsymbol{b}_{i}\otimes\boldsymbol{c}_{i}\rangle
|
\leq 
\sum_{i=1}^r |\lambda_r| 
|\langle 
\mathcal{X},
\boldsymbol{a}_{i}\otimes\boldsymbol{b}_{i}\otimes\boldsymbol{c}_{i}
\rangle|
\leq \|\mathcal{X}\|\|\mathcal{Y}\|_*.
\end{align*}
\end{proof}

\subsubsection{Proof of Lemma \ref{lem:tensor_nuclear_norm}}
\begin{lemma}
\label{lem:tensor_nuclear_norm} Let $\mathcal{X}\in\mathfrak{C}(r_{1},r_{2},r_{3},L_{0})$,
then 
\[
\|\mathcal{X}\|_{*}\leq\sqrt{\frac{(r_{1}\wedge d_{\Phi})r_{2}r_{3}}{\max\{(r_{1}\wedge d_{\Phi}),r_{2},r_{3})}}\|\mathcal{X}\|_{F}.
\]
\end{lemma}
\begin{proof}
Similar to the proof of Lemma 2 in \cite{wang2020learning}. Let $\boldsymbol{e}_{N},\boldsymbol{e}_{T}$, and $\boldsymbol{e}_{K}$
be the standard orthonormal basis of the three modes of $\mathcal{X}$.
Then 
\begin{align*}
\mathcal{X} & =\sum_{i\in[N],t\in[T]}\lambda_{K,i,t}^{*}\otimes\boldsymbol{e}_{N,i}\otimes\boldsymbol{e}_{T,t}\\
 & =\sum_{i\in\mathcal{I}_{1},t\in\mathcal{I}_{2}}\lambda_{K,i,t}\otimes\boldsymbol{e}_{N,i}\otimes\boldsymbol{e}_{T,t},
\end{align*}
where $\lambda_{K,i,t}^{*}$ and $\lambda_{K,i,t}$ are two third-mode
vectors of $\mathcal{X}$, the second inequality is based on $\mathcal{X}\in\mathfrak{C}(r_{1},r_{2},r_{3},\mu_{0},L_{0})$
and $(\mathcal{I}_{1},\mathcal{I}_{2})$ are two appropriate indices
of first and second modes with $|\mathcal{I}_{1}|=d_{\Phi}\wedge r_{1}$
and $|\mathcal{I}_{2}|=r_{2}$. Then, we have 
\begin{align*}
\|\mathcal{X}\|_{F}^{2} & =\sum_{i\in\mathcal{I}_{1},t\in\mathcal{I}_{2}}\langle\lambda_{K,i,t}\otimes\boldsymbol{e}_{N,i}\otimes\boldsymbol{e}_{T,t},\lambda_{K,i,t}\otimes\boldsymbol{e}_{N,i}\otimes\boldsymbol{e}_{T,t}\rangle=\sum_{i\in\mathcal{I}_{1},t\in\mathcal{I}_{2}}\|\lambda_{K,i,t}\|^{2},\\
\|\mathcal{X}\|_{*} & =\sum_{i\in\mathcal{I}_{1},t\in\mathcal{I}_{2}}\|\lambda_{K,i,t}\|.
\end{align*}
By Cauchy-Schwarz inequality, we have 
\[
\|\mathcal{X}\|_{*}=\sum_{i\in\mathcal{I}_{1},t\in\mathcal{I}_{2}}\|\lambda_{K,i,t}\|\leq\sqrt{(d_{\Phi}\wedge r_{1})r_{2}\sum_{i\in\mathcal{I}_{1},n\in\mathcal{I}_{2}}\|\lambda_{K,i,t}\|^{2}}=\sqrt{(d_{\Phi}\wedge r_{1})r_{2}}\|\mathcal{X}\|_{F}.
\]
Similar inequalities hold for $\lambda_{N,t,l}$ and $\lambda_{T,i,l}$
as the first and second mode vector of $\mathcal{X}$. Thus, we have
\[
\|\mathcal{X}\|_{*}\leq\sqrt{(d_{\Phi}\wedge r_{1})r_{3}}\|\mathcal{X}\|_{F},\quad\|\mathcal{X}\|_{*}\leq\sqrt{r_{2}r_{3}}\|\mathcal{X}\|_{F}.
\]
To sum up, the upper bound of the nuclear norm of a tensor $\mathcal{X}\in\mathfrak{C}(r_{1},r_{2},r_{3},L_{0})$
is 
\[
\|\mathcal{X}\|_{*}\leq\sqrt{\frac{(r_{1}\wedge d_{\Phi})r_{2}r_{3}}{\max\{(r_{1}\wedge d_{\Phi}),r_{2},r_{3}\}}}\|\mathcal{X}\|_{F},
\]
which completes the proof.
\end{proof}

\subsubsection{Proof of Lemma \ref{lem:G_nuclear_norm}}
\begin{lemma}
\label{lem:G_nuclear_norm} For any core tensor $\mathcal{G}\in\mathbb{R}^{r_{1}\times r_{2}\times_{3}}$
and $\mathcal{M}_{1}(\mathcal{G})=\sum_{i=1}^{r_{1}}\sigma_{i}\boldsymbol{x}_{i}\otimes\boldsymbol{z}_{i}$
being its singular value decomposition, we have 
\[
\|\mathcal{G}\|_{*}\leq r_{1}\sqrt{r_{2}\wedge r_{3}}\|\mathcal{M}_{1}(\mathcal{G})\|.
\]
Similar results hold for matrix flattening in modes $2$ and $3$.
The inequality becomes equality when $r_{1}=1$.
\end{lemma}
Let $Z_{i}\in\mathbb{R}^{r_{2}\times r_{3}}$ be the matrix reformulation
of $\boldsymbol{z}_{i}\in\mathbb{R}^{r_{2}r_{3}}$, which has the
singular value decomposition as 
\[
Z_{i}=\sum_{j=1}^{r_{2}\wedge r_{3}}\mu_{i,j}\boldsymbol{v}_{i,j}\otimes\boldsymbol{w}_{i,j},\quad i=1,\cdots,r_{1}.
\]
Since $\mathcal{M}_{1}(\mathcal{G})=\sum_{i=1}^{r_{1}}\sigma_{i}\boldsymbol{x}_{i}\otimes\boldsymbol{z}_{i}$,
under the isomorphism with the matricization in mode $1$ (Lemma 3.1
in \citet{hu2015relations}), we have 
\[
\mathcal{G}=\sum_{i=1}^{r_{1}}\sigma_{i}\boldsymbol{x}_{i}\otimes\left\{ \sum_{j=1}^{r_{2}\wedge r_{3}}\mu_{i,j}\boldsymbol{v}_{i,j}\otimes\boldsymbol{w}_{i,j}\right\} .
\]
Then, followed by the definition of nuclear norm, we have 
\begin{align*}
\|\mathcal{G}\|_{*} & \leq\sum_{i=1}^{r_{1}}\sigma_{i}\sum_{j=1}^{r_{2}\wedge r_{3}}\mu_{i,j}\leq\left(r_{1}\max_{i}\sigma_{i}\right)\left(\max_{i}\sum_{j=1}^{r_{2}\wedge r_{3}}\mu_{i,j}\right)\\
 & \leq r_{1}\|\mathcal{M}_{1}(\mathcal{G})\|\|\vee Z_{i}\|_{*}\leq r_{1}\sqrt{r_{2}\wedge r_{3}}\|\mathcal{M}_{1}(\mathcal{G})\|,
\end{align*}
where $\|\vee Z_{i}\|_{*}=\max\{\|Z_{i}\|_{*}:1\leq i\leq r_{1}\}$.
The last inequality is based on Lemma \ref{lem:nuclear_F_matrix}
and $1=\boldsymbol{z}_{i}^{\intercal}\boldsymbol{z}_{i}=\text{trace}(Z_{i}^{\intercal}Z_{i})=\|Z_{i}\|_{F}^{2}$
for any $i=1,\cdots,r_{1}$.

\subsubsection{Proof of Lemma \ref{lem:nuclear_F_matrix}}

\begin{lemma}
\label{lem:nuclear_F_matrix} Let $Z\in\mathbb{R}^{r_{2}\times r_{3}}$,
we have $\|Z\|_{*}\leq\sqrt{r_{2}\wedge r_{3}}\|Z\|_{F}$.
\end{lemma}
Let $Z=\sum_{i=1}^{r_{2}\wedge r_{3}}\sigma_{i}\boldsymbol{u}_{i}\otimes\boldsymbol{v}_{i}$
be the singular value decomposition of $Z$, we know that 
\[
\|Z\|_{*}=\sum_{i=1}^{r_{2}\wedge r_{3}}\sigma_{i},\quad\|Z\|_{F}=\sqrt{\sum_{i=1}^{r_{2}\wedge r_{3}}\sigma_{i}^{2}}.
\]
By Cauchy-Schwarz inequality, we have 
\[
\frac{\|Z\|_{*}}{r_{2}\wedge r_{3}}\leq\sqrt{\frac{\|Z\|_{F}^{2}}{r_{2}\wedge r_{3}}},
\]
which completes the proof after rearrangement.

\subsubsection{Proof of Lemma \ref{lem:bound-P(A)}}
\begin{lemma}
\label{lem:bound-P(A)}
$p_{\min}<\mathbb{P}(A_{i,(t-k+1):t}=a_{(t-k+1):t}\mid H_{i,t}))\leq p_{\max}$ and $p_{\min}<\mathbb{P}(A_{i,(t-k+1):t}=a_{(t-k+1):t})\leq p_{\max}$, where $p_{\min}=\delta^k$ and $p_{\max}=(1-\delta)^k$.
\end{lemma}
\begin{proof}
The first statement is straightforward under Assumption \ref{assum:positivity}. For the second one, we have by the law of expectation
\begin{align*}
\mathbb{P}(A_{i,t-k+1:t}=a) & =\E\{\mathbb{P}(A_{i,t-k+1:t}=a\mid H_{i,t})\}\\
 & =\int_{H_{i,t}}\mathbb{P}(A_{i,t-k+1:t}=a\mid H_{i,t})dF(H_{i,t}),
\end{align*}
where $F(H_{i,t})$ is the joint cumulative probability
function of $H_{i,t}$. Under the positivity assumption
in Assumption \ref{assum:positivity} with constant $\delta$, we
have $\delta^{k}<\mathbb{P}(A_{i,t-k+1:t}=a\mid H_{i,t})<(1-\delta)^{k}$
for any $a\in\mathbb{A}$ and the desired bound follows.
\end{proof}

\subsubsection{Proof of Lemma \ref{lem:Delta-Inequality}}
\begin{lemma}
\label{lem:Delta-Inequality} Define the constraint set
\[
\mathfrak{B}(\theta,\tau)=\{\Delta\in\mathbb{R}^{N\times T\times K}:\|\Delta\|_{L^{2}(\Omega)}^{2}\geq\theta,\|\Delta\|_{*}\leq\sqrt{\tau}\|\Delta\|_{F},\|\Delta\|_{\max}\leq1\}
\]
Define 
\[
\vartheta=\frac{(r_{1}\wedge d_{\Phi})r_{2}r_{3}}{\max\{(r_{1}\wedge d_{\Phi}),r_{2},r_{3}\}p_{\min}}k^2(N\vee T)\log(N+T+K).
\]
Whenever $\Delta\in\mathfrak{B}(\theta,\tau)$ and there exist small
enough constant $C_{4}$ (e.g., $C_{4}=0.005$) s.t. $C_{4}\theta>T$
\[
\mathbb{P}_{\mathcal{A}}\left\{ \frac{1}{2}\|\Delta\|_{L^{2}(\Omega)}^{2}\geq\|P_{\Omega}\Delta\|_{F}^{2}+2\vartheta\right\} \leq2\exp\left(-\frac{C_{4}\theta}{T}\right).
\]
\end{lemma}
\begin{proof}
Our proof is based on the standard peeling argument. Let $\xi$ be
a constant larger than $1$ (e.g., $\xi=2$) and define for every
$\rho\geq0$
\[
\mathfrak{B}_{\rho}(\theta,\tau)=\{\Delta\in\mathfrak{B}(\theta,\tau):\rho\leq\|\Delta\|_{L^{2}(\Omega)}^{2}\leq\rho\xi\}.
\]
And we have $\mathfrak{B}(\theta,\tau)=\bigcup_{l=1}^{\infty}\mathfrak{B}_{\theta\xi^{l-1}}(\theta,\tau)$
, therefore, for some $l\geq1$ and $\Delta\in\mathfrak{B}_{\theta\xi^{l-1}}(\theta,\tau)$,
we have
\[
\|\Delta\|_{L^{2}(\Omega)}^{2}-\|P_{\Omega}\Delta\|_{F}^{2}\geq\frac{1}{2}\|\Delta\|_{L^{2}(\Omega)}^{2}+2\vartheta\geq\frac{1}{2\xi}\theta\xi^{l}+2\vartheta.
\]
Define this as the bad event 
\[
\mathcal{E}(\Omega)_{l}=\left\{ \exists\Delta\in\mathfrak{B}_{\theta\xi^{l-1}}(\theta,\tau)\mid\left|\|P_{\Omega}\Delta\|_{F}^{2}-\|\Delta\|_{L^{2}(\Omega)}^{2}\right|\geq\frac{1}{2\xi}\theta\xi^{l}+2\vartheta\right\} ,
\]
and $\mathcal{E}(\Omega)=\left\{ \exists\Delta\in\mathfrak{B}(\theta,\tau)\mid\|\Delta\|_{L^{2}(\Omega)}^{2}-\|P_{\Omega}\Delta\|_{F}^{2}\geq\frac{1}{2}\|\Delta\|_{L^{2}(\Omega)}^{2}+2\vartheta\right\} \subset\bigcup_{l=1}^{\infty}\mathcal{E}(\Omega)_{l}$,
that is, if $\mathcal{E}(\Omega)$ holds for some tensor $\Delta$,
then this $\Delta$ must belong to $\mathfrak{B}_{\theta\xi^{l-1}}(\theta,\tau)$
for some $l$. Hence, by Lemma \ref{lem:Rademacher}, we have
\[
\mathbb{P}_{\mathcal{A}}\{\mathcal{E}(\Omega)_{l}\}\leq\exp\left(-\frac{C_{3}\xi^{l}\theta}{T}\right)\leq\exp\left(-\frac{C_{3}l\log(\xi)\theta}{T}\right)\leq\exp\left(-\frac{0.5C_{3}l\theta}{T}\right),
\]
since $\xi^{l}\geq l\log(\xi)$ for $\xi=2$. Then, let $C_{4}=0.5C_{3}$
and by union bound, we have
\[
\mathbb{P}_{\mathcal{A}}\{\mathcal{E}(\Omega)\}\leq\sum_{l=1}^{\infty}\left\{ \exp\left(-\frac{C_{4}\theta}{T}\right)\right\} ^{l}=\frac{\exp\left(-\frac{C_{4}\theta}{T}\right)}{1-\exp\left(-\frac{C_{4}\theta}{T}\right)}\leq2\exp\left(-\frac{C_{4}\theta}{T}\right),
\]
if $C_{4}\theta>T$ and 
\[
\mathbb{P}_{\mathcal{A}}\left\{ \frac{1}{2}\|\Delta\|_{L^{2}(\Omega)}^{2}\geq\|P_{\Omega}\Delta\|_{F}^{2}+2\vartheta\right\} \leq \mathbb{P}_{\mathcal{A}}(\mathcal{B})\leq2\exp\left(-\frac{C_{4}\theta}{T}\right),
\]
which completes the proof.
\end{proof}

\subsubsection{Proof of Lemma \ref{lem:Rademacher}}
\begin{lemma}
\label{lem:Rademacher} Let
\[
\tilde{\mathcal{Z}}_{\rho}=\sup_{\Delta\in\mathfrak{B}_{\rho}(\theta,\tau)}\left|\|P_{\Omega}\Delta\|_{F}^{2}-\|\Delta\|_{L^{2}(\Omega)}^{2}\right|=\sup_{\Delta\in\mathfrak{B}_{\rho}(\theta,\tau)}\left|\sum_{i=1}^{N}\|\Delta_{\Omega}^{(i)}\|_{F}^{2}-\|\Delta\|_{L^{2}(\Omega)}^{2}\right|.
\]
where $\Delta_{\Omega}^{(i)}=\sum_{t=1}^{T}\Delta_{i,t,(A_{i,(t-k+1):t})_{(10)}}\cdot e_{i}(N)\otimes e_{j}(T)\otimes e_{(A_{i,(t-k+1):t})_{(10)}}(K)\in\mathbb{R}^{N\times T\times K}$.
Then, there exists a small enough constant $C$ (e.g., $C=0.01$) s.t.
\[
\mathbb{P}_{\mathcal{A}}(\tilde{\mathcal{Z}}_{\rho}\geq\frac{1}{2\xi}\rho\xi+2\vartheta)\leq\exp\left(-\frac{C\rho\xi}{T}\right),
\]
for any $\Delta\in\mathfrak{B}_{\rho}(\theta,\tau)=\{\Delta\in\mathfrak{B}(\theta,\tau):\rho\leq\|\Delta\|_{L^{2}(\Omega)}^{2}\leq\rho\xi\}$.
\end{lemma}
\begin{proof}
We know
\begin{align*}
\sigma^{2}=\sup_{\Delta\in\mathfrak{B}_{\rho}(\theta,\tau)}\sum_{i=1}^{N}\text{var}(\|\Delta_{\Omega}^{(i)}\|_{F}^{2}) & \leq\sup_{\Delta\in\mathfrak{B}_{\rho}(\theta,\tau)}\sum_{i=1}^{N}\mathbb{E}(\|\Delta_{\Omega}^{(i)}\|_{F}^{4})\\
 & \leq\sup_{\Delta\in\mathfrak{B}_{\rho}(\theta,\tau)}\sum_{i=1}^{N}\left\{ \mathbb{E}(\|\Delta_{\Omega}^{(i)}\|_{F}^{2})\cdot\mathbb{E}(\|\Delta_{\Omega}^{(i)}\|_{F}^{2})\right\} \\
 & \leq T\sup_{\Delta\in\mathfrak{B}_{\rho}(\theta,\tau)}\sum_{i=1}^{N}\mathbb{E}(\|\Delta_{\Omega}^{(i)}\|_{F}^{2})\\
 & =T\sup_{\Delta\in\mathfrak{B}_{\rho}(\theta,\tau)}\|\Delta\|_{L^{2}(\Omega)}^{2}\leq T\rho\xi,
\end{align*}
where the second inequality is by Cauchy–Schwarz inequality, the third
inequality is by $\|\Delta\|_{\max}\leq1,\|\Delta_{\Omega}^{(i)}\|_{F}^{2}\leq T$
as each subject must be observed once at each time, and the last inequality
is derived by $\|\Delta\|^2_{L^{2}(\Omega)}\leq\rho\xi$. By the symmetrization
argument in Lemma 6.3 of \citet{gigli2013log} and Lemma 14, \citet{klopp2014noisy}
\begin{align*}
\mathbb{E}_{\mathcal{A}}(\tilde{\mathcal{Z}}_{\rho}) & \leq2\mathbb{E}_{\mathcal{A},\zeta}\left\{ \sup_{\Delta\in\mathfrak{B}_{\rho}(\theta,\tau)}\left|\sum_{i=1}^{N}\zeta_{i}\|\Delta_{\Omega}^{(i)}\|_{F}^{2}\right|\right\} \\
 & =2\mathbb{E}_{\mathcal{A},\zeta}\left\{ \sup_{\Delta\in\mathfrak{B}_{\rho}(\theta,\tau)}\left|\langle\sum_{i=1}^{N}\zeta_{i}\Delta_{\Omega}^{(i)},\Delta\rangle\right|\right\} \\
 & \leq2\mathbb{E}_{\mathcal{A},\zeta}\left\{ \sup_{\Delta\in\mathfrak{B}_{\rho}(\theta,\tau)}\|\sum_{i=1}^{N}\zeta_{i}\Delta_{\Omega}^{(i)}\|\|\Delta\|_{*}\right\} \\
 & \leq4\sqrt{\frac{(r_{1}\wedge d_{\Phi})r_{2}r_{3}}{\max\{(r_{1}\wedge d_{\Phi}),r_{2},r_{3}\}}}\mathbb{E}_{\mathcal{A},\zeta}\left\{ \sup_{\Delta\in\mathfrak{B}_{\rho}(\theta,\tau)}\|\sum_{i=1}^{N}\zeta_{i}\Delta_{\Omega}^{(i)}\|\|\Delta\|_{F}\right\} \\
 & \leq4\sqrt{\frac{(r_{1}\wedge d_{\Phi})r_{2}r_{3}}{\max\{(r_{1}\wedge d_{\Phi}),r_{2},r_{3}\}p_{\min}}}\mathbb{E}_{\mathcal{A},\zeta}\left\{ \sup_{\Delta\in\mathfrak{B}_{\rho}(\theta,\tau)}\|\sum_{i=1}^{N}\zeta_{i}\Delta_{\Omega}^{(i)}\|\|\Delta\|_{L^{2}(\Omega)}\right\} \\
 & \leq4\sqrt{\frac{(r_{1}\wedge d_{\Phi})r_{2}r_{3}\rho\xi}{\max\{(r_{1}\wedge d_{\Phi}),r_{2},r_{3}\}p_{\min}}}\mathbb{E}_{\mathcal{A},\zeta}\left\{ \sup_{\Delta\in\mathfrak{B}_{\rho}(\theta,\tau)}\|\Delta_{\Omega}^{\text{Rad}}\|\right\} \\
 & \leq C_{1}\rho\xi+\frac{(r_{1}\wedge d_{\Phi})r_{2}r_{3}U_{\text{Rad}}^{2}}{C_{1}\max\{(r_{1}\wedge d_{\Phi}),r_{2},r_{3}\}p_{\min}},
\end{align*}
where $\{\zeta_{i}\}_{i=1}^{N}$ are i.i.d. Rademacher random variables,
$U_{\text{Rad}}=\mathbb{E}_{\mathcal{A},\zeta}\left\{ \sup_{\Delta\in\mathfrak{B}_{\rho}(\theta,\tau)}\|\Delta_{\Omega}^{\text{Rad}}\|\right\} $
is the Rademacher complexity, and $\Delta_{\Omega}^{\text{Rad}}=\sum_{i=1}^{N}\zeta_{i}\Delta_{\Omega}^{(i)}$, with its $(i,t,l)$-entry as $\zeta_{i}\Delta_{i,t,l}$ if $\Omega_{i,t,l}=1$
and zero otherwise. By Lemma \ref{lem:Rademacher-complexity}, we
invoke the Theorem 3 in \citet{massart2000constants} with $\varepsilon=1$:
\[
\mathbb{P}_{\mathcal{A}}\left\{ \tilde{\mathcal{Z}}_{\rho}\geq2\mathbb{E}_{\mathcal{A}}(\tilde{\mathcal{Z}}_{\rho})+2\sigma\sqrt{x}+34.5Tx\right\} \leq\exp(-x),
\]
where $\sigma^{2}=\sup_{\Delta\in\mathcal{S}_{\theta,\tau,\gamma}(\rho)}\sum_{i=1}^{N}\text{var}(\|\Delta_{\Omega}^{(i)}\|_{F}^{2})\leq T\rho\xi$.
Let $x=C_{3}\rho\xi/T$ with small constant $C_{3},C_{4}$,
\begin{align*}
 & \mathbb{P}_{\mathcal{A}}\left\{ \tilde{\mathcal{Z}}_{\rho}\geq\frac{1}{2\xi}\rho\xi+2\vartheta\right\} \\
 & = \mathbb{P}_{\mathcal{A}}\left\{ \tilde{\mathcal{Z}}_{\rho}\geq\frac{1}{2\times2}\rho\xi+\frac{2(r_{1}\wedge d_{\Phi})r_{2}r_{3}}{\max\{(r_{1}\wedge d_{\Phi}),r_{2},r_{3}\}p_{\min}}k^2(N\vee T)\log(N+T+K)\right\} \\
 & \leq \mathbb{P}_{\mathcal{A}}\left\{ \tilde{\mathcal{Z}}_{\rho}\geq2\sqrt{C_{3}}\rho\xi+34.5C_{3}\rho\xi+2C_{4}\rho\xi+2\frac{(r_{1}\wedge d_{\Phi})r_{2}r_{3}}{\max\{(r_{1}\wedge d_{\Phi}),r_{2},r_{3}\}p_{\min}}U_{\text{Rad}}^{2}\right\} \\
 & \leq \mathbb{P}_{\mathcal{A}}\left\{ \tilde{\mathcal{Z}}_{\rho}\geq2\sqrt{TK\rho\xi}\sqrt{C_{3}\rho\xi/T}+34.5T\cdot C_{3}\rho\xi/T+2\mathbb{E}_{\mathcal{A}}(\tilde{\mathcal{Z}}_{\rho})\right\} \\
 & \leq \mathbb{P}_{\mathcal{A}}\left\{ \tilde{\mathcal{Z}}_{\rho}\geq2\sigma\sqrt{x}+34.5Tx+2\mathbb{E}_{\mathcal{A}}(\tilde{\mathcal{Z}}_{\rho})\right\} \\
 & \leq\exp\left(-\frac{C_{3}\rho\xi}{T}\right),
\end{align*}
which completed the proof. In order to get a bound in close form,
we need to obtain a suitable upper bound on $U_{\text{Rad}}=\mathbb{E}\left\{ \sup_{\Delta\in\mathfrak{B}_{\rho}(\theta,\tau)}\|\Delta_{\Omega}^{\text{Rad}}\|\right\} $,
which can be obtained in the case of sub-Gaussian noise. Most of the subsequent proof is adapted from Lemma 5 and 6, \citet{klopp2014noisy}. 
\end{proof}

\subsubsection{Proof of Lemma \ref{lem:Rademacher-complexity}}
\begin{lemma}
\label{lem:Rademacher-complexity}
For any $\Delta\in\mathfrak{B}_{\rho}(\theta,\tau)$,
we have
\[
\mathbb{E}\left\{ \sup_{\Delta\in\mathfrak{B}_{\rho}(\theta,\tau)}\|\Delta_{\Omega}^{\text{Rad}}\|\right\} \leq C_{5}k\sqrt{(N\vee T)}\|\Delta\|_{\max}\{\log(N+T+K)\}^{1/2},
\]
for large enough constant $C_{5}.$
\end{lemma}
\begin{proof}
For any $\Delta\in\mathfrak{B}_{\rho}(\theta,\tau)$, we separate the $\Delta_{\Omega}^{\text{Rad}}$ by 
$\Delta_{\Omega}^{\text{Rad}} = \sum_{h=0}^{k-1}\zeta_{i}\Delta_{\Omega}^{(i),h}$, where
\begin{align*}
\|\sum_{i=1}^{N}\E\{D(\zeta_{i}\Delta_{\Omega}^{(i),h})\overline{\square}D(\zeta_{i}\Delta_{\Omega}^{(i),h})\}\| & \leq\max_{i}\sum_{t,l}\Delta_{i,t,l_{t}}^{2}\lesssim T\|\Delta\|_{\max}^{2}\log(N+T+K),\\
\|\sum_{i=1}^{N}\E\{D(\zeta_{i}\Delta_{\Omega}^{(i),h})\underline{\square}D(\zeta_{i}\Delta_{\Omega}^{(i),h})\}\| & \leq\max_{t,l}\sum_{i}\Delta_{i,t,l_{t}}^{2}\lesssim N\|\Delta\|_{\max}^{2}\log(N+T+K),
\end{align*}
with probability greater than $1-(N+T+K)^{-2}$ under the restriction for $\delta$. For each $D(\zeta_{i}\Delta_{\Omega}^{(i)})$,
its spectral norm is 
\begin{align*}
\max_i\|D(\zeta_{i}\Delta_{\Omega}^{(i),h})\| \leq\max_i\max_{\|a\|=\|b\|=\|c\|}\langle D(\zeta_{i}\Delta_{\Omega}^{(i),h}),a\otimes b\otimes c\rangle\leq k^{-1/2}T^{1/2}\|\Delta\|_{\max}\cdot \log(N).
\end{align*}
Let $Z_{i}=\zeta_{i}\Delta_{\Omega}^{(i),h}$ and we know that $\E(Z_{i})=0$,
$\max_i\|Z_{i}\|\leq k^{-1/2}T^{1/2}\|\Delta\|_{\max}\log(N+T+K)$ for all $i\in[D]$, and $\sigma_{Z}^{2}\leq(N\vee T)\|\Delta\|_{\max}^{2}\log(N+T+K)$.
Then for any $\alpha\geq0$, we have
\begin{align*}
\mathbb{P}(\|D(\Delta_{\Omega}^{\text{Rad},h})\|\geq\alpha) & \leq(N+T+K)\exp\left\{ \frac{-\alpha^{2}}{2\sigma_{Z}^{2}+(2k^{-1/2}T^{1/2}\|\Delta\|_{\max}\log(N)\alpha)/3}\right\} .\\
 & \leq(N+T+K)\exp\left[-\frac{3}{4}\min\left\{ \frac{\alpha^{2}}{\sigma_{Z}^{2}},\frac{\alpha}{k^{-1/2}T^{1/2}\|\Delta\|_{\max}\log(N)}\right\} \right]\\
 & \leq(N+T+K)\exp\left(-\frac{3}{4}\frac{\alpha^{2}}{(N\vee T)\|\Delta\|_{\max}^{2}\log(N+T+K)}\right).
\end{align*}
By H{\"o}lder's inequality, we get for any $\Delta\in\mathfrak{B}_{\rho}(\theta,\tau)$
and large enough constant $C_{5}$,
\begin{align*}
 & \mathbb{E}\|D(\Delta_{\Omega}^{\text{Rad},h})\|\\
 & \leq\left\{ \mathbb{E}\|D(\Delta_{\Omega}^{\text{Rad}})\|^{2\log(N+T+K)}\right\} ^{1/\{2\log(N+T+K)\}}\\
 & \leq(N+T+K)^{1/\{2\log(N+T+K)\}}\log(N+T+K)\nu_{1}^{-1/2}2^{1/\{2\log(N+T+K)\}-1/2}\\
 & \leq\sqrt{2ek^{-1}(N\vee T)}\|\Delta\|_{\max}\log(N+T+K),
\end{align*}
where 
\begin{align*}
 & \mathbb{E}\|D(\Delta_{\Omega}^{\text{Rad}},h)\|^{2\log(N+T+K)}\\
 & \leq\int_{0}^{+\infty}\mathbb{P}(\|D(\Delta_{\Omega}^{\text{Rad},h})\|>t^{1/\{2\log(N+T+K)\}})dt\\
 & \leq(N+T+K)\int_{0}^{+\infty}\exp\left\{-t^{1/\log(N+T+K)}\nu_{1}\right\}dt\\
 & \leq(N+T+K)\log(N+T+K)\nu_{1}^{-\log(N+T+K)}\Gamma\{\log(N+T+K)\}\\
 & \leq(N+T+K)\log(N+T+K)^{\log(N+T+K)}\nu_{1}^{-\log(N+T+K)}2^{1-\log(N+T+K)},
\end{align*}
where $\nu_{1}=1/\{(N\vee T)\|\Delta\|_{\max}^{2}\log(N+T+K)\}$, $$\int_0^{+\infty}\exp\left\{-t^{1/\log(N+T+K)}\nu_{1}\right\}dt = \log(N+T+K)\nu_{1}^{-\log(N+T+K)}\Gamma\{\log(N+T+K)\},$$
and $\Gamma(x)\leq(x/2)^{x-1}$ for $x\geq2$. Therefore, 
\begin{align*}
  \mathbb{E}\left\{ \sup_{\Delta\in\mathfrak{B}_{\rho}(\theta,\tau)}\|\Delta_{\Omega}^{\text{Rad}}\|\right\} &\leq\mathbb{E}\left\{ \sup_{\Delta\in\mathfrak{B}_{\rho}(\theta,\tau)}\sum_{h=0}^{k-1}\|D(\Delta_{\Omega}^{\text{Rad},h})\|\right\} \\
  &\leq C_{5}k\sqrt{(N\vee T)}\|\Delta\|_{\max}\{\log(N+T+K)\}^{1/2}  
\end{align*}
for large enough constant $C_{5}.$
\end{proof}

\end{document}